\documentclass[12pt]{mycuphd}
\usepackage[english]{babel}
\begin{document}

\newcommand{\be}{\begin{equation}}
\newcommand{\ee}{\end{equation}}
\newcommand{\bea}{\begin{eqnarray*}}
\newcommand{\bean}{\begin{eqnarray}}
\newcommand{\eea}{\end{eqnarray*}}
\newcommand{\eean}{\end{eqnarray}}
\newcommand{\beal}{\begin{large}\begin{eqnarray*}}
\newcommand{\eeal}{\end{eqnarray*}\end{large}}
\newcommand{\beanl}{\begin{large} \begin{eqnarray}} \newcommand{\eeanl}{\end{eqnarray}\end{large}}
\newcommand{\bel}{\begin{large}\begin{equation}}
\newcommand{\eel}{\end{equation}\end{large}}
\newcommand{\nn}{\nonumber}
\newcommand{\x}{{\rm x}}
\newcommand{\y}{{\rm y}}
\newcommand{\z}{{\rmz}}
\newcommand{\spos}{\sum_{n=1}^{+\infty}}
\newcommand{\snneg}{\sum_{n=0}^{+\infty}}
\newcommand{\sneg}{\sum_{n=-\infty}^{-1}}
\newcommand{\som}{\sum_{n\neq 0}}
\newcommand{\somma}{\sum_{n=-\infty}^{+\infty}}
\newcommand{\pl}{\frac{\pi}{L}}
\newcommand{\nor}{\frac{1}{\sqrt{2L}}}
\newcommand{\sq}{\sqrt{2}}
\newcommand{\half}{\frac{1}{2}}
\newcommand{\lines}{\doublespacing}
\newcommand{\arr}{\baselineskip20pt}
\newcommand{\0}{|0\rangle}
\newcommand{\1}{|\Omega^{(1)}\rangle}
\newcommand{\2}{|\Omega^{(2)}\rangle}
\newcommand{\om}{|\Omega^{(2)}\rangle}
\newcommand{\factor}{e^{-\frac{i\pi\sq}{g\sqrt{L}}b_0^3}}
\newcommand {\psir}{\psi_R (0,\x)}
\newcommand{\psil}{\psi_L (0,\x)}
\doublespacing

\pagenumbering{roman}
\pagestyle{empty}

\title{ Nonperturbative Effects from Perturbation Theory in Adjoint $QCD_{1+1}}
\author{Eliana Vianello}\date{7 August 1998}\renewcommand{\topfraction}{0.80}\renewcommand{\bottomfraction}{0.80}\renewcommand{\textfraction}{0.10}\renewcommand{\floatpagefraction}{0.70}

\newpage

\vskip2in
\setcounter{page}{1}


\begin{titlepage}       
 \title{ NONPERTURBATIVE EFFECTS FROM PERTURBATION\\   THEORY IN ADJOINT $QCD_{1+1}$}                               
\author{Eliana Vianello}       
\conferraldate{May 17,}{2003}        
\maketitle
\end{titlepage}
\conferraldate{May  17,}{2003}

\newpage
\pagestyle{empty}
\vspace*{0.0in}
\vskip 37pt
\begin{center}
{ {\normalsize  NONPERTURBATIVE EFFECTS FROM PERTURBATION \\THEORY  IN ADJOINT $QCD_{1+1}$} \par}
\vspace{2.5in}

\hbox to0in {\vbox to0in {
\makebox[5.0in][r]{\large }
\vskip .2in       \makebox[5.4in][r]{Approved by:}  \vskip 20pt       \makebox[5.4in][r]{\leaders\hrule\hskip 2.5in}       \vskip -.2in       \makebox[5.4in][l]{\hspace{2.85in} Prof. Gary McCartor}   \vskip 16pt       \makebox[5.4in][r]{\leaders\hrule\hskip 2.5in}       \vskip -.2in       \makebox[5.4in][l]{\hspace{2.85in} Prof. Kent Hornbostel} \vskip 16pt              \makebox[5.4in][r]{\leaders\hrule\hskip 2.5in}       \vskip -.2in       \makebox[5.4in][l]{\hspace{2.85in} Prof. Roberto Vega} \vskip 16pt       \makebox[5.4in][r]{\leaders\hrule\hskip 2.5in}       \vskip -.2in       \makebox[5.4in][l]{\hspace{2.85in} Prof. George Reddien}
 \vskip .2in
\makebox[5.0in][r]{ }}}
\vskip -9pt
\end{center}

\newpage

\baselineskip 24pt
\leftmargin 1.5in
\rightmargin1.0in

%

\newpage

\pagestyle{plain}
\hspace{-0.3in}
Vianello, Eliana \hfill \parbox[t]{3.2in}{\hfill M. S., University of Padua, 1997.
}
\vskip-20pt
\hspace{-0.3in}
\parbox[t]{7in}{\mbox{} \\
\underline{Nonperturbative Effects from Perturbation }
}\\
\underline{Theory in Adjoint $QCD_{1+1}$ }


Advisor:
 \hspace{0.1in} Professor Gary McCartor
\\Doctor of Philosophy degree conferred May 17, 2003 \\
Thesis completed April  2003 \vspace{0.5in}%

\begin{center}
ABSTRACT
\end {center}

$SU(2)$ gauge theory coupled to massless fermions in the adjoint representation  is quantized
in light-cone gauge
by imposing the equal-time canonical algebra.  The theory is defined on  a space-time cylinder with "twisted"  boundary conditions, periodic for one colour component (the diagonal 3- component) and  antiperiodic for the other two.  The focus of the study is on the non-trivial vacuum structure and
the fermion condensate. It is shown that the indefinite-metric quantization of free gauge bosons is not compatible with the residual gauge symmetry of the interacting theory. A suitable quantization of the unphysical modes of the gauge field is necessary in order to guarantee the consistency of the subsidiary condition and allow the quantum representation of the residual gauge symmetry of the classical Lagrangian: the 3-colour component of the gauge field must be quantized in a space with an indefinite metric while the other two components require a positive-definite metric. The contribution of the latter to the  free  Hamiltonian becomes highly pathological  in this representation, but a larger portion of the interacting Hamiltonian can be diagonalized, thus  allowing  perturbative calculations to be performed. The vacuum is evaluated through second order in perturbation theory and this result is used for an approximate determination of the fermion condensate.

\singlespacing

\renewcommand{\contentsname}{TABLE OF CONTENTS}
\tableofcontents
\pagestyle{plain}
\def\thechapter       {\arabic{chapter}}

\doublespacing

\newpage
\setcounter{page}{6}
\chapter*{NOTATIONS AND CONVENTIONS}
\begin{itemize}
%
\item $x$ denotes space-time cohordinates ; $\x^{1}\equiv \x$ and $\x^{0}\equiv t $
are the corresponding space and, respectively, time components.
\item Light-cone cohordinates are often used 
 :\bea&&\x^{-}=\frac{t-\x}{\sqrt{2}} \; \;\;\;\;\; ,\; \;\;\;\;\; \x^{+}=\frac{t+\x}{\sqrt{2}} \\ \\&&\partial_{-}=\frac{\partial}{\partial \x^{-}}=\frac{\partial_{0}-\partial_{1}}{\sqrt{2}}\;\;\;\; \; \;,\; \;\;\;\;\;\partial_{+}=\frac{\partial}{\partial \x^{+}}=\frac{\partial_{0}+\partial_{1}}{\sqrt{2}} \;.\eea  
\item Greek indeces are Lorentz indeces while Latin indeces are colour indeces.
To avoid confusion when the symbols $+$ and $-$ are used as both Lorentz and colour indeces, it may help to remember that colour indeces are always upper indeces.

\end{itemize} 


\newpage
\pagestyle{plain}
\vspace{0.125in}
\begin{acknowledgements}
\noindent
This work would not have been possible without the guidance of my adviser 
Gary McCartor. I am grateful to him and to everybody in the SMU 
Physics Department.

\end{acknowledgements}

\setcounter{page}{1}
\pagenumbering{arabic}

\renewcommand{\bibname}{REFERENCES}


\renewcommand{\l}{\leftmargin1.5in}
\renewcommand{\r}{\rightmargin1.0in}
\l
\r

%
%


\chapter{INTRODUCTION}
The vacuum of QCD is believed to contain a condensate of quark-antiquark pairs with opposite
helicities and a net chiral charge. This is a sign of spontaneous breaking of the chiral symmetry.
The complexity of realistic field theories like QCD makes it convenient to investigate simpler models
with analogous features. A great simplification can be achieved by considering theories in one space and one time dimensions. 
Because of the absence of transverse dimensions, there is no spin in two dimensions and the chiral nature of massless fermions refers to their direction of motion. 
The simplest model possessing a non-trivial vacuum structure and a chiral condensate (given
by the nonzero vacuum expectation value of the operator $\bar{\psi}\psi$) is the Schwinger model \cite{schwinger},
which is the two-dimensional version of QED with massless fermions, an exactly soluble abelian theory. 
It is interesting to consider the more complex case of a two-dimensional QCD model. 
It is known that in non-abelian theories in two dimensions  a complex vacuum structure, reminiscent of the $\theta$-vacuum of QCD, is present when the Fermi field transforms according to the adjoint representation of $SU(N)$
\cite{witten}\cite{paniak}\cite{smilga}. Nonvanishing chiral condensates have been found
for two-dimensional Yang-Mills theories with adjoint fermions \cite{kogan} \cite{lenz}
\cite{paper}.
We shall use the canonical Hamiltonian formalism to obtain a perturbative evaluation of the vacuum and the condensate for $SU(2)$ gauge theory coupled to massless fermions in the adjoint representation. 
The quantum Hamiltonian will be determined after quantizing  the fields in a Fock space
at $t=0$. The theory will be studied on the space interval $-L\leq \x \leq L$ with ``twisted" boundary conditions \cite{thooft} \cite{shifman}, periodic for one colour component (the diagonal 3-component) and  antiperiodic for the other two. Compactifying the space dimension leads to a discrete set of allowed momenta , which provides a regularization of the infrared divergences affecting two-dimensional theories.  
The gauge condition that will be used to quantize the theory is the  light-cone gauge  $A_- =0 $, belonging to the family of algebraic  gauges. These gauges have the important advantage  that in the path-integral quantization of Yang-Mills theories the Faddeev-Popov
determinant is independent of the gauge field and  is absorbed by the normalization factor in the
generating functional of Green's functions.  The decoupling of Faddeev-Popov ghosts brings about
significant simplifications in the formalism. 
Two different methods are commonly used to impose the light-cone
gauge condition. In the light-front procedure the coordinate $ x^+$ plays the role of time: the canonical commutation relations are imposed at equal $x^+$,  instead of equal time, and Heisenberg
fields evolve in $x^+$. The gauge condition is then compatible with Dirac's procedure \cite{dirac}  for the quantization of constrained systems. All the constraints involving  the canonical variables can be imposed in
a strong sense and the redundant degrees of freedom associated with gauge invariance can therefore be eliminated from the formalism. The main advantage of the light-front representation is that the vacuum is very simple and can be described completely, which makes it particularly suitable for the non-perturbative study of field theories. 
In this approach, however, the condensate turns out to be proportional to $1/L$ and therefore vanishes in the continuum limit \cite{paper}. 
We shall consider the equal-time quantization of the model. In order to apply Dirac's procedure in this case, it is necessary to impose the gauge constraint by means of a Lagrange-multiplier field \cite{bassetto}, which introduces unphysical degrees of freedom that have to be dealt with by requiring that the expectation value of the Lagrange multiplier vanishes for physical states.
The Lagrange multiplier evolves as an $x^-$-independent free field, which guarantees the stability
of the physical subspace. 
By solving constraint equations for the colour components of this field,
we shall express them  in terms of the other fields and determine the algebra they satisfy. This will
allow us to show that the three subsidiary conditions are not independent, and that we only need to require that the 3-colour component of the Lagrange multiplier have zero expectation value between physical states.
As a consequence of the introduction of spurious degrees of
freedom,  the non-zero modes of $A_ +$, which are not physical in
two dimensions, are present in the Hamiltonian and need to be
quantized. 
Once the condition $A_- =0$ is imposed, the Lagrangian
is still invariant under a class of gauge transformations that do
not affect $A_-$ . The quantization of the gauge field must allow
a quantum representation of these transformations. The existence
of an operator implementing the residual gauge symmetry of the
classical Lagrangian is closely related to the existence of a
non-trivial vacuum structure. 
We shall show that, in order to
guarantee both the consistency of the subsidiary condition and the
possibility to implement the residual symmetry, the 3-colour
component of the gauge field must be quantized in a Fock space
with indefinite metric, while for the other two components the
definition of creation and annihilation operators must be
compatible with a positive-definite metric.
 Although with an
indefinite metric the Hamiltonian for the free gauge theory cannot
be fully diagonalized, it does annihilate the Fock vacuum. But
when two colour components of the gauge field are quantized
according to the positive-metric scheme, their contribution to the
Hamiltonian is highly pathological, containing terms that alter
the number of particles.  Although unphysical in the pure gauge
theory, the modes causing this pathology participate in the
interaction with fermions. That raises the problem of finding a
vacuum state to perturb about. As we shall see, the pathology can
be cured by adding to the unperturbed Hamiltonian a term coming
from the gauge-invariant ultraviolet regularization of the fermion
currents and kinetic term. This piece is quadratic in the gauge
field and the coupling constant. It has the form of a mass term,
where the mass is $m= \frac{g}{\sqrt{\pi}}$ (the well-known mass
acquired by the gauge field in the Schwinger model). The same
procedure will be followed for the zero mode of $A^3$.
This procedure allows standard Hamiltonian perturbation theory to
be applied for an approximate evaluation of the physical vacuum,
starting from  a vacuum that has some non-perturbative information
built in (the ultraviolet renormalization giving the ``mass term"
for the gauge field is nonperturbative). 
It will be shown that the state obtained 
in this way satisfies the subsidiary condition, as must be required of the vacuum 
state for consistency reasons.
 Although the condensate
is believed to be a non-perturbative effect we shall manage to
treat it in a perturbative framework.
Chapter 2 reviews some basic  facts about the theory under study, the definition of bilinear products of Fermi fields and the bosonization of the fermion degrees of freedom.
The Lagrange multipliers and the issues concerning the quantization of $A_+$ are studied in Chapter 3.
In Chapter 4 the vacuum is evaluated through second order.
The calculation of the condensate is presented in Chapter 5. \\
In order to test the method on a model with a known solution, we started by performing perturbative
calculation for the Schwinger model  and found some interesting results. In spite of the fact that
the Hamiltonian for the free gauge theory does not have a complete set of eigenstates, perturbation theory \emph{can} be applied. Perturbing about the states with one gauge boson satisfying the subsidiary condition in free theory, which are the only eigenstates, gives rise to a bifurcation: two states with different energies can be obtained from one unperturbed state. Thus, the complete set of one-particle eigenstates of the interacting Hamiltonian can be obtained by perturbing about the incomplete set of one-particle eigenstates of the free Hamiltonian.
The calculations for the Schwinger model are reported in Appendix A. 
\newpage

\chapter{SU(2) GAUGE THEORY COUPLED TO ADJOINT FERMIONS}

\doublespacing
\section{\textbf{Basics.}}

The lagrangian density for the theory is\footnote{The notation used here is similar to that of ref.\cite{paper}.}
\[
{\cal L} =-\frac{1}{2}\rm{Tr}(F_{\mu\nu}F^{\mu\nu})+\rm{Tr}(\bar{\Psi}\gamma^{\mu}D_{\mu}\Psi)
\]
where  \ \
\[F_{\mu\nu}=\partial_{\mu}A_{\nu}-\partial_{\nu}A_{\mu} + ig[A_{\mu},A_{\nu}] \quad , \quad D_{\mu}=\partial_{\mu}+ig[A_{\mu},\quad] \;.  \]
%
%
%
%
$A_{\mu}$ and $\Psi$ are matrices in the adjoint representation of $SU(2)$:
\[ A_{\mu}=A_{\mu}^{a}\tau^{a} \quad , \quad \Psi=\Psi^{a}\tau^{a} \qquad a=1,2,3
\]
where $ \tau^{a}=\frac{\sigma^{a}}{2}$
and $\sigma^{a}$ are the Pauli matrices, so that
\[
\left[\tau^a, \tau^b \right]=i\epsilon^{abc}\qquad ,
\qquad {\rm Tr} \left(\tau^a \tau^b \right)=\half \delta^{ab}\;.
\]
$\gamma^{0}$ and $\gamma^{1}$ are $2\times 2$ matrices satisfying the Dirac algebra
\[
\{\gamma^\mu, \gamma^\nu\} =2g^{\mu\nu}\,.
\]
We shall use the following representation
\[
\gamma^{0}=\left(\begin{array}{cc}0 & -i\\ i & 0\end{array}\right)\; \;, \; \;\gamma^{1}=\left(\begin{array}{cc}0 & i\\ i & 0\end{array}\right) \; .
\]
$\Psi^{a}$ is a 2-component Dirac field :
\[\Psi^{a}=   \left(\begin{array}{cc}\Psi^{a}_{R}\\  \Psi^{a}_{L}
\end{array}\right) \]
The lagrangian is invariant under the gauge transformation
\[
\Psi _{R/L}^\prime = U \Psi _{R/L} U^{-1}\; ,
\]
\[
A^\prime_{\mu} = UA_\mu U^{-1}+\frac{i}{g}\partial_\mu U U^{-1}
\]
where $U$ is a spacetime-dependent element of SU(2).\\
Note that $F_{\mu \nu}$ and $D_\mu$ transform covariantly under gauge
transformations:
\[
F_{\mu \nu}^\prime = U F_{\mu\nu}U^{-1} \qquad,
\qquad D_\mu ^\prime = U D_\mu U^{-1}\;.
\]
The equations of motion for the gauge fields are
\be\label{eqofmotion}
D_\mu F^{\mu \nu}=g J^\nu
\ee
where the fermion current $J^\mu$ is defined as
\be\label{currents}
J^\mu \equiv i{\bar \Psi}^b \gamma^\mu \Psi^a \left[ \tau^a, \tau^b \right] \; .
\ee
The conservation law  associated with the gauge invariance is
\[
\partial_\nu \left( J^\nu -i \left[ A_\mu, F^{\mu\nu }\right]   \right)=0 \; ,
\]
the fermion current being conserved in the covariant sense:
\[
D_\mu J^\mu =0 \; .
\]

\section{\textbf{The light-cone gauge}}

The light-cone gauge condition
\be \label{gauge}
  n^\mu A_\mu ^a=0 \quad \qquad {\rm with}\quad n=\frac{1}{\sqrt{2}}(1,-1)
\ee
or, equivalently,
\be
 A_ - ^a  \equiv \frac{1}{\sqrt{2}} (A_0 ^a -A_1 ^a )=0 \: ,
\ee
can be enforced by means of a Lagrange multiplier $\lambda(x)=\lambda^a (x)\tau^a$ , by adding
the gauge-fixing term \cite{bassetto}
\[
{\cal L}_{gf}=2{\rm Tr}\left(\lambda \, n^\mu A_\mu \right)
 \]
to the Lagrangian.
The theory defined by the Lagrangian
\[{\cal L}'={\cal L}+{\cal L}_{gf} \]
can be consistently quantized by means of Dirac's procedure. The gauge conditions
(\ref{gauge}) can be obtained as the Euler-Lagrange equations associated to the
fields $\lambda^a (x)$.

The quantum commutators corresponding to the classical Dirac's
brackets are
\[
[A_0^a (t,\x), (\pi^1)^b (t, \y)]= [A_1^a (t,\x), (\pi^1)^b (t, \y)]= \delta_{ab}(\x-\y)
\]
where $( \pi^1)^b=F_{01}^b$. We can see that the gauge constraint $A_- ^a=0 $ can be imposed in a strong sense while
$A_+^a$ satisfies
\[
[A_+ ^a (t,\x), F_{01}^b (t, \y)]= \sq \delta_{ab}(\x-\y) \;.
\]
This procedure introduces spurious degrees of freedom into the theory. The
Euler-Lagrange equations associated to the gauge fields
\be
D_\mu F^{\mu \nu}+\lambda n^{\nu}=g J^\nu
\ee
are not equivalent to eqs.(\ref{eqofmotion}) owing to the presence of the Lagrange multiplier. Equivalence with
the original theory can be recovered by imposing the subsidiary condition
$\lambda=0$. However, since the commutators of $\lambda$ with the other
fields are not zero, such
condition is incompatible with the quantization of the theory and
cannot be imposed in a strong sense. As in the standard Gupta-Bleuler quantization
of QED in the Feynman gauge, the subsidiary  condition will have to be imposed as a weak condition selecting
the physical subspace ${\cal V}_{phys}$ of the theory:
\be
|phys \rangle \in  {\cal V}_{phys} \Leftrightarrow
\langle phys | \lambda |phys \rangle =0
\ee
The stability of the physical subspace under time evolution is guaranteed
by the  fact that, as we shall see,  $\lambda$ satisfies a free-field equation of motion and has, therefore, a well defined decomposition into positive and negative frequency parts, so that one can equivalently state the subsidiary condition as
\be \label{subsidiary}
|phys \rangle \in {\cal V}_{phys}\Leftrightarrow
\lambda^{(+)} |phys \rangle =0
\ee
where $\lambda^{(+)}$ denotes the annihilation, or positive frequency, component
of the field $\lambda(x)$\,.

\section{\textbf{Dynamics}}

It is convenient to introduce the helicity basis \cite{paper}
\[\tau^{+}=\frac{\tau^{1}+i\,\tau^{2}}{\sqrt{2}}\qquad , \qquad \tau^{-}
=\frac{\tau^{1}-i\,\tau^{2}}{\sqrt{2}}
\]
These satisfy
\be \label{commtau}
[\tau^{+}, \tau^{-}]=\tau^{3} \qquad , \qquad [\tau^{3}, \tau^{\pm }]=\pm \tau^{\pm}
\ee
and
\be
\rm{Tr}(\tau^{+} \tau^{-})=\rm{Tr}(\tau^{3})^{2}=\frac{1}{2} \quad ,
\quad \rm{Tr}(\tau^{\pm})^{2}=\rm{Tr}(\tau^{3}\tau^{\pm})=0 \; .
\ee
With respect to this basis $A_{\mu}$ and $\Psi$ are decomposed as
\be
A_\mu = A_\mu ^3 \tau ^3 +A_\mu ^- \tau ^+ +A_\mu ^+ \tau ^-
\ee
where $A_\mu ^\pm \equiv \frac{1}{\sqrt{2}}(A_\mu ^1 \pm A_\mu ^2)$ ,
\be
\Psi _{ R/L}=
\phi_{R/L}\tau ^3+  \psi _{R/L}\tau ^+ +  \psi_{R/L} ^\dag \tau ^-
\ee
where  $\phi_{R/L} \equiv \Psi_{R/L}^3$ and $\psi_{R/L} \equiv \frac{1}{\sqrt 2}\left(\Psi ^1 _{R/L}- i \Psi ^2 _{R/L}\right)$\, ,
 $\psi^\dag _{R/L} \equiv \frac{1}{\sqrt 2}(\Psi ^1 _{R/L}+ i \Psi ^2 _{R/L})$\, .\\

We shall restrict  the space
variable to the interval $-L \leq \x \leq L$  and impose ``twisted"
boundary conditions: the fields $\psi_R$ and $\psi_L$ will
be taken to be antiperiodic; it will be convenient, however, to take
$\phi_R$ and $\phi_L$ to be periodic.  For
consistency, then, $A_\mu^\pm$ must be taken to be antiperiodic
while $A_\mu^3$ is periodic.

With the above definitions the Lagrangian density can be written as

\baselineskip15pt
\bean
\!\!\!{\cal L}&\!=&\! \!-\half (F_{01}^3 )^2 -\half F_{01}^+ F_{01}^- -\half F_{01}^- F_{01}^+
+\frac{i}{\sqrt{2}}\left[\phi_{R}\partial_{+}\phi_{R}+ \phi_{L}\partial_{-}\phi_{L}\right]
\nn \\ \nn \\
&&\! +\frac{i}{\sqrt{2}}\left[\psi_{R}^\dag \partial_{+}\psi_{R}+\psi_{R} \partial_{+}\psi_{R}^\dag +\psi_{L}^\dag \partial_{-}\psi_{L}+\psi_{L}\partial_{-}\psi_{L}^\dag  \right]\nn \\ \\
&&\! -{g\over\sqrt{2}}\left[A^3 _+  J^3 _R  +A^- _+  J^+ _R   +A_+ ^+  J^- _R +A^3 _-  J^3 _L  +
A^- _-  J^+ _L+ A_- ^+  J^- _L \right] \nn \\ \nn \\
&&+\lambda^3 A_- ^3+\lambda^+ A_- ^- +\lambda^- A_- ^+  \nn
\eean
\lines
where
\[
A_- ^{3,\pm} \equiv \frac{1}{\sqrt{2}}\left(A_0 ^{3,\pm}-A_1 ^{3, \pm}\right) \qquad , \qquad
A_+^{3,\pm} \equiv \frac{1}{\sqrt{2}}\left(A_0 ^{3,\pm}+A_1 ^{3, \pm}\right)
\]
and
\[
J^{3,\pm}_L =\frac{1}{\sqrt{2}}\left(J^{3,\pm}_0 + J^{3,\pm}_1\right) \qquad , \qquad
J^{3,\pm}_R =\frac{1}{\sqrt{2}}\left(J^{3,\pm}_0 -J^{3,\pm}_1\right)
\]
The equations of motion for the gauge fields  take the form
\bean
&&\partial_- F^3 +J^3 _R =0 \\
&&\partial_+ F^3 +i g (F^+ A^- - F^- A^+) -J^3 _L +\lambda^3 =0 \\
&&\partial_- F^- +J^- _R =0\\
&&\partial_+ F^- +i g (F^- A^3-F^3 A^-  ) -J^- _L +\lambda^- =0 \\
&&\partial_- F^+ +J^+ _R =0 \\
&&\partial_+ F^+ +i g (F^3 A^+-F^+ A^3  ) -J^+ _L +\lambda^+ =0\\
&& A_- =0
\eean
where $A^{3,\pm}\equiv A^{3,\pm}_+ $ and $F^{3,\pm}\equiv
F^{3,\pm}_{01}$, which is the only non-vanishing component of the
antisymmetric tensor $F^{3,\pm}_{\mu\nu}$ in two dimensions. The
condition $A_-=0$ implies
\[
F^{3,\pm}=\partial_0 A^{3,\pm} _1 -\partial_1 A^{3,\pm}_0 =\partial_- A^{3,\pm}
\]
From the expression of the energy-momentum tensor
\[
\Theta^{\mu \nu }=\frac{\partial\cal{L}}{\partial(\partial_{\mu}\varphi_{\alpha})}\partial^{\nu}\varphi_{\alpha}-{\cal L}g^{\mu \nu}
\]
one obtains the canonical Hamiltonian
\[
 P^{0}\equiv H=\int_{-L}^{L}d\x\, \Theta^{00}(x) .
\]
where
\bea
\Theta^{00}&= &F^3 \partial_0 A_1 ^3 +F^- \partial_0 A_1 ^+ +F^+ \partial_0 A_1 ^-
+\frac{i}{2}\left(  \phi_R \partial_0 \phi_R +  \psi^\dag _R \partial_0 \psi_R  + \psi _R \partial_0 \psi^\dag _R  \right) \\
&& +\frac{i}{2}\left(  \phi_L \partial_0 \phi_L + \psi^\dag _L \partial_0 \psi_L +\psi _L \partial_0 \psi^\dag _L \right)-{\cal L}
\eea
With some manipulations and  using the constraint $A_- =0$ one gets
\bea
H&=&\int_{-L}^{L} d\x \, \Bigg\{ \half  \left(F^3 \right)^2 +F^+ F^-  -
\frac{1}{\sqrt{2}}\left(\partial_1 F^3 A^3 +\partial_1 F^+ A^- +\partial_1 F^- A^+  \right) \\
&&+\frac{i}{2}\left( \phi_{L}\partial_{1}\phi_{L} +\psi_{L}^\dag \partial_{1}\psi_{L}+\psi_{L}\partial_{1}\psi_{L}^\dag -
\phi_{R}\partial_{1}\phi_{R}-\psi_{R}^\dag  \partial_{1}\psi_{R}-\psi_{R} \partial_{1}\psi_{R}^\dag  \right)\\
&&+ g  \left(A^3   J^3 _R  +A^-  J^+ _R   +A^+  J^- _R \right) \Bigg\}\;.
\eea

\section{\textbf{Quantization of the Fermi field.}}
%
%
%
%
%

The Fock representation for the fermionic degrees of freedom at $t=0$ is
obtained by Fourier expanding $\Psi_{R/L}(0,\x)$ .  We have
\begin{eqnarray}
\phi _R(0,\x) &=& {1 \over \sqrt {2L}} \sum_{N=1}^\infty
\left(r_N e^{ik _N \x} + r{_N ^\dag} e^{-ik_N \x} \right)
+{\stackrel {\;o} {\phi}}_R  \label{phiR}\\
\psi _R(0,\x)&=& {1 \over \sqrt {2L}} \sum_{n={1\over2}}^\infty
\left(b_n e^{ik _n \x} + d{_n ^\dag} e^{-ik_n \x} \right)
\label{psiR} \\
\phi _L(0,\x) &=& {1 \over  \sqrt {2L}} \sum_{N=1}^\infty
\left(\rho_N e^{-ik_N \x} + \rho{_N ^\dag}
e^{ik_N \x} \right) +{\stackrel {\;o} {\phi}}_L \label{phiL}\\
\psi _L(0,\x) &=& {1 \over \sqrt {2L}} \sum_{n={1\over2}}^\infty
\left(\beta_n e^{-ik _n \x} + \delta{_n ^\dagger}
e^{ik_n \x} \right) \; ,
\end{eqnarray}
where ${\stackrel {\;o} {\phi}}_{R/L}$ are the zero modes of $\phi_{R/L}$.
The lower-case (upper-case) indices run over positive
half-odd integers (integers) and $k_n= n \pi /L$, $k_N= N \pi /L$ .

The canonical anti-commutation relations for the Fermi fields are
\bea
\left\{\phi_{R} (0,\x),\phi_{R}(0,y)\right\}
&=&  \delta _{P}(x-y)  \\
\left\{\phi_{L} (0,\x),\phi_{L}(0,y)\right\}
&=&  \delta_{P} (x-y) \; ,
\eea
where $\delta_{P}$ denotes the periodic delta function, which can be expanded
in the interval $[-L, +L]$ as
 \[
\delta_P (\x-\y)=\frac{1}{2L}\sum_{N=-\infty}^\infty e^{i \frac{\pi}{L}N(\x-\y)} \;,
\]
and
\bean
\left\{\psi_{R} (0,\x),\psi^\dag_{R}(0,y)\right\}
&=&  \delta _{A}(x-y)  \label{commpsiR}\\
\left\{\psi_{L} (0,\x),\psi_{L}^\dag (0,y)\right\}
&=&  \delta_{A} (x-y) \label{commpsiL}\;,
\eean
where $\delta_{A}$ denotes the anti-periodic delta function
\[
 \delta_A (\x-\y)=\frac{1}{2L}\somma e^{i \frac{\pi}{L}n(\x-\y)} \;.
\]
All the other anti-commutators vanish.\\
These induce the following algebra for the Fourier modes:
\be
\{\rho{^\dagger _N}, \rho_M \} = \{r{^\dagger _N}, r_M \}
= \delta_{N,M}
\label{rhccrs}
\end{equation}
\begin{equation}
\{ b{_n ^\dagger}, b_m \} = \{ d{_n^\dagger}, d_m \}
= \{ \beta{_n ^\dagger},\beta _m \}
= \{\delta {_n^\dagger},\delta_m\} = \delta _{n, m}
\label{lhccrs}
\end{equation}
\begin{equation}
\{{\stackrel {\;o} {\phi}}_R,{\stackrel {\;o} {\phi}}_R \} =
\{{\stackrel {\;o} {\phi}}_L,{\stackrel {\;o} {\phi}}_L \} =
{1\over2L}\; ,
\label{zmccrs}
\end{equation}
all other anti-commutators vanishing. \\
The fermionic Fock space is generated in the usual way by the action of the
creation operators on a vacuum state $|0\rangle$.



\section{\textbf{The currents.}}

A rigorous definition of quantum fields as operators in a Hilbert
space requires smearing with  test functions. As mathematical
objects quantum fields are operator-valued distributions and the
product of fields at the same point is not defined. Quantum field
theories are affected by ultraviolet singularities arising from ill-defined
operator products. A renormalization procedure is necessary to
remove such singularities. The most simple example
is the normal ordering prescription for  free fields, which can be defined
by means of a space-like point splitting as
\[
{\bf :} A(x)B(x) {\bf :} \equiv \lim_{{\tiny \begin{array}{c}\varepsilon\rightarrow 0 \\\varepsilon^{2}< 0 \end{array}}}\Big(A(x+\varepsilon)B(x)- \langle0| A(x+\varepsilon)B(x) |0\rangle \Big) \;.
\]
Here the diverging vacuum expectation value (v.e.v.) is removed before
the limit $\varepsilon \rightarrow 0$ is taken. \\
Starting from the expression for the currents (\ref{currents}) we
can define the free fermion currents for this theory at $t=0$ by means of a
point-splitting in the space direction
\be\label{freecurrR}
\tilde{J} _R(0, \x) \equiv \lim_{\epsilon \rightarrow 0}\frac{1}{\sqrt{2}}
 {\Psi}^b  _R (0, \x+\epsilon)  \Psi^a _R (0, \x) \left[ \tau^a, \tau^b \right]
\ee
\be\label{freecurrL}
\tilde{J} _L(0, \x) \equiv \lim_{\epsilon \rightarrow 0} \frac{1}{\sqrt{2}}
 \Psi^b  _L (0, \x+\epsilon)  \Psi^a _L (0, \x) [ \tau^a, \tau^b ]
\ee
or, for each colour component in the helicity basis,
\bea
\tilde{J}^3 _R&\equiv &\lim_{\epsilon \rightarrow 0} \frac{1}{\sqrt{2}}
\left(\psi^\dagger_{R}(0,\x+\epsilon) \psi_{R}(0,\x)+\psi_{R}^\dag (0,\x) \psi_{R} (0,\x+\epsilon) \right)\\
\tilde{J}^+ _R &\equiv &\lim_{\epsilon \rightarrow 0} \frac{1}{\sqrt{2}}
\left(\phi_{R} (0,\x+\epsilon)\psi^\dagger_{R}(0, \x)+\phi_R(0,\x)\psi^\dagger_{R} (0,\x+\epsilon) \right)\\
\tilde{J}^-  _R &\equiv &\lim_{\epsilon \rightarrow 0}\frac{1}{\sqrt{2}}
\Big( \psi_{R}(0,\x+\epsilon) \phi_{R}(0,\x)+\psi_{R}(0,\x) \phi_{R}(0,\x+\epsilon) \Big)
\eea
and analogous expressions for the left currents.\\
It is easy to see that the definition of $\tilde{J}^3$ is  equivalent to the normal
ordering prescription. As a matter of fact by evaluating
 $\langle 0|{\psi}^{\dag}_R (0, \x+\epsilon)\psi_R (0,\x)|0\rangle$
one sees that in the normal ordering procedure the ultraviolet divergence
$\frac{i}{2\pi \epsilon}$  is subtracted. This purely imaginary singularity is also removed
by means of the hermitian point-splitting as in the definition of $\tilde{J}^3$.  More explicitly,
 using the Fock representation (\ref{psiR})
we have
\bea
\psi_R ^\dag (0, \x+\epsilon)\psi_R (0,\x)=\frac{1}{2L}\sum_{m,n=\half}^{\infty}
\bigg(d_n b_m e^{i(k_n +k_m )\x}e^{ik_n \epsilon}
+d_n d_m ^\dag e^{i(k_n -k_m )\x}e^{ik_n \epsilon}\qquad \quad \\
+b_n ^\dag b_m e^{-i(k_n -k_m )\x}e^{-ik_n \epsilon}+
b_n ^\dag d_m ^\dag e^{-i(k_n +k_m )\x}e^{-ik_n \epsilon}\bigg) \; .
\eea
Since
\[
\frac{1}{2L}\sum_{m,n=\half}^{\infty}
d_n d_m ^\dag e^{i(k_n -k_m )\x}e^{ik_n \epsilon}=
-\frac{1}{2L}\sum_{m,n=\half}^{\infty}d_m^ \dag d_n e^{i(k_n -k_m )\x}e^{ik_n \epsilon}+\frac{1}{2L}\sum_{n=\half}^{\infty}e^{ik_n \epsilon}
\]
and
\[
\frac{1}{2L}\sum_{n=\half}^{\infty}e^{ik_n \epsilon}=
\frac{1}{2L}\sum_{N=0}^{\infty}
e^{iN \pl \epsilon}e^{i\frac{\pi}{2L}\epsilon}=\frac{1}{2L}
\frac{1}{1-e^{i\frac{\pi}{L}\epsilon}}\,e^{i\frac{\pi}{2L}\epsilon}=
\frac{i}{2\pi\epsilon}+o(\epsilon)
\]
we can write
\be \label{pointsplittingpsiR}
\psi_R ^\dag (0, \x+\epsilon)\psi_R (0,\x)=
\textbf{:}\psi_R ^\dag (0, \x)\psi_R (0,\x) \textbf{:}+\frac{i}{2\pi\epsilon}+o(\epsilon)
\ee
and
\bea
\tilde{J}_R ^3 (0, \x)=\sqrt{2}\, \textbf{:}\psi_R ^\dag (0, \x)\psi_R (0,\x) \textbf{:}=
 \frac{1}{\sqrt{2}L}\sum_{m,n=\half}^{\infty}
\bigg(d_n b_m e^{i(k_n +k_m )\x}
-d_m ^\dag d_n e^{i(k_n -k_m )\x}\qquad \quad \\
+b_n ^\dag b_m e^{-i(k_n -k_m )\x}+
b_n ^\dag d_m ^\dag e^{-i(k_n +k_m )\x}\bigg) \; .
\eea
In a similar manner one gets:
\be \label{pointsplittingpsiL}
\psi_L ^\dag (0, \x+\epsilon)\psi_L (0,\x)=
\textbf{:}\psi_L ^\dag (0, \x)\psi_L (0,\x) \textbf{:}-\frac{i}{2\pi\epsilon}+o(\epsilon)
\ee
and
\[
\tilde{J}_L ^3 (0, \x)=\sqrt{2}\, \textbf{:}\psi_L ^\dag (0, \x)\psi_L (0,\x) \textbf{:}
\]
From the above expression for $\tilde{J}_R ^3 $ it is not hard to see that
we can write
\be\label{JR3}
\tilde{J}_R ^3 (0, \x)=\frac{1}{\sq L}
\sum_{N=1}^{\infty}\left(C_N ^3 e^{ik_N \x}+{C_N ^3 }^\dag e^{-ik_N \x} \right)+\frac{C_0 ^3}{\sq L}
\ee
where
\[
C_N^3 =\sum_{n={1 \over 2}}^\infty \left(b^\dagger_n b_{N+n}
-d^\dagger_n d_{N+n}\right)
-\sum_{n={1 \over 2}}^{N-{1 \over 2}} b_n d_{N-n}
\]
and
\begin{equation}
C^3_0 = \sum_n (b _n ^\dagger b_n - d_n ^\dagger d_n ) \; .
\end{equation}
Analogously one can see that
\be \label{JL3}
\tilde{J}_L ^3 (0, \x)=\frac{1}{\sq L}
\sum_{N=1}^{\infty}\left(D_N ^3 e^{-ik_N \x}+{D_N ^3 }^\dag e^{ik_N \x} \right)
+\frac{D_0 ^3}{\sq L}
\ee
where
\[D_N^3 = \sum_{n={1 \over 2}}^\infty \left(\beta^\dagger_n \beta_{N+n}
-\delta^\dagger_n \delta_{N+n}\right)
-\sum_{n={1 \over 2}}^{N-{1 \over 2}} \beta_n \delta_{N-n}
\]
and
\begin{equation}
D^3_0 = \sum_n (\beta _n ^\dagger \beta_n - \delta_n ^\dagger \delta_n ) \; .
\end{equation}
As for
$J^+$ and $J^-$ it is easy to see that the operator products are not singular and
the limit can be taken without subtractions:
\bea
\tilde{J}^+ _R (0,\x)&=&\frac{1}{2L}\sum_{N=0}^{\infty}\sum_{n=\half}^{\infty}
\Big( r_N d_n e^{i(N+n)\pl}- b_n ^\dag r_Ne^{i(N-n)\pl}\\ &&\hspace{40pt}+
r_N ^\dag d_n e^{-i(N-n)\pl}+r_N ^\dag b_n ^\dag e^{-i(N+n)\pl}\Big)
\eea
\bea
\tilde{J}^- _R (0,\x)&=&\frac{1}{2L}\sum_{N=0}^{\infty}\sum_{n=\half}^{\infty}
\Big(b_n r_N  e^{i(N+n)\pl \x}-r_N ^\dag b_n  e^{-i(N-n)\pl \x}\\
&&\hspace{60pt}
+d_n^ \dag  r_N e^{i(N-n)\pl \x}+d_n^ \dag r_N ^\dag  e^{-i(N+n)\pl \x}\Big)
\eea
where we have set
\be \label{zeromode}
{\stackrel {\;o} {\phi}}_R =\frac{r_{0}+r_{0}^\dag}{\sqrt{2L}}\;.
\ee
We can write
\be \label{JRpm}
\tilde{J}_R ^\pm (0, \x)=\frac{1}{\sq L}
\sum_{n=\half}^{\infty}\left(C_n ^\pm e^{ik_n \x}+{C_n ^\mp }^\dag e^{-ik_n \x} \right)
\ee
where
\bean
C_n ^+ &=& \sum_{M=0}^\infty r^\dagger_M d_{n+M}
-\sum_{m={1 \over 2}}^\infty b^\dagger_m r_{n+m}
-\sum_{m={1 \over 2}}^{n} d_m r_{n-m} \label{Cn+}\\
C_n ^- &=& \sum_{m={1 \over 2}}^\infty d^\dagger_m r_{n+m}
-\sum_{M=0}^\infty r^\dagger_M b_{M+n}
-\sum_{m={1 \over 2}} ^{n} r_{n-m} b_m \label{Cn-}\; .
\eean
Similar expressions can be found for the operators $D_n ^\pm$ such that
\be\label{JLpm}
\tilde{J}_L ^\pm (0, \x)=\frac{1}{\sq L}
\sum_{n=\half}^{\infty}\left(D_n ^\pm e^{-ik_n \x}+{D_n ^\mp }^\dag e^{ik_n \x} \right)
\ee
Using  the fundamental anti-commutators (\ref{rhccrs}),
(\ref{lhccrs}) and  ( \ref{zmccrs}) one can  verify that these
operators satisfy the commutation relations\footnote{Note that these relations do not hold in this form if the zero mode of $\phi_{R/L}$ is discarded, as in ref. \cite{paper}}\cite{goddard}
\begin{eqnarray}
\left [ C^3_N , C^3_M         \right ] &=& N \delta_{N,-M}\label{commc3} \\
\left [ C_n^{\pm} , C_m^{\pm} \right ] &=& 0 \label{commcpm}\\
\left [ C^3_N , C_m^{\pm}     \right ] &=& \pm C_{N+m}^{\pm}
\label{commc3cpm}\\
\left [ C^+_n , C^-_m  \right ] &=& C^3_{n+m} + n \delta_{n,-m}\; . \label{commcpcm}
\end{eqnarray}
where we have defined
\[
C^3_{-N}\equiv (C^3_N)^\dag \ , \qquad C_{-n}^{\pm}\equiv (C_n^{\mp})^\dag
\]
 The algebra satisfied by the $D$s is of course identical.

Definitions (\ref{freecurrR}) and (\ref{freecurrL}) are not suitable for regularizing
the currents in the
interacting case because the gauge transformation properties of
the currents are not preserved. We want a prescription such that the regularized currents still transform
covariantly \cite{paper}:
\[
J_{\mu}^\prime (x) = U(x)J_\mu (x)U^\dag(x)
\]
In order to achieve this we modify def.(\ref{freecurrR}-\ref{freecurrL}) by suitably inserting factors
of the form:
\[
e^{ ig \int_x^{x + \varepsilon} A \cdot dx}
\]
which, under gauge transformations behave  according to
\[
e^{ ig \int_x^{x + \varepsilon} A \cdot dx}\longrightarrow U(x)e^{ ig \int_x^{x + \varepsilon} A \cdot dx} U^\dag (x+\varepsilon)
\]
for $\varepsilon$ infinitesimal. As a consequence we have:
\[
e^{ ig \int_x^{x + \varepsilon} A \cdot dx}\Psi(x+\varepsilon)e^{- ig \int_x^{x + \varepsilon} A \cdot dx}
\longrightarrow
U(x) e^{ ig \int_x^{x + \varepsilon} A \cdot dx}\Psi(x+\varepsilon)
e^{- ig \int_x^{x + \varepsilon} A \cdot dx}U^\dag (x)
\]
and the regularized interacting current $\hat{J}(0,\x)$ defined by
\[
\hat{J} _R(0, \x) \equiv -\lim_{\epsilon \rightarrow 0}
\Psi^a _R(0,\x+\epsilon)
 \Psi^b _R (0, \x) \, \big[ e^{ ig \int_\x^{\x + \epsilon} A_1 \cdot d\x}\,\tau^a\,
e^{- ig \int_\x^{\x + \epsilon} A_1 \cdot d\x}, \tau^b \big]
\]
transforms covariantly.\\
Using
\[
e^{ ig \int_\x^{\x + \epsilon} A_1 \cdot d\x}\tau^a e^{- ig \int_\x^{\x + \epsilon} A_1 \cdot d\x}=
\tau^a +ig \epsilon[ A_1 ,
\tau^a  ] + o(\epsilon^2)
\]
we have
\[
\hat{J}_R (0,\x) =\tilde{J}_R (0,\x)-\lim_{\epsilon \rightarrow 0}\left(\frac{i}{2}g\epsilon \Psi_R ^a (0, \x+\epsilon)\Psi^b _R (0,\x)
A^c (0,\x)\left[ [\tau^c, \tau^a], \tau^b \right] \right)\, .
\]
Since \ $ \left[ [\tau^c, \tau^a], \tau^b \right]=i\epsilon^{cad}i\epsilon^{dbf}\tau^f = \delta^{ab}\tau^c
-\delta^{cb}\tau^a $ \ ,  we can write
\bea
\hat{J}_R (0,\x)=\tilde{J}_R (0,\x)-ig \lim_{\epsilon \rightarrow 0}\bigg[\epsilon{\rm Tr}\Big\{\Psi_R  (0, \x+\epsilon)\Psi _R (0,\x)\Big\}A(0,\x)\\
-\epsilon\Psi_R  (0, \x+\epsilon){\rm Tr}\Big\{\Psi _R
(0,\x)A(0,\x)\Big\} \bigg] \eea so that, going back to the
helicity basis we have
\bea 
\hat{J}_R ^3 = \tilde{J}_R ^3
-\frac{i}{\sqrt{2}}g \lim_{\epsilon \rightarrow 0}\!\!\!&\bigg[
&\!\!\! \epsilon \Big\{\psi_R ^\dag (0,\x+\epsilon) \psi_R
(0,\x)+\psi_R (0,\x+\epsilon)
\psi_R^\dag (0,\x)\Big\}A^3_ 1(0,\x) \\
&&\hspace{-15pt}-\epsilon\phi_R (0,\x+\epsilon)\psi_R ^\dag (0,\x)A^- _1 (0,\x)-\epsilon\phi_R (0,\x+\epsilon)\psi_R (0,\x)A^+ _1 (0,\x) \bigg]\\ \\
\hat{J}_R ^+ = \tilde{J}_R ^+  -\frac{i}{\sqrt{2}}g \lim_{\epsilon \rightarrow 0}\!\!\!&\bigg[&\!\!\! \epsilon \Big\{
\phi_R  (0,\x+\epsilon)\phi_R (0,\x)+\psi_R  (0,\x+\epsilon)\psi_R ^\dag(0,\x)\Big\} A^+ _1 (0,\x)\\
&& \hspace{-15pt}-\epsilon\psi_R ^\dag (0,\x+\epsilon)\phi_R (0,\x)A^3 _1 (0,\x) -\epsilon\psi_R ^\dag (0,\x+\epsilon)\psi_R ^\dag (0,\x)A^- _1 (0,\x)  \bigg]
\eea
\bea
\hat{J}_R ^- = \tilde{J}_R ^-  -\frac{i}{\sqrt{2}}g \lim_{\epsilon \rightarrow 0}\!\!\!&\bigg[&\!\!\! \epsilon \Big\{
\phi_R (0,\x+\epsilon)\phi_R  (0,\x)+\psi_R ^\dag (0,\x+\epsilon)\psi_R (0,\x)\Big\} A^- _1 (0,\x) \\
&& \hspace{-15pt}-\epsilon\psi_R (0,\x+\epsilon)\phi_R (0,\x)A^3 _1 (0,\x) -\epsilon\psi_R (0,\x+\epsilon)\psi_R (0,\x) A^+ _1 (0,\x) \bigg]
\eea
As we have seen (eq. \ref{pointsplittingpsiR}), the operator product $\psi_R ^\dag (0,\x+\epsilon)
\psi_R (0,\x)$ has a pole in $\epsilon=0$, which
cancels the $\epsilon$ in the numerator giving a non-zero limit. The same singularity
affects $\psi_R  (0,\x+\epsilon)
\psi_R ^\dag (0,\x)=-\left(\psi_R ^\dag (0,\x+\epsilon)
\psi_R (0,\x)\right)^\dag$. Moreover we have
\bea
\phi_R  (0,\x+\epsilon)\phi_R  (0,\x)&=&\frac{1}{2L}\sum_{M,N=0}^{\infty}
\Big( r_N r_M e^{i(k_N+k_M)\x}e^{ik_N \epsilon}
+r_N ^\dag r_M e^{-i(k_N-k_M)\x}e^{-ik_N \epsilon}\\
&&\hspace{-20pt}+r_N ^\dag  r_M ^\dag e^{-i(k_N+k_M)\x}e^{-ik_N \epsilon}
-r_M ^\dag r_N e^{i(k_N-k_M)\x}e^{ik_N \epsilon}\!
+\! \frac{1}{2L}\sum_{N=0}^{\infty}e^{iN\pl\epsilon}
\Big)\\
&\!\!\!=&\!\!\! \frac{1}{2L}\sum_{M,N=0}^{\infty}
\Big( r_N r_M e^{i(k_N+k_M)\x}+r_N ^\dag  r_M ^\dag e^{-i(k_N+k_M)\x}\Big)
+ \frac{i}{2\pi\epsilon}+o(\epsilon)
\eea
All the other operator products appearing in the above expressions for the regularized
currents  do not have poles in $\epsilon=0$ and give no contribution, so that, after taking the limit $\epsilon\rightarrow 0$,
 one is left with
\bean
&&\hat{J}_R ^3 = \tilde{J}_R ^3 +\frac{g}{\sqrt{2}\pi}A^3 _1 \nn \\
&&\hat{J}_R ^+ = \tilde{J}_R ^+ +\frac{g}{\sqrt{2}\pi}A^+ _1 \label{Rcurr}\\
&&\hat{J}_R ^- = \tilde{J}_R ^- + \frac{g}{\sqrt{2}\pi}A^-  _1\; . \nn
\eean
For the left currents we have non-zero contributions in the small $\epsilon$
limit from
\bea
&&\psi_L ^\dag (0,\x+\epsilon)
\psi_L (0,\x) \sim  -\frac{i}{2\pi\epsilon}\; ,\\
&&\psi_L  (0,\x+\epsilon)
\psi_L ^\dag (0,\x) \sim -\frac{i}{2\pi\epsilon}\; ,\\
&&\phi_L  (0,\x+\epsilon)\phi_L (0,\x) \sim -\frac{i}{2\pi\epsilon}\; .
\eea
and, therefore
\bean
&&\hat{J}_L ^3 = \tilde{J}_L ^3 -\frac{g}{\sqrt{2}\pi}A^3 _1\nn \\
&&\hat{J}_L ^+ = \tilde{J}_L ^+ -\frac{g}{\sqrt{2}\pi}A^+ _1\\
&&\hat{J}_L ^- = \tilde{J}_L ^-  - \frac{g}{\sqrt{2}\pi}A^-  _1\; .\nn
\eean

\section{\textbf{The kinetic term}}

As a gauge-covariant regularization of the kinetic term let us consider
\bea
&& \!\!\!\!\!\!\!\!\left[\rm{Tr} (i \Psi_R(0,\x) \partial_1 \Psi_R(0,\x)\right]_{reg} \equiv\\
&&\qquad \equiv \lim_{\epsilon \rightarrow 0}\left \{\rm{Tr}\!\left(i e^{ig\int_\x ^{\x+\epsilon} A_1 (0,\y)d\y} \Psi_R(0, \x+\epsilon) e^{-ig\int_\x ^{\x+\epsilon} A_1 (0,\y) d\y}
\partial_1 \Psi_R (0,\x)\right)
-\rm{v.e.v.}\right \}
\eea
Using
\bea
e^{ig\int_\x ^{\x+\epsilon} A_1 (0,\y)d\y} \Psi_R(0, \x+\epsilon) e^{-ig\int_\x ^{\x+\epsilon} A_1(0,\y)d\y}
=\Psi_R (0, \x+\epsilon)+ig\epsilon \left[ A_1 (0,\x), \Psi_R (0,\x+\epsilon) \right]&&\\
+\frac{ i}{2}g \epsilon^2 \! \left[\partial_1 A_1 (0,\x), \Psi_R (0,\x+\epsilon) \right]-\frac{g^2}{2} \epsilon^2 \left[A_1 (0,\x), \left[A_1 (0,\x), \Psi_R (0,\x+\epsilon)
\right]\, \right]+o(\epsilon^3)&&
\eea
and disregarding terms that vanish in the limit $\epsilon\rightarrow 0$, one gets
\bea
\rm{Tr}\left(i e^{ig\int_\x ^{\x+\epsilon} A_1 d\x} \Psi_R(0, \x+\epsilon) e^{-ig\int_\x ^{\x+\epsilon} A_1 d\x}
\partial_1 \Psi_R \right) \simeq  \rm{Tr} \left(i \Psi_R (0, \x+\epsilon)\partial_1 \Psi_R(0, \x) \right )&&\\
+\left[\left(\half \epsilon g A_1 ^3(0, \x)
+\frac{1}{4}\epsilon^2 g \partial_1 A_1 ^3 (0, \x)\right)\psi_R ^\dag (0, \x+\epsilon) \partial_1 \psi_R(0, \x)+h. c. \right]&&\\
-\left[\frac{i}{4} \epsilon^2 g^2 \left( A_1 ^+ (0, \x) A_1 ^- (0, \x) +(A_1 ^3)^2 (0, \x) \right)\psi_R ^\dag (0, \x+\epsilon) \partial_1 \psi_R(0, \x)+h. c. \right]&&\\
 -\left[\frac{i}{4} \epsilon^2 g^2 A_1 ^+(0, \x)A_1 ^-(0, \x) \phi_R(0, \x+\epsilon) \partial_1 \phi_R (0, \x)+ h.c. \right]&&
\eea
One finds that
\bea
\psi_R ^\dag (0,\x+\epsilon)\partial_1 \psi_R (0,\x)& \simeq & \textbf{:}\,\psi_R ^\dag \partial_1 \psi_R\ \! \textbf{: }-\frac{1}{2i\pi\epsilon^2}+
\rm{const.}\\
\phi_R(0,\x+\epsilon)\partial_1 \phi_R (0,\x) &\simeq & \textbf{:}\,\phi_R \partial_1 \phi_R\ \! \textbf{: }-\frac{1}{2i\pi\epsilon^2}+
\rm{const.}
\eea
and therefore
\be\label{Rkinetic}
\left[\rm{Tr} (i \Psi_R(0,\x) \partial_1 \Psi_R(0,\x)\right]_{reg} =
\frac{i}{2} \, \textbf{:}\,\psi_R ^\dag \buildrel \leftrightarrow \over
\partial_1 \psi_R  \textbf{: }+\frac{i}{2}\,\textbf{:}\,\phi_R \partial_1 \phi_R  \textbf{: }
+\frac{g^2}{4\pi}\left(2 A^+ _1  A^- _1 +(A^3 _1 )^2\right)\;.
\ee
Analogously, using
\bea
\psi_L  ^\dag (0,\x+\epsilon)\partial_1 \psi_L (0,\x)& \simeq & \textbf{:}\,\psi_L ^\dag \partial_1 \psi_L\ \! \textbf{:}+\frac{1}{2i\pi\epsilon^2}+
\rm{const.}\\
\phi_L(0,\x+\epsilon)\partial_1 \phi_L (0,\x) &\simeq & \textbf{:}\,\phi_L \partial_1 \phi_L\ \! \textbf{:}+\frac{1}{2i\pi\epsilon^2}+
\rm{const.}
\eea
one gets
\be\label{Lkinetic}
\left[\rm{Tr} (i \Psi_L(0,\x) \partial_1 \Psi_L(0,\x)\right]_{reg} =
\frac{i}{2} \, \textbf{:}\,\psi_L ^\dag \buildrel \leftrightarrow \over
\partial_1 \psi_L \textbf{: }+\frac{i}{2}\,\textbf{:}\,\phi_R \partial_1 \phi_L \textbf{: }
-\frac{g^2}{4\pi}\left(2 A^+ _1  A^- _1 +(A^3 _1 )^2\right)\; .
\ee

\section{\textbf{Bosonization.}}
It is convenient to consider an alternative representation of the anticommutation
relations (\ref{commpsiR}) and (\ref{commpsiL}),  known as \emph{bosonization}.
This procedure will allow us to express
the fields $\psi_{R/L}$ and $\psi_{R/L}^\dag$ at $t=0$ in terms of the bosonic
operators $C_N^3$ and $D_N^3$ appearing in the Fourier decomposition of the currents   $\widetilde{J}^3_R$ and  $\widetilde{J}^3_L$ (eqs. \ref{JR3}-\ref{JL3}).
 Following  \cite{nakawaki} we define
\bea
\varphi_{R}^{(+)}(0,\x)&=&-\sum_{N=1}^{\infty}\ \frac{1}{N}{C_N^3}e^{ik_N \x} \\
\varphi_{R}^{(-)}(0,\x)&=&\sum_{N=1}^{\infty}\frac{1}{N}{(C_N^3)^\dag}e^{-ik_N\x} \\
\varphi_{L}^{(+)}(0,\x)&=&-\sum_{N=1}^{\infty} \frac{1}{N}{D_N^3}e^{-ik_N\x} \\
\varphi_{L}^{(-)}(0,\x)&=&\sum_{N=1}^{\infty}\ \frac{1}{N}{(D_N^3)^\dag}e^{ik_N\x}
\eea
and
\be \label{sigmaRL}
\sigma_{R/L}(0, \x)=\sqrt{2L}e^{\varphi_{R/L}^{(-)}(0,\x)}\psi_{R/L}(0,\x)e^{\varphi_{R/L}^{(+)}(0,\x)}\;.
\ee
The following relations hold \cite{nakawaki}
\be \label{sigmasigma+R}
\sigma_{R}^{+}(0,\x)\sigma_{R}(0,\x)=\sigma_{R}(0,\x)\sigma_{R}^{+}(0,\x)=1\; ,
\ee
\be \label{sigmasigma+L}
\sigma_{L}^{+}(0,\x)\sigma_{L}(0,\x)=\sigma_{L}(0,\x)\sigma_{L}^{+}(0,\x)=1\; ,
\ee
\be \{\sigma_{R}(x),\sigma_{L}(y)\}=\{\sigma_{R}(x),\sigma^{+}_{L}(y)\}=0 \; ,
\ee
\be
[C^3_0\, , \, \sigma_{L}]=[D^3_0\, , \, \sigma_{R}]=0
\ee
\be
[C^3_0\, , \, \sigma_{R}]=-\sigma_{R}\quad, \qquad [D^3_0\, , \, \sigma_{L}]=-\sigma_{L}
\ee
\be \label{1.28}
[C_N^3 ,\sigma_{R/L}]=[D_N^3,\sigma_{R/L}]=0\;.
\ee
The operators
\be\label{spurions}
\sigma_{R/L}\equiv \sigma_{R/L}(0,0)
\ee
are called \emph{spurions}. \\
Let us consider the states
\[
|M,N\rangle = \sigma_L^M \sigma_R^N |0\rangle \; ,
\]
where
\[ \sigma_{R/L}^{-N}\equiv (\sigma_{R/L}^\dag)^{N}\; .
\]
One can show that, for $M,N>0$ ,
\bea
&&|M,N\rangle =\delta_{M-\half}^\dag \cdots \delta_{\half}^\dag \,
 d_{N-\half}^\dag \cdots d_{\half}^\dag |0\rangle \\
&&|-M,N\rangle = \beta_{M-\half}^\dag \cdots \beta_{\half}^\dag \,
 d_{N-\half}^\dag \cdots d_{\half}^\dag |0\rangle \\
&&|M,-N\rangle = \delta_{M-\half}^\dag \cdots \delta_{\half}^\dag \,
 b_{N-\half}^\dag \cdots b_{\half}^\dag |0\rangle \\
&&|-M,-N\rangle = \beta_{M-\half}^\dag \cdots \beta_{\half}^\dag \,
 b_{N-\half}^\dag \cdots b_{\half}^\dag |0\rangle \; .
\eea
It is easy to see that the states  $|M,N\rangle $ are eigenstates of $C_0^3$ and $D_0^3$ :
\bea
C_0^3|M,N\rangle &=&-N|M,N\rangle\\
D_0^3|M,N\rangle &=&-M|M,N\rangle \; .
\eea
One can verify that, for any $P>0$
\[
C_P^3|M,N\rangle =0 \quad , \qquad D_P^3|M,N\rangle =0
\]
and since
\[
[C_0^3, {C_P^3}^\dag ]=[D_0^3, {C_P^3}^\dag ] =0 \\
\]
\[
[C_0^3, {D_P^3}^\dag ]=[D_0^3, {D_P^3}^\dag ]=0
\]
the action of ${C_P^3}^\dag$ and ${D_P^3}^\dag$ does not modify the eigenvalues
of $C_0^3$ and $D_0^3$.\\
It can be shown  \cite{uhlenbrock} that the fermion Fock space ${\cal {F}}$, generated
by the action of the creation operators $b_n^\dag$ , $d_n^\dag$ , $\beta_n^\dag$ , $\delta_n^\dag$
on the vacuum $|0\rangle$,
can be decomposed as an infinite  direct sum of irreducible representations of the
bosonic algebra satisfied by the operators $C^3_P$  and $D^3_P$ ($P\neq 0$), each
representation corresponding to an eigenspace of $C_0^3$ and $D_0^3$ . More
explicitly we have
\[
{\cal{F}}=\oplus_{M,N}{\cal{F}}_{MN} \qquad M,N=0,\pm1,\pm 2,\ldots
\]
where ${\cal{F}}_{MN}$  is the Fock space generated by applying products of the operators
${C_P^3}^\dag$ and ${D_P^3}^\dag$ to the vacuum $|M,N\rangle $ and
\[
 \forall \, |\Phi_{MN}\rangle \in {\cal{F}}_{MN}\; : \quad C_0^3|\Phi_{MN}\rangle =-N|\Phi_{MN}\rangle  \quad , \qquad D_0^3|\Phi_{MN}\rangle =-M|\Phi_{MN}\rangle  \;.
\]
The free fermion hamiltonian
\bean
H_{\psi}&=&\frac{i}{2} \int_{-L}^{L}d\x \left( \psi_L^\dag (0,\x)  \buildrel \leftrightarrow \over
\partial_1
\psi_L (0,\x)-\psi_R^\dag (0,\x)  \buildrel \leftrightarrow \over
\partial_1 \psi_R (0,\x) \right) \nn \\
&=&\sum_{n=\half}^{\infty} k_n (\beta_n^\dag \beta_n + \delta_n^\dag \delta_n
+b_n^\dag b_n + d_n^\dag d_n )
\eean
and the momentum operator
\bean
P_{\psi}&=&\frac{i}{2} \int_{-L}^{L}d\x \left( \psi_L^\dag (0,\x)  \buildrel \leftrightarrow \over
\partial_1 \psi_L (0,\x)+\psi_R^\dag (0,\x)  \buildrel \leftrightarrow \over
\partial_1 \psi_R (0,\x) \right) \nn \\
&=&\sum_{n=\half}^{\infty} k_n (\beta_n^\dag \beta_n + \delta_n^\dag \delta_n
-b_n^\dag b_n - d_n^\dag d_n) \label{p1}
\eean
can be expressed in terms of the boson operators by means of the following identities,
known as Kronig's identities:
\bean
&&H=\frac{\pi}{2L}\Big((C_0^3)^2+ (D_0^3)^2 \Big)+\frac{\pi}{L}\sum_{N=1}^\infty
\left({C_N^3}^\dag  C_N^3 +{D_N^3}^\dag  D_N^3 \right)
\; \\
&& P_{1}=\frac{\pi}{2L}\Big((D_0^3)^2- (C_0^3)^2 \Big)+\pl \sum_{N=1}^\infty
\left({D_N^3}^\dag  D_N^3 -{C_N^3}^\dag  C_N^3\right)
\label{p2}\;.
\eean
Finally, using
\[ \sigma_{R/L}(0,\x)=e^{-iP_{\psi}\x} \sigma_{R/L} e^{P_{\psi}\x}
\]
it is easy to see that
\bea
\sigma_R (0,\x)&=&e^{ \frac{i\pi}{2L}C_0^3\x }\sigma_R e^{ \frac{i\pi}{2L}C_0^3\x }\\
\sigma_L (0,\x)&=&e^{ -\frac{i\pi}{2L}D_0^3\x }\sigma_R e^{ -\frac{i\pi}{2L}D_0^3\x }
\eea
and  the operators $\psi_{R/L}$  at $t=0$ can then be written in terms of bosonic operators
as
\bean
\psi_R (0,\x)&=& \nor \, e^{-\varphi_R^{(-)} (0,\x)}e^{ \frac{i\pi}{2L}C_0^3\x }\sigma_R e^{ \frac{i\pi}{2L}C_0^3\x }e^{-\varphi_R^{(+)} (0,\x)} \label{psiRbos}\\
\psi_L (0,\x)&=& \nor \, e^{-\varphi_L^{(-)} (0,\x)}e^{- \frac{i\pi}{2L}D_0^3\x }\sigma_L e^{- \frac{i\pi}{2L}D_0^3\x }e^{-\varphi_L^{(+)} (0,\x)}\; .\label{psiLbos}
\eean

\section{\textbf{The Hamiltonian}}

Using the regularized expressions (\ref{Rcurr}) and
(\ref{Rkinetic}) for the currents and the fermion kinetic terms we
obtain the regularized quantum hamiltonian \bean \hat
{H}&=&\int_{-L}^{L} d\x \, \Bigg\{ \half  \left(F^3 \right)^2 +F^+
F^-  -
\frac{1}{\sqrt{2}}\left(\partial_1 F^3 A^3 +\partial_1 F^+ A^- +\partial_1 F^- A^+  \right) \nn \\
&&+\frac{i}{2}\left({\bf :}  \phi_{L}\partial_{1}\phi_{L}{\bf :} +{\bf :} \psi_{L}^\dag \buildrel \leftrightarrow \over
\partial_{1}\psi_{L}{\bf :} +-{\bf :}
\phi_{R}\partial_{1}\phi_{R}{\bf :} -{\bf :} \psi_{R}^\dag  \buildrel \leftrightarrow \over
\partial_{1}\psi_{R}{\bf :}  \right)\nn \\
&&+ g  \left(A^3   J^3 _R  +A^-  J^+ _R   +A^+  J^- _R \right)
+ \frac{g^2}{4\pi}\left[ (A^3)^2 + 2A^+ A^- \right] \Bigg\}
\label{hamiltonian}
\eean
where the products of gauge fields will also have to be
defined.\\
Starting from the Fourier expansions of $A^a _1= \frac{1}{\sqrt {2}}A^a  \,$ and $F^a$ at $t=0$ in the space interval
$[-L,L]$ with the chosen boundary conditions :
\bea
A^3 _1(0,\x)&=&\nor \sum_{N} a^3 _N e^{-ik_N \x}\\
A^{1,2} _1(0,\x)&=&\nor \sum_{n} a^{1,2} _n e^{-ik_n \x}\\
F^3 (0,\x)&=&\nor \sum_{N}b^3 _N e^{-ik_N \x}\\
F^{1,2} (0,\x)&=&\nor \sum_{n}b^{1,2} _n e^{-ik_n \x}
\eea
and using $a^3 _{-N }= {a^3 _N}^ \dag$ , $b^3 _{-N} = {b^3 _N}^ \dag$ , $a^\pm _{-n}= {a_n ^1}^\dag
\pm i {a_n^2}^\dag = {a^\mp _n}^\dag$ , $b^\pm _{-n}={b^\mp _n}^\dag$ , we can write
\bean
A^3 (0,\x)&=&\frac{1}{\sqrt{L}}\,
\sum_{N=1}^{\infty}\left( a^3 _N e^{-ik_N \x}+{a^3 _N }^\dag e^{ik_N \x}\right) +
\frac{1}{\sqrt{L}}\, a_0 ^3 \\
A^\pm (0,\x)&=&\frac{1}{\sqrt{L}} \,\sum_{n=-\infty}^{\infty}a_n^\pm e^{-ik_n\pl\x}=\frac{1}{\sqrt{L}} \,
\sum_{n=\frac{1}{2}}^{\infty}\left( a^\pm _N e^{-ik_n \x}+{a^\mp _n}^\dag e^{ik_n \x}\right) \label{Apm}\\
F^3 (0,\x)&=& \nor \,
\sum_{N=1}^{\infty}\left( b^3 _N e^{-ik_N \x}+{b^3 _N }^\dag e^{ik_N \x}\right)
+ \nor \, b_0 ^3 \\
F^\pm (0,\x)&=&\nor \,\sum_{n=-\infty}^{\infty} b^\pm _n e^{-ik_n \x}=\nor \,\sum_{n=\frac{1}{2}}^{\infty}\left( b^\pm _n e^{-ik_n \x}+{b^\mp _n }^\dag e^{ik_n \x}\right) \label{Fpm}\; .
\eean
From the commutation relations
\[
[ A^a (0,\x), F^b(0,\y)]=i \sqrt{2}\delta^{ab }\delta(\x-\y)
\]
we see that we must have
\bea
&&[ A^\pm (0,\x), F^\mp (0,\y)]=i \sqrt{2}\delta(\x-\y)\\
&&[ A^3 (0,\x), F^3 (0,\y)]=i \sqrt{2}\delta(\x-\y)
\eea
and
\be\label{commab}
[a^3 _M, {b^3 _N}^\dag]=i\delta_{MN}
\quad, \quad
[a^\pm _m, {b^\pm _n}^\dag]=i\delta_{mn}
\quad , \qquad
\ee
all the other commutators vanishing.\\
\lines
We can now write
\bean
\hat{H}&=& H^{0}_F +\half \left(b_0 ^3\right)^2 +\frac{g^2}{2\pi}\left(a_0 ^3\right)^2+
g \sqrt{\frac{2}{L}}    C_0^3 a_0 ^3 \nn \\
&&\!\! \!+\sum_{N=1}^{\infty}\bigg\{ {b^3 _N}^\dag b^3 _N
+i k_N {a^3_N}^\dag b^3 _N -i k_N { b^3 _N}^\dag
 a^3_N +\frac{g^2}{\pi}{a^3_N}^\dag a^3_N +
g\sqrt{\frac{2}{L}}\left(C^3 _N a^3 _N +{C^3 _N}^\dag {a^3 _N}^\dag\right)\!\! \bigg\}\nn\\
&&\!\!\!+ \sum_{n=\half}^{\infty}\bigg\{ {b^+ _n}^\dag b^+ _n +{b^- _n}^\dag b^- _n
+i k_n {a^+_n}^\dag b^+ _n -i k_n { b^- _n}^\dag a^-_n
+i k_n {a^-_n}^\dag b^- _n -i k_n { b^+ _n}^\dag a^+_n   \nn \\
&&\!\!\!\quad+\frac{g^2}{\pi}\left({a^+_n}^\dag a^+_n + {a^-_n}^\dag a^-_n \right)
+g\sqrt{\frac{2}{L}}\left( C_n ^+ a_n^- + C_n ^- a_n^+ + {C_n ^+}^\dag {a_n^-}^\dag
+ {C_n ^-}^\dag {a_n^+}^\dag \right)\!\!\bigg\}\!
\eean
where ${H}^0_F$ is the free fermion Hamiltonian
\be \label{HF}
{H}^0_F=\sum_{n=\half}^{\infty} k_n \left(\beta_n^\dag \beta_n + \delta_n^\dag \delta_n
+b_n^\dag b_n + d_n^\dag d_n \right)+\sum_{N=0}^{\infty}\left(\rho_N^\dag \rho_N +r_N^\dag r_N \right)
\ee
which, as we have seen, can also be expressed as
\be \label{HFbos}
H_F^0=
\frac{\pi}{2L}\Big((C_0^3)^2+ (D_0^3)^2 \Big)+\frac{\pi}{L}\sum_{N=1}^\infty
\left({C_N^3}^\dag  C_N^3 +{D_N^3}^\dag  D_N^3 \right)
+\sum_{N=0}^{\infty}\left(\rho_N^\dag \rho_N +r_N^\dag r_N \right) .
\ee 
\chapter{THE QUANTIZATION OF THE GAUGE FIELD}


\section{\textbf{The Lagrange multipliers and the subsidiary condition.}}
The Lagrange multipliers $\lambda^3$, $\lambda^\pm$ are given in terms of the
other fields by
\bea
\lambda^{3}&=&-\sqrt{2}\partial_1 F^3 -igF^+ A^- +ig F^- A^+ +\sqrt{2}g\tilde{J}^3 _{0}\\
\lambda^{+}&=&-\sqrt{2}\partial_1 F^+ -igF^3 A^+ +ig F^+ A^3 +\sqrt{2}g\tilde{J}^+ _{0}\\
\lambda^{-}&=&-\sqrt{2}\partial_1 F^--igF^- A^3 +ig F^3 A^- +\sqrt{2}g\tilde{J}^- _{0}\; .
\eea
Consider the Fourier expansion
\bea
\lambda^3 (0,\x)&=&\nor \sum_{N=-\infty}^{\infty} \lambda ^3 _N e^{-ik_N \x}\; ,
\\
\lambda^{\pm}(0,\x)&=&\nor \sum_{n=-\infty}^{\infty}\lambda_n^\pm e^{-ik_n \x}\; .
\eea
Note that  from   $\lambda^3=(\lambda^3)^\dag $ and
 $\lambda^\pm=(\lambda^\mp)^\dag $ it follows that
$\lambda^3_{-N}=(\lambda^3_N)^\dag$ and
 $\lambda^\pm_{-n}=(\lambda^\mp_n)^\dag$.
We want to show that the time evolution of the Lagrange multipliers in the Heisenberg
picture is that of free fields satisfying the simple equation $\partial_- \lambda=0$.
In order to see this let us evaluate their commutators with the Hamiltonian.\\
$
[\hat{H} , \lambda^3 (0,\y)]
$
consists of the following terms:

\vspace{-40pt}
\baselineskip36pt
\bea
&1.&ig\int_{-L}^{L}d\x [F^+ (0,\x) F^- (0,\x)\,,\, F^- (0,\y) A^+ (0,\y) -F^+ (0,\x)A^- (0,\y)]\\
&2.& -\frac{ig}{\sq}\int_{-L}^{L}d\x \,[ \, \partial_1 F^+ (0,\x) A^- (0,\x)+\partial_1 F^- (0,\x) A^+ (0,\x)\, , \,
\\
&& \hspace{90pt}
F^- (0,\y) A^+ (0,\y) -F^+ (0,\y)A^- (0,\y)\, ]\\
&3.&\int_{-L}^{L}d\x \,[\, \partial_1 F^3 (0,\x)A^3  (0,\x) \,, \, \partial_1 F^3  (0,\y) \,] \\
&4.& g\pl\sum_{N=1}^{\infty}\, \left[\,{C_N ^3}^\dag C_N ^3+ {D_N ^3}^\dag D_N ^3\, , \, \tilde{J}^3_R (0,\y)+
\tilde{J}^3_L(0,\y)\, \right]\\
&&- g\sq \int_{-L}^{L}d\x \, \left[ \,A^3(0,\x)\tilde{J}_R^3 (0,\x)\,,\, \partial_1 F^3(0,\y)\, \right] \\
&5.&
g^2 \! \int_{-L}^{L}\! d\x \left[A^3(0,\x)\tilde{J}_R^3 (0,\x)\,,\tilde{J}_R^3 (0,\y)\right]\!
-\frac{g^2}{2\sq \pi}\int_{-L}^{L}\! d\x \left[(A^3)^2(0,\x) ,  \partial_1 F^3(0,\y)\right]
\\ &6.&
g^2 \int_{-L}^{L}d\x \, \left[\, A^- (0,\x) \tilde{J}^+_R(0,\x)\, , \, -iF^+ (0,\y) A^- (0,\y)+\tilde{J}^3_R (0,\y)\, \right]
\\ &7.&
g^2 \int_{-L}^{L}d\x \,\left[\,A^+ (0,\x) \tilde{J}^-_R(0,\x)\, , \, iF^- (0,\y) A^+ (0,\y)+\tilde{J}^3_R (0,\y)\, \right]
\\ &8.&
\frac{ig^3}{2\pi} \int_{-L}^{L}d\x \, \left[\,A^+ (0,\x) A^- (0,\x), F^- (0,\y)A^+ (0,\y)-F^+ (0,\y)A^- (0,\y)\, \right]
\eea
We have
\baselineskip24pt
\bea
1.&=&g\sq  \int_{-L}^{L}d\x \left(-F^+ (0,\y) F^- (0,\x) \delta_A (\x-\y)+
F^+ (0,\x) F^- (0,\y) \delta (\x-\y)\right)=0 \\ \\
2. &=&g \int_{-L}^{L}d\x \left(-\partial_{\x}F^+ (0,\x)A^-(0,\y)\delta_A (\x-\y)+F^+ (0,\y)A^-(0,\x)\partial_{\x}\delta_A (\x-\y)\right)\\
& &+g \int_{-L}^{L}d\x \left(\partial_{\x}F^- (0,\x)A^+(0,\y)\delta(\x-\y)-F^- (0,\y)A^+(0,\x)\partial_{\x}\delta(\x-\y)\right)\\
& &=g\, \partial_{\y}\Big(F^- (0,\y)A^+(0,\y)-F^+ (0,\y)A^- (0,\y)\Big)\\ \\
3. &=& i\sq \int_{-L}^{L}d\x \,\partial_{\x}F^3 (0,\x)\partial_{\y}\delta_P(\x-\y)=
i\sq\, \partial_{y}^2 F^3 (0,\y)\\ \\
4. &=&\frac{g}{2L}\sum_{N=1}^{\infty}\left( -k_N C_N^3 e^{ik_N \x}+
k_N {C_N^3}^\dag e^{-ik_N \x} -k_N D_N^3 e^{-ik_N \x}+
k_N {D_N^3}^\dag e^{ik_N \x}\right)+\\
&&-2ig\int_{-L}^{L}d\x \tilde{J}^3(0,\x)\partial_{\y}\delta_P (\x-\y)\\
&=&ig\partial_{\y}\tilde{J}^3_R(0,\y)-ig\partial_{\y}\tilde{J}^3_L(0,\y)-2ig\partial_{\y}\tilde{J}^3_R(0,\y)
=-ig\, \partial_{\y}\Big(\tilde{J}^3_R(0,\y)+\tilde{J}^3_L(0,\y)\Big)\\ \\
5. &=& \int_{-L}^{L}d\x \left[ \frac{g^2}{2L^2}A^3 (0,\x)\sum_{N=1}^{\infty}N \left(e^{ik_N(\x-\y)}-e^{-ik_N(\x-\y)}\right)-\frac{ig^2}{\pi}A^3(0,\x)\partial_{\y}\delta_P (\x-\y)\right]\\
&=& \int_{-L}^{L}d\x \left[ \frac{ig^2}{\pi}A^3(0,\x)\partial_{\y}\delta_P (\x-\y)-\frac{ig^2}{\pi}A^3(0,\x)\partial_{\y}\delta_P (\x-\y)\right]=0\\ \\
6. &=&\sq g^2 \tilde{J}^+_R(0,\y) A^- (0,\y)-\frac{g^2}{2L^2}\int_{-L}^{L}d\x A^- (0,\x) \sum_{n=-\infty}^{\infty}
\sum_{N=-\infty}^{\infty}C^+ _{n+N} e^{ik_n \x}e^{ik_N \y}\\
&=&\sq g^2 \tilde{J}^+_R(0,\y) A^- (0,\y)-\frac{g^2}{2L^2}\int_{-L}^{L}d\x A^- (0,\x)\sum_{n=-\infty}^{\infty}C^+ _{n} e^{ik_n \x}\sum_{N=-\infty}^{\infty}e^{-ik_N (\x-\y)}\\
&=&\sq g^2 \tilde{J}^+_R(0,\y) A^- (0,\y)-g^2\sq\int_{-L}^{L}d\x A^- (0,\x)
\tilde{J}^+_R (0,\x)\delta_P (\x-\y)=0\\ \\
7. &=&\sq g^2 \tilde{J}^-_R(0,\y) A^+ (0,\y)-g^2\sq\int_{-L}^{L}d\x A^+ (0,\x)
\tilde{J}^-_R (0,\x)\delta_P (\x-\y)=0 \\ \\
8. &=& \frac{g^2}{\sq\pi}\int_{-L}^{L}d\x \Big(-A^- (0,\x)A^+ (0,\y)\delta_A (\x-\y)+
A^+ (0,\x)A^- (0,\y)\delta_A (\x-\y)\Big)=0
\eea
\baselineskip30pt
so that
\[
[\hat{H} , \lambda^3 (0,\y)]=-i\partial_{\y} \lambda^3 (0,\y)
\]
and
\[
[\hat{H} , \lambda^3_N]=\nor
\int_{-L}^{L}d\y \, e^{ik_N \y}[\hat{H} , \lambda^3 (0,\y)]=-k_N \lambda^3_N \;.
\]
The time evolution of $\lambda^3$ is given by
\[
e^{i\hat{H}t}\lambda^3 (0,\x)e^{-i\hat{H}t}=\nor \sum_{N=1}^{\infty}\Big( \lambda ^3 _N e^{-ik_N (t+\x)}+ {\lambda ^3 _N }^\dag e^{ik_N (t+\x)}+\lambda_0 ^3\Big)
\]
Let us consider $[\hat{H} , \lambda^+ (0,\y)]$ . A similar calculation gives
\bea
\int_{-L}^{L} d\x \, &\! \textbf{\Big[}&\! \half  \left(F^3 (x)\right)^2 +F^+ (x)F^-  (x) -
\frac{1}{\sqrt{2}}\Big\{\partial_1 F^3 (x) A^3 (x) +\partial_1 F^+ (x)A^- (x)\\
&&+\partial_1 F^- (x)A^+ (x)  \Big\}
+ \frac{g^2}{4\pi}\left\{(A^3)^2 (x)+ 2A^+ (x)A^- (x)\right\}\, , \,
-\sqrt{2}\partial_1 F^+ (y)\\
&&-igF^3 (y) A^+ (y)
+ig F^+ (y) A^3 (y) \,\textbf{ \Big]}_{x_0 =y_0 =0}=\\
&=& -i\partial_{\y} \Big\{ -\sqrt{2}\partial_1 F^+ (0,\y)-igF^3 (0,\y) A^+ (0,\y)+ig
 F^+ (0,\y)A^3 (0,\y) \Big\}\\
&& -\frac{ig^2}{\pi}\partial_{\y}A^+ (0,\y)
\eea
and
\[
g^2 \int_{-L}^{L}d\x \Big\{[A^3 (0,\x) \tilde{J}^3_R(0,\x)\, , \, \tilde{J}^+ _R (0,\y)]+i [A^-  (0,\x)\tilde{J}^+_R (0,\x)\, , \, A^3 (0,\y)F^+ (0,\y) ]\Big\}=0
\]
while
\bea
 &&g^2 \int_{-L}^{L}d\x  \Big\{A^+ (0,\x)\left[\tilde{J}^- _R (0,\x)\, , \,\tilde{J}^+ _R (0,\y)\right]
-i  \tilde{J}^3 _R (0,\x)[A^3 (0,\x)\, , \,
F^3 (0,\y)] A^+ (0,\y) \Big\}\quad  \; \\
&&=g^2 \int_{-L}^{L}d\x  \Big\{\frac{A^+ (0,\x)}{2L^2}\sum_{m,n=-\infty}^\infty e^{ik_n \x}e^{ik_m \y}
[C_n ^-\, , \, C_m ^+] +\sq \tilde{J}^3 _R (0,\x)A^+ (0,\y) \delta (\x-\y) \Big\}\quad\\
&&=-\frac{g^2}{2L^2} \int_{-L}^{L}d\x A^+ (0,\x) \sum_{m,n=-\infty}^\infty e^{ik_n \x}e^{ik_m \y}\left(C^3_{m+n}+m \delta_{m,-n}\right)+g^2 \sq \tilde{J}^3 _R (0,\y)A^+ (0,\y) \\
&&=-\frac{g^2}{L} \int_{-L}^{L}d\x A^+ (0,\x)  \Big\{ \sum_{N=-\infty}^\infty C^3_{N}e^{ik_N \x}
\delta(\x-\y)+\frac{iL}{\pi}\partial_{\x}\delta(\x-\y)\Big\}\hspace{95pt}\\
&&\quad+g^2 \sq \tilde{J}^3 _R (0,\y)A^+ (0,\y) \hspace{290pt}\\
&&=\frac{ig^2}{\pi}\partial_{\y}A^+ (0,\y)\hspace{330pt}
\eea
and
\[
-g \sq  \int_{-L}^{L}d\x  \tilde{J}^+ _R (0,\x) [A^- (0,\x)\, , \, \partial_{\y} F^+ (0,\y)]=
-2ig\partial_{\y}\tilde{J}^+ _R (0,\y)
\]
so that
\bea
[\hat{H} , \lambda^+ (0,\y)]&=&-i\partial_{\y} \Big\{ -\sqrt{2}\partial_1 F^+ (0,\y)-igF^3 (0,\y) A^+ (0,\y)+ig F^+ (0,\y) A^3 (0,\y) \Big\}\\
&&-2ig\partial_{\y}\tilde{J}^+ _R (0,\y) +g \left[ H_F \, , \, \tilde{J}^+_R (0,\y)+  \tilde{J}^+_L (0,\y)\right]\; .
\eea
Let us evaluate $\left[ H_F \, , \, \tilde{J}^+_R (0,\y)\right]$ and $\left[ H_F \, , \, \tilde{J}^+_L (0,\y)\right]$ . We have
\bea
\left[ H_F \, , \,  C_n ^+\right] = \Big[ \sum_{m=\half}^{\infty}k_m (b_m^\dag b_m+d_m^\dag d_m)
+\sum_{M=1}^{\infty}k_M r_M^\dag r_M \; , \,  \sum_{N=0}^\infty r^\dagger_N d_{n+N}\\
\quad -\sum_{j={1 \over 2}}^\infty b^\dagger_j r_{n+j}
-\sum_{j={1 \over 2}}^{n} d_j r_{n-j} \Big]
\eea
and, using the relation
\[
[AB\, , \, CD]= A \{B\, ,C\}D-AC \{B\, , D\}+\{A\, ,  C\} DB-C\{A\, ,  D\} B
\]
we get, for positive $n$,
\bea
\left[ H_F \, , \,  C_n ^+\right] &=&
- \sum_{m,j=\half}^{\infty} k_m b_m^\dag r_{n+j}\delta_{mj}
- \sum_{m=\half}^{\infty}\sum_{N=0}^\infty k_m r^\dagger_N d_m \delta_{m, \, n+N}\\
&&- \sum_{m=\half}^{\infty}\sum_{j={1 \over 2}}^{n}k_m r_{n-j}d_m \delta_{mj}
+\sum_{M=1}^{\infty}  \sum_{N=0}^\infty k_M r_M^\dag  d_{n+N}\delta_{MN}\\
&&+\sum_{M=1}^{\infty}
\sum_{j={1 \over 2}}^\infty k_M b^\dagger_j r_M \delta_{M,\, n+j}
+\sum_{M=1}^{\infty} \sum_{j={1 \over 2}}^{n}k_M d_j  r_M \delta_{M,\, n-j}\\
&=&- \sum_{j=\half}^{\infty}( k_j -k_{n+j}) b_j^\dag r_{n+j}
- \sum_{N=0}^\infty (k_{n+N}-k_N ) r^\dagger_N d_{n+N}\\
&&- \sum_{j={1 \over 2}}^{n}(k_j +k_{n-j})d_j r_{n-j}\\
&=&-k_n C^+_n \;. \label{commutatorofHandC}
\eea
Analogously one obtains
\bea
&&\left[ H_F \, , \,  (C_n ^-)^\dag \right]=k_n (C^-_n)^\dag \\
&&\left[ H_F \, , \,  (D_n ^+) \right]=-k_n D^+_n \\
&&\left[ H_F \, , \,  (D_n ^-)^\dag \right]=k_n (D^-_n)^\dag \; .
\eea
Therefore we have
\[
\left[ H_F \, , \, \tilde{J}^+_R (0,\y)\right]=\frac{1}{\sq L}
\sum_{n=\half}^{\infty}\left(-k_nC_n ^+ e^{ik_n \y}+k_n{C_n ^- }^\dag e^{-ik_n \y} \right)
=i\partial_{\y}\tilde{J}^+_R (0,\y)
\]
and
\[
\left[ H_F \, , \, \tilde{J}^+_L (0,\y)\right]=
\frac{1}{\sq L}
\sum_{n=\half}^{\infty}\left(-k_n D_n ^+ e^{-ik_n \x}+k_n(D_n ^- )^\dag e^{ik_n \x} \right)
=-i\partial_{\y}\tilde{J}^+_L (0,\y) \;.
\]
Finally we can write
\bea
[\hat{H} , \lambda^+ (0,\y)]&=&-i\partial_{\y} \Big\{ -\sqrt{2}\partial_1 F^+ (0,\y)-igF^3 (0,\y) A^+ (0,\y)+ig F^+ (0,\y) A^3 (0,\y) \Big\}\\
&&-ig\partial_{\y}\left(\tilde{J}^+ _R (0,\y) +  \tilde{J}^+_L (0,\y)\right)\\
&=&-i\partial_{\y} \lambda^+ (0,\y)
\eea
and, obviously,
\bea
[\hat{H} , \lambda^- (0,\y)]=-i\partial_{\y} \lambda^- (0,\y)
\eea
so that
\bea
[\hat{H} , \lambda^\pm_n]=\nor \int_{-L}^{L}d\y e^{ik_n\y}[\hat{H} , \lambda^\pm (0,\y)]=
-k_n\lambda_n^\pm \;.
\eea
As a consequence we have
\[\lambda^\pm (t,\x)=
e^{i\hat{H}t}\lambda^\pm (0,\x)e^{-i\hat{H}t}
=\nor \sum_{n=\half}^{\infty}\Big( \lambda ^\pm _n e^{-ik_n (t+x)}
+ (\lambda ^\pm _n )^\dag e^{ik_n (t+x)}\Big)
\]
\baselineskip24pt
We have thus shown that
the Heisenberg field $\lambda(t, \x)$ has a free-field decomposition into positive
and negative frequency components, which is fundamental for a consistent quantization
of the theory. This result guarantees that the decomposition of $\lambda$ into
Fock creation and annihilation operators and the definition of the physical subspace
by means of the subsidiary condition are stable under time evolution. 

Another important result is
\[ [\hat{H}, \lambda_0^3 ]=0\,.
\]
Indeed, the zero mode of $\lambda^3$ is a conserved charge.
In order to satisfy the subsidiary condition,
we shall require that its physical eigenstates have zero eigenvalue.

To further investigate the structure of the physical subspace let us
consider the algebra of the Lagrange multipliers.
Using the canonical commutation relations we get

\vspace{-35pt}
\baselineskip30pt
\bea
[\lambda^3 (0,\x),\lambda^+ (0,\y)]&=&-2g F^+ (0,\y) \partial_{\x} \delta_P (x-y)
-2gF^+ (0,\x) \partial_{\y}\delta_A (\x-\y) \\
&&+ i\sq g^2 F^+ (0,\x) A^3 (0,\y) \delta_A (x-y)\\
&&-i\sq g^2 F^3 (0,\y) A^+ (0,\x) \delta_A (x-y)\\
&& +g^2 [\tilde{J}_R ^3  (0,\x)+ \tilde{J}^3_L (0,\x)\, , \tilde{J}_R ^+ (0,\y)+ \tilde{J}^+_L (0,\y)]
\eea
and
\bea
[\lambda^3 _N, \lambda^+ _m] &=&\frac{1}{2L} \int _{-L}^{L} d\x \, e^{ik_N \x}
 \int _{-L}^{L}d\y \, e^{ik_m \y}\; [\lambda^3 (0,\x),\lambda^+ (0,\y)]\\
&=&\frac{ g}{\sq L} \int _{-L}^{L} d\x \, e^{i(k_N +k_m)\x}
\bigg[ i(k_N+k_m)\sq F^+ (0,\x)+igF^+ (0,\x)A^3 (0,\x)\\
&& \qquad \qquad  \qquad \qquad-ig F^3 (0,\x)A^+ (0,\x)\bigg]+ \frac{g^2}{L} [ {C_{-N} ^3} + D_N^3  ,
C_{-m} ^+  + D_m^+ ]\\
&=&\frac{ g}{\sq L} \int _{-L}^{L} d\x \, e^{i(k_N +k_m)\x}
\bigg[ -\sq \partial_1 F^+ (0,\x)+igF^+ (0,\x)A^3 (0,\x) \\
&&\qquad \qquad \qquad  -ig F^3 (0,\x)A^+ (0,\x)\bigg]+ \frac{g^2}{L}\left( C^+_{-N-m} + D_{N+m}^+  \right) \;
\eea
where  (\ref{JR3}),  (\ref{JL3}), (\ref{JRpm}), (\ref{JLpm}) and (\ref{commc3cpm})
were used.
Finally one has
\[
[\lambda^3 _N, \lambda^+ _m]=\frac{g}{\sqrt{L}}\; \lambda^+ _{N+m}
\]
and, analogously,
\[
[\lambda^3 _N, \lambda^- _m]=-\frac{g}{\sqrt{L}}\; \lambda^- _{N+m} \; .
\]
In particular, for $N=0$
\[
[\lambda^3 _0, \lambda^\pm _m]=\pm \frac{g}{\sqrt{L}}\; \lambda^\pm _{m} \; .
\]
\doublespacing
which shows that $\lambda^\pm$ are charged fields. This result has the important consequence that the subsidiary conditions involving
$\lambda^\pm$ are identically satisfied for states with zero eigenvalue of the charge $\lambda_0^3$ :
\[
\langle phys |\lambda^\pm _n |phys \rangle =\pm\frac{g}{\sqrt{L}}
\langle phys |[\lambda_0^3 \, , \,\lambda^\pm _n ]|phys \rangle = 0
\]
as long as
\[ \lambda_0 ^3 |phys \rangle =0 \; .
\]
Therefore we only need to require that physical states satisfy the conditions
\bea
&& \lambda_N^3 |phys \rangle =0 \quad {\rm for \ N>0}\;, \\
&&\lambda_0^3 \,|phys \rangle =0 \; .
\eea
One can also show that
\bea
[\lambda_n ^+ , \lambda_m ^- ]&=&\frac{ g}{\sq L} \int _{-L}^{L} d\x \, e^{i(k_n +k_m)\x}
\bigg[ -\sq \partial_1 F^3 (0,\x)+igF^- (0,\x)A^+ (0,\x) \\
&&\qquad \qquad \qquad  -ig F^+ (0,\x)A^- (0,\x)\bigg]+ \frac{g^2}{L} [ {C_{-n} ^+} + D_n^+  , C_{-m} ^-  + D_m^- ]
\eea
which, using \ $\left [ C^+_{-n} , C^-_{-m}   \right ] = C^3_{-n-m} - n \delta_{n,-m}$
\  and\   $\left [ D^+_n , D^-_m   \right ] =D^3_{n+m} + n \delta_{n,-m}$,  gives
\[
[\lambda^+ _n, \lambda^- _m]=\frac{g}{\sqrt{L}}\; \lambda^3 _{n+m}
\]
and

\baselineskip30pt
\bea
[\lambda^3(0,\x) \, , \, \lambda^3(0,\y)] &=& g^2[\tilde{J}^3_R(0,\x)\,,\,\tilde{J}^3_R(0,\y)]+ g^2[\tilde{J}^3_L(0,\x)\,,\,\tilde{J}^3_L(0,\y)]\\
&=&\frac{g^2}{2L^2}\sum_{N,M=-\infty}^{\infty}\Big([C^3_N\, , \, C^3_M]  e^{ik_N \x}e^{ik_M \y}
+[D^3_N\, , \, D^3_M]  e^{-ik_N \x}e^{-ik_M \y}\Big)\\
&=&\frac{g^2}{2L^2}\sum_{N=-\infty}^{\infty}\big(Ne^{ik_N( \x- \y)}+Ne^{-ik_N( \x- \y)}\big)=0
\eea
\doublespacing

These relations imply that the Lagrange multipliers generate zero norm states
when applied to physical states. This is consistent with the expectation that
modes of the Lagrange multipliers can be found in zero norm physical states,
in analogy with the Gupta-Bleuler quantization of QED, where zero norm
combinations of the unphysical scalar and longitudinal photons are present
in the physical subspace.

\section{\textbf{Quantization.}}

 As we have seen, the representation chosen for the Fermi
field at $t=0$ is such that the free fermion Hamiltonian is diagonal, or, in other words,
the action of the creation operators $\, b^\dag _n \, , \, \beta ^\dag _n \,, \, d^\dag _n \, , \,
\delta ^\dag _n\, , \, r^\dag _N \, , \, \rho^ \dag _N$ on the Fock vacuum generates
eigenstates of the free Hamiltonian.  Basis states in the free fermion Fock space
are  a suitable starting point for standard perturbative calculations. 

The quantization of the gauge field is more delicate. Let us consider the part of the unperturbed Hamiltonian which involves the non-zero unphysical modes  of the gauge field:
\bean
H_G&=&\sum_{n=\half}^{\infty}\bigg\{ {b^+ _n}^\dag b^+ _n +{b^- _n}^\dag b^- _n
+i k_n {a^+_n}^\dag b^+ _n -i k_n { b^- _n}^\dag a^-_n
+i k_n {a^-_n}^\dag b^- _n -i k_n { b^+ _n}^\dag a^+_n \bigg\} \nn \\
&&+\sum_{N=1}^{\infty}\bigg\{ {b^3 _N}^\dag b^3 _N
+i k_N {a^3_N}^\dag b^3 _N -i k_N { b^3 _N}^\dag
 a^3_N \bigg\} \label{HG}
\; .
\eean
We are naturally led to a Fock representation with a vacuum state defined as the state  $|0\rangle$  such that  $a^{3}_N  |0\rangle = b^{3}_N  |0\rangle =0 $
and $a^{\pm}_n  |0\rangle = b^{\pm}_n  |0\rangle =0 $
for $n, N >0$ . Creation and annihilation operators can then be defined, for $n, N >0$  , as\cite{eliana}:
\bean
A_N ^3 &\equiv & \frac{1}{\sqrt{2L}}\,a^3 _N +i\sqrt {\frac{L}{2}}\,
b^3 _N  \\
A_{-N} ^3 &\equiv & \frac{1}{\sqrt{2L}}\,a^3 _N-i\sqrt {\frac{L}{2}}\,
b^3 _N  \\
A_n ^\pm &\equiv & \frac{1}{\sqrt{2L}}\,a^\pm _n +i\sqrt {\frac{L}{2}}\,
b^\pm _n  \label{Fock+n} \\
A_{-n} ^\pm &\equiv & \frac{1}{\sqrt{2L}}\,a^\mp _n-i\sqrt {\frac{L}{2}}\,
b^\mp _n  \label{Fock-n}\;
\eean
so that
\bea
a_N ^3 =\sqrt{\frac{L}{2}}\left(A^3 _N +A^3 _{-N}\right)\quad &,& \qquad
b_N ^3 = \frac{A^3 _N -A^3 _{-N}}{i\sqrt{2L}}\\
a_n ^\pm =\sqrt{\frac{L}{2}}\left(A^\pm _n +A^\mp _{-n}\right)\quad &,& \qquad
b_n ^\pm = \frac{A^\pm _n -A^\mp _{-n}}{i\sqrt{2L}}
\eea
The commutation relations
\bea
&&[A^3 _M, (A^3 _N)^\dag] =\delta_{MN} \quad , \qquad
[A^3 _{-M}, (A^3 _{-N})^\dag] =-\delta_{MN}\\
&& [A_m ^\pm , (A_ n ^\pm)^\dag]= \delta_{mn} \quad , \qquad
[A^\pm _{-m}, (A^\pm _{-n})^\dag] =-\delta_{mn} \eea can be
represented in a Fock space endowed with an indefinite metric. As
a consequence of the unphysical nature of the degrees of freedom
we are considering, the presence of an indefinite metric is not
surprising and we know that it can be dealt with consistently
provided that its restriction to the physical subspace is positive
semidefinite. Note that (\ref{HG}) is not diagonal in this
representation, nor can it be diagonalized. The vacuum and the
states created out of it by repeated action of the operators
$(b^3_N)^\dag$ and $(b^\pm_n)^\dag$ provide an incomplete set of
eigenstates. This anomalous situation  is related to the fact that
the metric is not positive definite. A similar situation occurs in
the Schwinger model where it can, nonetheless,  be shown that  the
\emph{complete} set of one-particle eigenstates of the full
Hamiltonian, as given by the known solution of the model, can be
obtained perturbatively starting from the
\emph{incomplete }set of unperturbed one-particle eigenstates  (see Appendix A). 

The above quantization of the unphysical non-zero modes of the gauge
field is required for  the non-interacting gauge theory, where the subsidiary
conditions can be expressed as $b_N^3 |phys\rangle=0$ , $b_n^\pm |phys\rangle=0$ (as can easily be
seen by setting $g=0$ in the expressions for the Lagrange multipliers).
The physical subspace can be defined  by the 3 independent conditions 
\[ \left(A_N^3\! -\!A_{-N}^3\right)\!|phys\rangle \!=\!\left(A_n^\pm\! -\!A_{-n}^\pm \right)\!\!|phys\rangle\!=\!0\,,\]
expressed in terms of annihilation operators.
It is then possible to follow the  Gupta-Bleuler procedure and show that the physical subspace has a positive semi-definite metric, with zero-norm states being the ones containing ghost-like
modes, and  \\ $\langle phys| H_G| phys\rangle =0$,  so that unphysical modes do not contribute to the energy spectrum. 

As is characteristic of two-dimensional pure Yang-Mills theories in light-cone gauge, the Hamiltonian (\ref{HG}) has no interaction terms and  coincides with that of free gauge bosons. The interaction is carried by the
Lagrange multipliers and has the effect of modifying the subsidiary condition and the physical subspace.
We expect more restrictive conditions as a consequence of the g-dependent terms in $\lambda$.
The colour components of $\lambda$ do not commute with one another and the subsidiary conditions
are not independent. As a matter of fact, the Lagrange multipliers satisfy the same algebra as in the previously considered case
where fermions  are present. The conditions that need to be imposed are $\widetilde{\lambda}^3_0|phys\rangle=0$
and $\widetilde{\lambda}^3_N |phys\rangle=0$, for $N>0$ where
\bea
\widetilde{\lambda}^3_0 &=&
\frac{ig}{\sqrt{2}L}\,\sum_{m=\half}^{\infty}\bigg( (a^-_m)^\dag b^-_{m}+(b^+_m)^\dag a^+_{m}
-(a^+_m)^\dag b^+_{m}-(b^-_m)^\dag a^-_{m}\bigg)\\
\widetilde{\lambda}^3_N &=&\frac{i}{\sqrt{L}}\, k_N b_N^3
+\frac{ig}{\sqrt{2}L}\; \sum_{m=\half}^{N-\half}\left(b_m^- a^+_{N-m}-b_m^+ a^-_{N-m}\right)\nn \\
&+&\!\frac{ig}{\sqrt{2}L}\, \sum_{m=\half}^{\infty}\! \bigg( (a^-_m)^\dag b^-_{m+N}+(b^+_m)^\dag a^+_{m+N}
-(a^+_m)^\dag b^+_{m+N}-(b^-_m)^\dag a^-_{m+N}\bigg) \; .
\eea
We can  see that the eigenstates of (\ref{HG}) generated by the action of $ b_n^\pm$
are no longer physical states as in the free case. We need to require the more restrictive condition that no modes of $A^\pm$ be present in the physical subspace. Only physical states with modes of $A^3$  are now
zero-norm states and again one has  $\langle phys| H_G| phys\rangle =0.$ 

This indefinite metric representation of the gauge field, suggested by the free nature of the Hamiltonian
associated with it,  turns out to be unsuitable for the quantization of the full non-abelian gauge theory, on account
of its residual gauge invariance. To see this let us consider the operator
\be \label{U}
U(x)=e^{iN\pl (t+\x)\tau^3}
\ee
It satisfies the condition $U(t,-L)=U(t, L)$ and $A_- ^\prime =UA_-U^\dag +\frac{i}{g}\partial_-U U^\dag=0$.\\
It leaves the gauge-fixing condition invariant and it preserves
the boundary conditions. It is, therefore, a residual gauge
symmetry of the theory. Let us see its action on $A$.
 \\ Using $[\tau^+,\tau^-]=\tau^3$ and $[\tau^3, \tau^\pm]=\pm \tau^\pm$ we get:
\[
A ^\prime =UAU^\dag +\frac{i}{g}\partial_+U U^\dag = e^{iN\pl (t+\x)}A^- \tau^+ + e^{-iN\pl (t+\x)}A^+ \tau^-
-\frac{N\pi \sq}{gL}\tau^3
\]
or
\[
{A^-} ^\prime =e^{iN\pl (t+\x)}A^- \quad, \qquad {A^+} ^\prime =e^{-iN\pl (t+\x)}A^+ \quad, \qquad
{A^3} ^\prime=A^3-\frac{N\pi \sq}{gL}
\]
and from $F^\prime=UFU^\dag $
we have
\[
{F^-} ^\prime =e^{iN\pl (t+\x)}F^- \quad, \qquad {F^+} ^\prime =e^{-iN\pl (t+\x)}F^+ \quad, \qquad
{F^3} ^\prime=F^3 \;.
\]
Let us concentrate on the transformation properties of the + and $-$ colour components.  A quantum operator $T^N$ representing this symmetry in the space of states must be such that the quantum fields represented at
$t=0$ as in (\ref{Apm}) and (\ref{Fpm}) have the following transformation properties:
\bea
T^NA^- (0,\x) (T^N)^{\dag}&=&\frac{1}{\sqrt{L}} \,\sum_{n=-\infty}^{\infty}a_n^- e^{-in\pl\x}e^{iN\pl\x}=
\frac{1}{\sqrt{L}} \,\sum_{n=-\infty}^{\infty}a_{n+N}^- e^{-in\pl\x}\\
T^NA^+ (0,\x) (T^N)^{\dag}&=&\frac{1}{\sqrt{L}} \,\sum_{n=-\infty}^{\infty}a_n^+ e^{-in\pl\x}e^{-iN\pl\x}=
\frac{1}{\sqrt{L}} \,\sum_{n=-\infty}^{\infty}a_{n-N}^+ e^{-in\pl\x}\\
T^NF^-  (0,\x) (T^N)^{\dag}&=&\nor \,\sum_{n=-\infty}^{\infty} b^- _n e^{-in\pl \x}e^{iN\pl\x}=
\nor \,\sum_{n=-\infty}^{\infty} b^- _{n+N} e^{-in\pl \x}\\
T^NF^+  (0,\x) (T^N)^{\dag}&=&\nor \,\sum_{n=-\infty}^{\infty} b^+ _n e^{-in\pl \x}e^{-iN\pl\x}=
\nor \,\sum_{n=-\infty}^{\infty} b^+ _{n-N} e^{-in\pl \x}\;.
\eea
We see that must have
\bea
T^N a_n^+ (T^N)^{\dag}=a^+_{n-N} \quad, \qquad T^N b_n^+ (T^N)^{\dag}=b^+_{n-N} \\
T^N a_n^- (T^N)^{\dag}=a^-_{n+N} \quad, \qquad T^N b_n^- (T^N)^{\dag}=b^-_{n+N}
\eea
or, in terms of the Fock creation and annihilation operators defined in (\ref{Fock+n}--\ref{Fock-n}):
\bean
&&T^N A_n^+ (T^N)^{\dag}=A^+_{n-N} \qquad  \hspace{23pt} {\textstyle {\rm for} \ n\geq  N+\frac{1}{2}}\\
&&T^N A_{n}^+ (T^N)^{\dag}=(A^+_{n-N})^\dag
\hspace{10pt}\qquad {\textstyle {\rm for} \ \half \leq n \leq  N-\frac{1}{2}}\label{half1}\\
&&T^N A_{-n}^+ (T^N)^{\dag}=A^+_{-n-N}\qquad  \quad{\textstyle {\rm for} \ n\geq \frac{1}{2}}\\
&&T^NA_{-n}^- (T^N)^{\dag}=A^-_{-n+N}\hspace{12pt}\qquad {\textstyle {\rm for} \ n\geq  N+\frac{1}{2}}\\
&&T^N A_{-n}^- (T^N)^{\dag}=(A^-_{N-n})^\dag  \hspace{4pt}\qquad {\textstyle {\rm for} \ \half \leq n \leq  N-\frac{1}{2}}\label{half2}\\
&&T^N A_n^- (T^N)^{\dag}=A^-_{n+N}\hspace{11pt}\qquad  \quad{\textstyle {\rm for} \ n\geq \frac{1}{2}}
\eean
We can see from (\ref{half1}) and (\ref{half2}) that $T$ must turn annihilation operators into creation operators.
This does not allow the vacuum to be invariant. The transformed vacuum $T|0\rangle$
must be such that $(A^+_{-\half})^\dag T|0\rangle=0$, a condition which cannot be satisfied by a state in
the Fock space we are considering. The symmetry of the theory
under index-shifting at  a classical level suggests that in a Fock quantization the creation or annihilation
nature of the operators must be preserved  under index-shifting. Interpreting $(A^\pm_{-n})^\dag$ as creation operators generating negative-norm states when acting on the vacuum appears to be inconsistent with  this symmetry
transformation. As we shall see, implementing this symmetry as a unitary operator in the Hilbert space will be necessary to obtain the non-trivial vacuum
structure which is characteristic of this theory when coupled to fermions.
In the quantization of $A^\pm$ in the free gauge theory, as well as  of  $A^3$  in both the free and
the interacting case, the indefinite metric is necessary to express the subsidiary
condition in terms of annihilation operators and it allows to get rid of ghost-like modes, which are present in zero-norm physical states, by constructing a Hilbert space with positive definite metric as a quotient space.  But in the
quantization of the interacting $A^\pm$ the indefinite metric does not seem to play a crucial role. As a matter of fact, a standard definition of creation and annihilation operators with canonical commutators:
\[
 A_n ^\pm \equiv  \frac{1}{\sqrt{2L}}\,a^\pm _n +i\sqrt {\frac{L}{2}}\,
b^\pm _n \ ,   \qquad [A_m ^\pm , (A_ n ^\pm)^\dag]= \delta_{mn} \ ,
\]
for both positive and negative $n$, leads to the following expressions for $\widetilde{\lambda}_0^3 $ and
$\widetilde{\lambda}_N^3$:
\bea
\widetilde{\lambda}^3_0 &=&
\frac{g}{\sqrt{2}L}\sum_{m=-\infty}^{\infty}\bigg( (A_{m}^-)^\dag A_m^-
- (A_{m}^+)^\dag A_m^+ \bigg)\\
\widetilde{\lambda}^3_N &=&\frac{i}{\sqrt{L}}\, k_N b_N^3
+\frac{g}{\sqrt{2}L}\sum_{m=-\infty}^{\infty}\bigg( (A_{m-N}^-)^\dag A_m^-
- (A_{m-N}^+)^\dag A_m^+ \bigg)\; .
\eea
The subsidiary conditions can be satisfied by requiring that
\bea
&&(A_N^3-A_{-N}^3) |phys\rangle =0\ , \forall N>0 \\
&&A_n^\pm |phys\rangle=0
\eea
and (\ref{HG}) can be written as
\bea
H_G&=&\sum_{n=-\infty}^{\infty}\left\{ \left(k_n+\frac{1}{2L}\right) \bigg( (A_n^+)^\dag A_n^+ +(A_n^-)^\dag A_n^- \bigg)
-(A_n^+)^\dag (A_{-n}^-)^\dag -A_n^+ A_{-n}^- \right\} \\
&&+\sum_{N=1}^{\infty}\bigg\{ {b^3 _N}^\dag b^3 _N
+i k_N {a^3_N}^\dag b^3 _N -i k_N { b^3 _N}^\dag
 a^3_N \bigg\} \; .
\eea
The vacuum is the only physical state in the positive metric Fock representation of $A^+$ and $A^-$
and, although it is not an eigenstate of $H_G$,  we still have $\langle 0| H_G|0\rangle=0$.
The transformation $U$ can now be represented by an operator $T^N$ such that:
\bean
&&T^N A_n^+ (T^N)^{\dag}=A^+_{n-N}\quad, \quad \qquad
T^N (A_n^+)^\dag (T^N)^{\dag}=(A^+_{n-N})^\dag \label {TA+}\\
&&T^N A_n^- (T^N)^{\dag}=A^-_{n+N}\quad, \quad \qquad
T^N (A_n^+)^\dag (T^N)^{\dag}=(A^+_{n-N})^\dag \label{TA-}
\eean
for any positive or negative $n$.\\
One can check that (\ref{TA+}--\ref{TA-}) are satisfied for $N=1$ by
\newpage
\beal \label{operator}
\widetilde{T}=&\cdots &e^{\frac{\pi}{2}\left( A_n^ + (A^+_{n-1})^\dag - (A_n^+)^\dag  A^+_{n-1}\, +\,
A_{-n}^- (A^-_{-n+1})^\dag - (A_{-n}^-)^\dag A^-_{-n+1}\right)}\cdots \\
&\cdots& e^{\frac{\pi}{2}\left( A_{ 3/2}^+ (A^+_{1/2})^\dag - (A_{3/2}^+)^\dag  A^+_{1/2}\,+\,
A_{-3/2}^- (A^-_{-1/2})^\dag - (A_{-3/2}^-)^\dag A^-_{-1/2}\right)}\\
&& e^{\frac{\pi}{2}\left( A_{ 1/2}^+ (A^+_{-1/2})^\dag - (A_{1/2}^+)^\dag  A^+_{-1/2}\,+\,
A_{-1/2}^- (A^-_{1/2})^\dag - (A_{-1/2}^-)^\dag A^-_{1/2}\right)} \\
&& e^{\frac{\pi}{2}\left( A_{- 1/2}^+ (A^+_{-3/2})^\dag - (A_{-1/2}^+)^\dag  A^+_{-3/2}\,+\,
A_{1/2}^- (A^-_{3/2})^\dag - (A_{1/2}^-)^\dag A^-_{3/2}\right)} \cdots \\
&\cdots &e^{\frac{\pi}{2}\left( A_{-n}^ + (A^+_{-n-1})^\dag - (A_{-n}^+)^\dag  A^+_{-n-1}\, +\,
A_{n}^- (A^-_{n+1})^\dag - (A_{n}^-)^\dag A^-_{n+1}\right)}\cdots
\eeal

Although unphysical, the modes of $A^+$ and $A^-$ interact with fermions and can no longer be
eliminated from the theory when coupling to fermions is considered.
The fact that the gauge-invariant vacuum state associated with a positive-definite Fock
representation of $A^+$ and $A^-$ is not an eigenstate of the unperturbed Hamiltonian
prevents us from performing a standard perturbative calculation.  On the other hand,
 as a result of the gauge-invariant renormalization of the fermion products, the term
$\, \frac{g^2}{2\pi}A^+ A^- \,$ has been introduced into the theory. Being quadratic in
the gauge field it has the well known form of a mass term, the ``mass'' being $m\equiv \frac{g}{\sqrt{\pi}}$. This suggests treating it as part of the unperturbed Hamiltonian,
in spite of its dependence on $g^2$, leaving only  the order-g terms $g (A^+  \widetilde{J}^-_R +
A^-  \widetilde{J}^+_R) \, $,
which couple the gauge fields to the fermion currents, in the perturbation
Hamiltonian.  By doing so we can diagonalize the ``unperturbed" hamiltonian related to the $+$ and $-$ gauge fields. We can write
\bea
H_G=\sum_{n=-\infty}^{\infty}(k_n+m)\left( (A^+_n)^\dag A^+_n + (A^-_n)^\dag A^-_n\right)
+\sum_{N=1}^{\infty}\bigg\{ {b^3 _N}^\dag b^3 _N
+i k_N {a^3_N}^\dag b^3 _N -i k_N { b^3 _N}^\dag
 a^3_N  \bigg\}
\eea
where now the Fock operators $A_n^\pm$ are defined as
\[
A_n ^\pm  \equiv  \sqrt{\frac{m}{2}}\,a^\pm _n +\frac{i}{\sqrt{2m}}\,b^\pm _n
\]
for both positive and negative $n$, with  $m=\frac{g}{\sqrt{\pi}}$.\\
Inverting these relations we obtain
\bea
a_n ^\pm =\frac{A^\pm _n +(A^\mp _{-n})^\dag}{\sqrt{2m}}\quad &,& \qquad
b_n ^\pm =\sqrt{\frac{m}{2}}\, \frac{A^\pm _n -(A^\mp _{-n})^\dag}{i}\;.
\eea
From (\ref{commab}) we see that these operators satisfy the Fock algebra
\[
[A_m ^\pm , {A_ n ^\pm}^\dag]= \delta_{mn}
\]
all the other commutators vanishing.\\
Analogously, we quantize the zero mode of $A^3 $ by defining
\[
A_0^3 \equiv  \sqrt{\frac{m}{2}}\,a^3 _0 +\frac{i}{\sqrt{2m}}\,b^3 _0
\]
so that we can write
\[
\half \left(b_0 ^3\right)^2 +\frac{g^2}{2\pi}\left(a_0 ^3\right)^2=
m{A_0^3}^\dag A_0^3 \;.
\]
The Hamiltonian can now be written as
\[
\hat{H}= H_0+H_I
\]
where
\bea
H_0=H_F^0 +
\sum_{N=1}^{\infty}\bigg\{ {b^3 _N}^\dag b^3 _N
+i k_N {a^3_N}^\dag b^3 _N -i k_N { b^3 _N}^\dag
 a^3_N  \bigg\} +m (A_0^3)^\dag A_0^3\\
+\sum_{n=-\infty}^{\infty}
\Big(k_n+m\Big) \bigg( (A_n^+)^\dag A_n^+ + (A_n^- )^\dag A_n^- \bigg)
\eea
and
\bea
H_I=
\sum_{N=1}^{\infty}\bigg\{
g\sqrt{\frac{2}{L}}\left(C^3 _N a^3 _N +{C^3 _N}^\dag {a^3 _N}^\dag\right)
 +\frac{g^2}{\pi}{a^3_N}^\dag a^3_N\bigg\} \Bigg\} +\frac{g}{\sqrt{mL}}C_0^3 \Big((A_0^3)^\dag + A_0^3 \Big)\\
+ \frac{g}{\sqrt{mL}}\, \Bigg\{ \sum_{n=-\infty}^{\infty}\bigg( A_n^- C_n^+ +A_n^+ C_n^- +(A_n^-)^\dag (C_n^+)^\dag +(A_n^+)^\dag (C_n^-)^\dag \bigg)
\eea
$H_F^0$ is the free fermion Hamiltonian (\ref{HF}-- \ref{HFbos}).\\ 
\chapter{THE VACUUM}
\vspace{-12pt}
Let
\bea
|\Omega\rangle &=& |0\rangle + |\Omega^{(1)}\rangle + |\Omega^{(2)}\rangle +\dots\\
E&=&E_0 + E_1 + E_2 +\dots
\eea
We shall determine the corrections to the vacuum state, $ |\Omega^{(1)}\rangle$ and $ |\Omega^{(2)}\rangle$, and to the
vacuum energy, $E_1$ and $E_2$, by requiring that
\bean
&&H_0 |0\rangle = E_0  |0\rangle\\
&&H_0  |\Omega^{(1)}\rangle + H_1  |0\rangle = E_0  |\Omega^{(1)}\rangle + E_1  |0\rangle \label{first} \\
&&H_0  |\Omega^{(2)}\rangle +  H_1 |\Omega^{(1)}\rangle = E_0  |\Omega^{(2)}\rangle +  E_1 |\Omega^{(1)}\rangle +E_2  |0\rangle \;.\label{second}
\eean
$H_0$ is normal-ordered in such a way that
\[
H_0  |0\rangle =  0 \
\quad {\rm and}\quad  E_0=0
\]
while higher order corrections to the energy are not necessarily zero.\\
 We have
\bea
H_I |0\rangle=
\sum_{N=1}^{\infty}
g\sqrt{\frac{2}{L}}(C^3 _N)^\dag {a^3 _N}^\dag|0\rangle  
 +\frac{g}{\sqrt{mL}}\sum_{n=-\infty}^{\infty}\bigg((A_n^-)^\dag  (C_n^+)^\dag |0\rangle +(A_n^+)^\dag (C_n^-)^\dag |0\rangle \bigg)  \,.
\eea
One can immediately see that
\[
E_1=\langle 0|H_I |0\rangle =0\,.
\]
As we have seen (page\pageref{commutatorofHandC}),  $\ [H_F\, , \, C_n^+]=-k_n C_n^+$,  so that we have
\[
[H_0 \, , \, C_n^+]=-k_n C_n^+ \quad {\rm and } \quad [H_0 \, , \, (C_n^+)^\dag]=k_n (C_n^+)^\dag \, .
\]
Analogously one can prove that
\[
[H_0 \, , \, C_n^-]=-k_n C_n^- \quad {\rm and } \quad [H_0 \, , \, (C_n^-)^\dag]=k_n (C_n^-)^\dag \, .
\]
These relations, together with $H_0 |0\rangle=0$ , tell us that
\[
H_0 (C_n^\pm )^\dag |0\rangle = k_n (C_n^\pm )^\dag |0\rangle\, ,
\]
while from the expressions of $C_n^\pm$ (\ref{Cn+}--\ref{Cn-}) we see that
\[
C_n^\pm |0\rangle = (C_{-n}^\mp)^\dag |0\rangle = 0 \,.
\]
We also have $C_N^3 |0\rangle = (C_{-N}^3)^\dag \0 = 0$ and, from (\ref{HFbos}),
\[
H_0 (C_N^3)^\dag  |0\rangle = k_N (C_N^3)^\dag  |0\rangle \,.
\]
It is now easy to verify that the state
\bea
|\Omega^{(1)}\rangle
&=&
-\frac{g \sq}{\sqrt{L}}\sum_{N=1}^{\infty}\left(\frac{{a_N^3}^\dag}{2k_N}+\frac{i{b_N^3}^\dag}{(2k_N)^2}
\right){C_N^3}^\dag \0 \\
&&-\frac{g}{\sqrt{mL}}\sum_{n=\half}^{\infty}\frac{1}{2k_n+m}\bigg( (A_n^-)^\dag (C_n^+)^\dag |0\rangle +(A_n^+)^\dag (C_n^-)^\dag |0\rangle \bigg)
\eea
satisfies (\ref{first}) with $E_0=E_1=0$ .\\
In order to evaluate $E_2$ and $|\Omega^{(2)}\rangle$ we need $H_I |\Omega^{(1)}\rangle$. Using (\ref{commc3}), (\ref{commcpcm}) we get
\bea
H_I |\Omega^{(1)}\rangle \!\!\!&=&\!\!\!
\frac{2g^2}{L} \sum_{N=1}^{\infty}\frac{N}{(2k_N)^2}
-\frac{g^2}{mL}\sum_{n=\half}^{\infty}\frac{2n}{2k_n+m}|0\rangle  \\
\! \!\!&-&\!\!\!\frac{2g^2}{L} \sum_{M,N=1}^{\infty}{a_M^3 }^\dag {C_M^3}^\dag
\left(\frac{{a_N^3}^\dag}{2k_N}+\frac{i{b_N^3}^\dag}{(2k_N)^2}
\right){C_N^3}^\dag \0\\
\!\!\!&-&\!\!\! \frac{g^2 \sq}{\sqrt{m}L}\sum_{M=1}^{\infty}\sum_{n=\half}^{\infty}
\frac{{a_M^3}^\dag (C_M^3)^\dag \bigg( (A_n^-)^\dag  (C_n^+)^\dag +(A_n^+)^\dag (C_n^-)^\dag  \bigg)}{2k_n+m}|0\rangle\\
\!\!\!&-&\!\!\!  \frac{g^2 \sq}{\sqrt{m}L} \sum_{m=-\infty}^{\infty}\sum_{N=1}^{\infty}\bigg((A_m^-)^\dag   (C_m^+)^\dag +(A_m^+)^\dag (C_m^-)^\dag \bigg) \left(\frac{{a_N^3}^\dag}{2k_N}+\frac{i{b_N^3}^\dag}{(2k_N)^2}
\right){C_N^3}^\dag \0 \\
\!\!\!&-&\!\!\! \frac{g^2}{mL}\sum_{n=\half}^{\infty}\frac{ (A_0^3)^\dag
 C_0^3 \bigg( (A_n^-)^\dag  (C_n^+)^\dag +(A_n^+)^\dag (C_n^-)^\dag  \bigg)}{2k_n+m}|0\rangle\\
\!\!\!&-&\!\!\! \!\! \frac{g^2}{mL} \sum_{m=-\infty}^{\infty}\sum_{n=\half}^{\infty}\frac{\bigg((A_m^-)^\dag   (C_m^+)^\dag +(A_m^+)^\dag (C_m^-)^\dag  \bigg) \bigg((A_n^-)^\dag   (C_n^+)^\dag +(A_n^+)^\dag (C_n^-)^\dag  \bigg)}{2k_n+m}\0
 \;.
\eea
 We therefore have
\[
E_2=
\frac{2g^2}{L} \sum_{N=1}^{\infty}\frac{N}{(2k_N)^2}
-\frac{g^2}{mL}\sum_{n=\half}^{\infty}\frac{2n}{2k_n+m}|0\rangle
\]
a diverging quantity that has to be subtracted from the Hamiltonian. \\
It is not hard  to verify  that (\ref{second}) is satisfied if the state $|\Omega^{(2)}\rangle$ is given by

\baselineskip18pt
\bea \label{2}
|\Omega^{(2)}\rangle  \!\!\!&=&\!\!\!
\frac{g^2}{L}\, \sum_{M=1}^{\infty}\sum_{N=1}^{\infty}
\left(\frac{{a_M^3}^\dag}{2k_M}+\frac{i{b_M^3}^\dag}{(2k_M)^2}
\right)\left(\frac{{a_N^3}^\dag}{2k_N}+\frac{i{b_N^3}^\dag}{(2k_N)^2}
\right){C_M^3}^\dag {C_N^3}^\dag \0 \\ \\
&\;+&\!\!\!\! \frac{g^2 \sq}{\sqrt{m}L}\sum_{M=1}^{\infty}\sum_{n=\half}^{\infty}\Bigg(\frac{{a_M^3}^\dag}{2k_{M}+2k_n +m}\\
&&\hspace{70pt}+\frac{i{b_M^3}^\dag}{(2k_{n}+2k_M +m)^2} \Bigg){C_M^3}^\dag \, \frac{ (A_n^-)^\dag  (C_n^+)^\dag +(A_n^+)^\dag (C_n^-)^\dag }{2k_n+m}\0 \\ \\
 &\;+&\!\!\!\! \frac{g^2 \sq}{\sqrt{m}L}\sum_{N=1}^{\infty}\sum_{p=-\infty}^{\infty}
\Bigg\{ \frac{1}{2k_{p}+2k_N +m} \left(\frac{{a_N^3}^\dag}{2k_N}+\frac{i{b_N^3}^\dag}{(2k_N)^2}
\right) \\
&&\hspace{65pt}+ \frac{i{b_N^3}^\dag}{2k_N (2k_{p}+2k_N +m)^2}\Bigg\}
\bigg((A_p^-)^\dag   (C_p^+)^\dag +(A_p^+)^\dag (C_p^-)^\dag  \bigg){C_N^3}^\dag \0\\ \\
&\;+&\!\!\!\!\frac{g^2}{mL}\sum_{n=\half}^{\infty} \frac{ (A_0^3)^\dag
 \bigg({- (A_n^-)^\dag  (C_n^+)^\dag +(A_n^+)^\dag (C_n^-)^\dag }\bigg)}
{2(k_{n}+m) (2k_n+m)}|0\rangle\\ \\
&\,+&\!\!\!\! \!\frac{g^2}{mL} \sum_{p=\half}^{\infty}\sum_{n=\half}^{\infty} \left[\frac{(A_p^+)^\dag (A_n^+)^\dag(C_p^-)^\dag   (C_n^-)^\dag  }{2(2k_p +m)(2k_n+m)}\0
+\frac{(A_p^-)^\dag (A_n^-)^\dag(C_p^+)^\dag   (C_n^+)^\dag  }{2(2k_p +m)(2k_n+m)}\0\right]\\ \\
&+&\!\!\!\! \! \!\frac{g^2}{mL}\sum_{p=-\infty}^{\infty}\sum_{n=\half}^{\infty} \left[ \frac{(A_p^-)^\dag (A_n^+)^\dag(C_p^+)^\dag   (C_n^-)^\dag  }{2(k_{n}+k_p +m)(2k_n+m)}\0
+\frac{(A_p^+)^\dag (A_n^-)^\dag(C_p^-)^\dag   (C_n^+)^\dag  }{2(k_{n}+k_p +m)(2k_n+m)}\0\right]
.
\eea
\lines

\section{\textbf{The subsidiary condition.}}

Let us verify that the state $|\Omega \rangle = \0 +| \Omega^{(1)}\rangle +|\Omega^{(2)}\rangle $  satisfies the subsidiary condition.
It is easy to see that
\[ \lambda_0^3 |\Omega \rangle =0
\]
where
 \[
\lambda_0^3 =\frac{g}{\sq L}\left( C_0^3+ D_0^3 \right)+\frac{g}{\sq L}\sum_{m=-\infty}^{\infty}\bigg( (A_{m}^-)^\dag A_m^-
- (A_{m}^+)^\dag A_m^+ \bigg) \,.
\]
As a matter of fact we have
\bea
&&\lambda_0^3 \0 =  0 \\
&&\lambda_0^3 \1 = -\frac{g^2}{L \sqrt{2mL}} \, \Bigg\{ \sum_{n=\half}^{\infty}\frac{C_0^3}{2k_n+m}\bigg( (A_n^-)^\dag (C_n^+)^\dag |0\rangle +(A_n^+)^\dag (C_n^-)^\dag |0\rangle \bigg)\\
&& \hspace{60pt}+\sum_{n=\half}^{\infty}\frac{1}{2k_n+m}\bigg( (A_n^-)^\dag (C_n^+)^\dag |0\rangle -(A_n^+)^\dag (C_n^-)^\dag |0\rangle \bigg) \Bigg\}
\eea
Using  the relation $ [C_0^3, (C_n^\pm)^\dag ] =\mp (C_{n}^\pm)^\dag$ it is easy
to see that $\lambda_0^3 \1 =0$ and that the same holds for every term in $\2$
(p. \pageref{2}). Each term in the perturbative expansion of the physical vacuum
$  |\Omega \rangle $ is an eigenstate of  the conserved charge $\lambda_0^3  $ with
eigenvalue $0$. As we have seen, in order to be a physical state, the vacuum must also be annihilated by the positive frequency components of $\lambda^3$. This means that it must satisfy $ \lambda^3_N  |\Omega \rangle =0 $ where
\bea
\lambda^3_N &=&\frac{ik_N b_N^3}{\sqrt{L}}+ \frac{g}{\sq L}\left( (C_N^3)^\dag + D_N^3 \right)\\
&&+\frac{g}{\sq L}\sum_{m=-\infty}^{\infty}\bigg( (A_{m-N}^-)^\dag A_m^-
- (A_{m-N}^+)^\dag A_m^+ \bigg)\; .
\eea
Clearly this condition cannot be satisfied term by term in the expansion of
$  |\Omega \rangle $,  as is the case for $\lambda_0^3$. The action of   $\lambda_N^3$
mixes up the perturbative orders and the condition cannot be satisfied exactly
by our perturbative evaluation of the vacuum. We can only check that it holds for
the two lowest orders in the expansion of  $\lambda_N^3 |\Omega \rangle$.
We have
\bea
&&\lambda_N^3 \0 =\frac{g}{\sq L} (C_N^3)^\dag \0 \\
&&\lambda_N^3 \1 = -\frac{g}{\sq L}
(C_N^3)^\dag \0 -\frac{g^2}{L \sqrt{L}}\sum_{M=1}^\infty \left(\frac{{a_M^3}^\dag}{2k_M}+\frac{i{b_M^3}^\dag}{(2k_M)^2}
\right) (C_N^3)^\dag (C_M^3)^\dag \0 \\
&&\hspace{60pt}-\frac{g^2}{L \sqrt{2mL}}\sum_{n=\half}^\infty \frac{(C_N^3)^\dag}{2k_n +m}
 \left( (C_n^- )^\dag (A_n^+ )^\dag + (C_n^+ )^\dag (A_n^- )^\dag \right) \0 \\
&&\hspace{60pt}-\frac{g^2}{L \sqrt{2mL}}\sum_{n=\half}^\infty \frac{1}{2k_{n}+m}\left( (C_{n}^+)^\dag (A_{n-N}^-)^\dag - (C_{n}^-)^\dag (A_{n-N}^+)^\dag \right)\0  \,.
  \eea
Disregarding higher order terms we can write
\bea
\lambda_N^3 \1 \simeq -\frac{g}{\sq L} (C_N^3)^\dag \0
-\frac{g^2}{L \sqrt{L}}\sum_{M=1}^\infty \left(\frac{{a_M^3}^\dag}{2k_M}+\frac{i{b_M^3}^\dag}{(2k_M)^2}
\right) (C_N^3)^\dag (C_M^3)^\dag \0 \\
-\frac{g^2}{L\sqrt{2mL}}\Bigg\{
 \sum_{n=\half}^\infty \frac{(C_N^3)^\dag}{2k_n }
 \left( (C_n^- )^\dag (A_n^+ )^\dag + (C_n^+ )^\dag (A_n^- )^\dag \right) \0 \hspace{10pt}\\
\hspace{65pt}+\sum_{n=\half}^\infty \frac{1}{2k_{n}}\left( (C_{n}^+)^\dag (A_{n-N}^-)^\dag - (C_{n}^-)^\dag (A_{n-N}^+)^\dag \right)\0 \Bigg\}
\eea
and keeping only the lowest order terms in $\lambda_N^3 \2 $ we get
\bea
\lambda_N^3 \2 &\simeq &
\frac{g^2}{L \sqrt{L}}\sum_{M=1}^\infty \left(\frac{{a_M^3}^\dag}{2k_M}+\frac{i{b_M^3}^\dag}{(2k_M)^2}
\right) (C_N^3)^\dag (C_M^3)^\dag \0 \\
&&+\frac{g^2}{L\sqrt{2mL}}\Bigg\{
 \sum_{n=\half}^\infty \frac{k_N (C_N^3)^\dag \bigg( (A_n^+ )^\dag(C_n^- )^\dag +
(A_n^- )^\dag(C_n^+ )^\dag \bigg)}{2k_n k_{n+N}} \0 \\
&&\hspace{50pt}+ \sum_{n=-\infty}^\infty \frac{ \bigg( (A_n^+ )^\dag(C_n^- )^\dag +
(A_n^- )^\dag(C_n^+ )^\dag \bigg)(C_N^3)^\dag}{2 k_{n+N}} \0 \Bigg\} \,.
\eea
Using the relation $ [(C_n^\pm)^\dag  , (C_N^3)^\dag] =\pm (C_{n+N}^\pm)^\dag$ we get
\bea
 \lambda_N^3 \2 &\simeq &
\frac{g^2}{L \sqrt{L}}\sum_{M=1}^\infty \left(\frac{{a_M^3}^\dag}{2k_M}+\frac{i{b_M^3}^\dag}{(2k_M)^2}
\right) (C_N^3)^\dag (C_M^3)^\dag \0 \\
&&+\frac{g^2}{L\sqrt{2mL}}\Bigg\{
 \sum_{n=\half}^\infty \frac{1}{2k_n } (C_N^3)^\dag \bigg( (A_n^+ )^\dag(C_n^- )^\dag +
(A_n^- )^\dag(C_n^+ )^\dag \bigg) \0 \\
&&\hspace{50pt}+ \sum_{n=-N+\half}^\infty \frac{1}{2 k_{n+N}} \bigg( -(A_n^+ )^\dag(C_{n+N}^- )^\dag +
(A_n^- )^\dag(C_{n+N}^+ )^\dag \bigg) \0 \Bigg\}
\eea
and one can immediately see that
\[\lambda_N^3 \Big(\0 + \1 +\2 \Big) \simeq 0 \; . \]
Note that this condition would not be satisfied even for the two lowest orders
if $A^3$ had been quantized with a positive definite metric,
while no such inconsistency appears as a consequence of our quantization
of $A^+$ and $A^-$.

\section{\textbf{The degenerate vacua}.}

We have seen that under the residual gauge transformation
\be
U(x)=e^{iN\pl (t+\x)\tau^3}
\ee
the gauge fields transform according to
\bea
&&{A^-} ^\prime =e^{iN\pl (t+\x)}A^- \quad, \qquad {A^+} ^\prime =e^{-iN\pl (t+\x)}A^+ \quad, \qquad
{A^3} ^\prime=A^3-\frac{N\pi \sq}{gL}\\
&&{F^-} ^\prime =e^{iN\pl (t+\x)}F^- \quad, \qquad {F^+} ^\prime =e^{-iN\pl (t+\x)}F^+ \quad, \qquad
{F^3} ^\prime=F^3\;.
\eea
As a consequence, under the action of the operator $T^N$ representing this transformation, the
Fourier modes into which the fields are decomposed at $t=0$ transform as
\bean
&&T^N a_n^+ (T^N)^{\dag}=a^+_{n-N} \quad, \qquad \hspace{34pt} T^N b_n^+ (T^N)^{\dag}=b^+_{n-N} \label{ta+t}\\
&&T^N a_n^- (T^N)^{\dag}=a^-_{n+N} \quad, \qquad  \hspace{34pt}  T^N b_n^- (T^N)^{\dag}=b^-_{n+N}\label{ta-t}\\
&&T^N a_M^3 (T^N)^{\dag}=a^3_{M} \quad {\rm for}\ M\neq 0 \\
&&T^N a_0^3 (T^N)^{\dag}=a^3_{0}-\frac{N\pi \sq}{g\sqrt{L}}\label{ta0t}\\
&&T^N b_M^3 (T^N)^{\dag}=b^3_{M}\;.
\eean
From (\ref{ta+t}) and (\ref{ta-t}) we also get
\bean
&&T^N A_n^+ (T^N)^{\dag}=A^+_{n-N}\\
&&T^N A_n^- (T^N)^{\dag}=A^-_{n+N}
\eean
Let us consider the transformation of the Fermi field. It is easy to see that
\[
\Psi^\prime = U\Psi U^\dag = e^{iN\pl (t+\x)}\psi \tau^+ + e^{-iN\pl (t+\x)}\psi^\dag \tau^- +\phi \tau^3
\;.\]
Therefore, at $t=0$,
\[
\psi^\prime (0,\x) =e^{iN\pl \x}\psi (0,\x) \quad, \qquad \phi^\prime(0,\x) = \phi (0,\x)
\]
From the bosonized form of $\psi_{R/L}$ (eqs. \ref{psiRbos}, \ref{psiLbos}), using relations (\ref{sigmasigma+R}---\ref{1.28}) and the identity
\[
e^A B=Be^A e^c  \quad {\rm if}\  [A, B]=cB , \quad {\rm where}\ c \ {\rm is \ a}\ {\rm c-number}
\; ,\]
one can easily prove that
\[
\sigma_L^\dag \sigma_R e^{i\pi (C_0^3 + D_0^3)}\psi (0,\x) \left(\sigma_L^\dag \sigma_R e^{i\pi (C_0^3 + D_0^3)}\right)^\dag =e^{iN\pl \x}\psi (0,\x) \;.
\]
We also have
 \begin{large}
\[
e^{-\frac{i\pi\sq}{g\sqrt{L}}b_0^3}a_0^3\, e^{\, \frac{i\pi\sq}{g\sqrt{L}}b_0^3}
={\textstyle a_0^3-\frac{\pi\sq}{g\sqrt{L}} }\;.
\]
\end{large}
\hspace{-5pt}The operator $T$ representing the residual gauge transformation can therefore
be expressed as
\begin{large}
\[
T=\sigma_L^\dag \sigma_R e^{i\pi (C_0^3 + D_0^3)}e^{-\frac{i\pi\sq}{g\sqrt{L}}b_0^3}\,\widetilde{T}\;.
\]
\end{large}
\hspace{-7pt}
where $\widetilde{T}$ is the operator transforming $A^+$ and $A^-$
(p. \pageref{operator}) .
In order to determine the transformation properties of $C^\pm_n$ let us consider
$\psi^\prime_R$.  We can write
\bea
\psi^\prime _R(0,\x)&=& {1 \over \sqrt {2L}} \sum_{n=\half}^\infty
\left(b_n e^{in\pl \x}e^{iN\pl \x} + d{_n ^\dag} e^{-in\pl \x}e^{iN\pl \x} \right)\\
&=&{1 \over \sqrt {2L}}\Bigg( \sum_{n={N+\half}}^\infty
b_{n-N} e^{in\pl \x} + \sum_{n=-N+\half}^\infty d{_{n+N} ^\dag} e^{-in\pl \x}\Bigg) \;.
\eea
For $N>0$  $\psi^\prime _R$ can be written as
\[
\psi^\prime _R (0,\x)= {1 \over \sqrt {2L}}\left( \sum_{n={N+\half}}^\infty
b_{n-N} e^{in\pl \x} + \sum_{n=\half}^{N-\half} d{_{N-n} ^\dag} e^{in\pl \x}
+\sum_{n=\half}^\infty d{_{n+N} ^\dag} e^{-in\pl \x}\right)
\]
and for $N<0$
\[
\psi^\prime _R (0,\x)= {1 \over \sqrt {2L}}\left( \sum_{n=\half}^\infty
b_{n-N} e^{in\pl \x} + \sum_{n=\half}^{-N-\half} b_{-n-N}  e^{-in\pl \x}
+\sum_{n=-N+\half}^\infty d{_{n+N} ^\dag} e^{-in\pl \x}\right)\;.
\]
We can therefore see that we must have
\bea
T^N b_n (T^N)^{\dag}&=&b_{n-N}\quad \ {\rm for} \ N<n \\
T^N b_n (T^N)^{\dag}&=&d_{N-n}^\dag  \quad {\rm for} \ N>n \\
T^N d_n (T^N)^{\dag}&=&d_{n+N}^\dag  \quad \ {\rm for} \ N>-n \\
T^N d_n (T^N)^{\dag}&=&b_{-N-n}^\dag  \quad {\rm for} \ N<-n
\;.
\eea
We can now determine the transformation properties of $C_n^\pm$. \\
For $n>N>0$ we have
\bea
T^N C_n^- (T^N)^{\dag}&=&\sum_{m=\half}^{\infty}d_{m+N}^\dag r_{n+m}
-\sum_{M=0}^\infty r_M^\dag b_{M+n-N}\\
&&-\sum_{m=N+\half}^n r_{n-m}b_{m-N}-\sum_{m=\half}^{N-\half}r_{n-m}
d_{N-m}^\dag\\
&=&\sum_{m=N+\half}^{\infty}d_{m}^\dag r_{n-N+m}
-\sum_{M=0}^\infty r_M^\dag b_{M+n-N}\\
&&-\sum_{m=\half}^{n-N} r_{n-N-m}b_{m}+\sum_{m=\half}^{N-\half}
d_{m}^\dag r_{n-N+m}\\
&=&C_{n-N}^-
\eea
and for $N>n>0$
\bea
T^N C_n^- (T^N)^{\dag}&=&\sum_{m=\half}^{\infty}d_{m+N}^\dag r_{n+m}
-\sum_{M=0}^{N-n-\half} r_M^\dag d_{N-n-M}^\dag\\
&&-\sum_{M=N-n+\half}^\infty r_{M}^\dag b_{M+n-N}-\sum_{m=\half}^{n}r_{n-m}
d_{N-m}^\dag\\
&=&\sum_{M=n+\half}^{\infty}d_{M-n+N}^\dag r_{M}
+\sum_{M=0}^{n-\half}d_{M-n+N}^\dag r_M \\
&&-\sum_{m=\half}^{N-n} r_{N-n-m}^\dag d_{m}^\dag -\sum_{m=\half}^{\infty}
r_{m+N-n}^\dag b_{m} \\
&=&(C_{N-n}^+ )^\dag \equiv C_{n-N}^-
\eea
Analogously it is possible to show that for any positive or negative $N$
\bean
T^N C_n^- (T^N)^{\dag}&=&C_{n-N}^-  \label{tc-t}\\
T^N C_n^+ (T^N)^{\dag}&=&C_{n+N}^+ \;. \label{tc+t}
\eean
It follows from (\ref{ta+t}), ({\ref{ta-t}), (\ref{ta0t}) and (\ref{tc-t}--\ref{tc+t})
that the action of the transformation $T^N$ on the Hamiltonian (\ref{hamiltonian})
is given by
\[
T^N \hat{H}(T^N)^\dag =\hat{H}-\frac{g^2 N \pi}{2L}(C_0^3 +D_0^3 ) +\frac{ig^2 N\pi}{2L}
\sum_{n=\half}^\infty \left( (a_n^+)^\dag b_n^+ + ( b_n^-)^\dag a_n^-  -(a_n^-)^\dag b_n^-
-( b_n^+)^\dag a_n^+ \right)\,. \\
\]
$\hat{H}$ is not invariant under the action of $T^N$ but
\[
T^N \hat{H}(T^N)^\dag =\hat{H}-\frac{gN \pi}{\sq}\lambda_0^3
\]
which means that $\hat{H}$ is invariant when restricted to the physical subspace.
Note also that
\[
T^N \lambda_0^3 (T^N)^\dag =\lambda_0^3
\]
so that  if $ \lambda_0^3 |\varphi\rangle =0$ then also $ \lambda_0^3 T^N |\varphi\rangle =0$ and
\[
\hat{H}T^N |\varphi\rangle = T^N\hat{H}(T^N)^\dag T^N  |\varphi\rangle
+\frac{g N \pi}{\sq}\lambda_0^3  T^N  |\varphi\rangle =T^N \hat{H}|\varphi\rangle\;.
 \]
In particular this means that the states
\[
|\Omega_N\rangle \equiv T^N |\Omega\rangle \ , \quad N=0,\pm1, \pm 2, \dots
\]
are an infinite set of degenerate vacua. These states are clearly
not gauge-invariant.  Physically acceptable gauge-invariant vacua
can be obtained by constructing superpositions that diagonalize
the operators $T^N$. Let us consider the states
\[
|\theta\rangle \equiv \sum_{N=-\infty}^\infty e^{-iN\theta}|\Omega_N \rangle
\]
known as $\theta-vacua$. We have
\[
T^M |\theta\rangle =  \sum_{N=-\infty}^\infty e^{-iN\theta}|\Omega_{N+M} \rangle
=e^{iM\theta}|\theta\rangle
\]
so that $|\theta\rangle$ is invariant up to a phase factor under the action of $T^M$.

The theory is also invariant \cite{paper} under the transformation $R$ such that
\bea
&&R\psi R^{-1}=\psi^\dag \\
&&R\phi R^{-1}=-\phi \\
&&RA^\pm R^{-1}= A^\mp\\
&&RA^3 R^{-1}= -A^3
\eea
corresponding to the $SU(2)$ transformation $U$=\begin{large}$e^{i\pi\tau_1}$\end{large}.\\
The action of $R$ on the fermion Fock operators is
\bea
&&Rb_n R^{-1} =d_n \qquad   \\
&&R\beta_n R^{-1} =\delta_n \\
 &&Rr_N R^{-1} =-r_N   \qquad  \\
&&R\rho_N R^{-1} =-\rho_N \;.
\eea
As a consequence we have
\bea
&&R C_n^\pm R^{-1}= C_n^\mp  \\
&&RC_N^3 R^{-1}=-C_N^3 \;.
\eea
The gauge Fock operators transform as
\bea
&&RA_n^\pm R^{-1}= A_n^\mp\\
&&RA_N^3 R^{-1}= -A^3_N \;.
\eea
Note that the state $R \0$ is annihilated by all the destruction operators and, since $R^2=1$,
we must have
\[
R \0 = \pm \0 \;.
\]
Without loss of generality we may take $R \0=\0$. This relation, together with the transformation properties of the Fock operators, defines the action of $R$ on all states. \\
One can immediately see that the state $|\Omega\rangle$ is invariant under the action of $R$:
\[
R|\Omega\rangle =|\Omega\rangle \;.
\]
Let us consider now the action of $R$ on the other vacuum states $|\Omega_N \rangle\equiv T^N | \Omega\rangle$.
From the definition of the spurion operators (p. \pageref{sigmaRL}) it is not hard to
 see that
\[
R\sigma_{R/L}R^{-1}= \sigma_{R/L}^\dag
\]
and it is straightforward to verify that
\[
RTR^{-1}= -T^\dag
\]
and
\[
RT^N R^{-1}= (-1)^N (T^\dag)^N \equiv  (-1)^N T^{-N}\;.
\]
As a consequence, $R$ interchanges $|\Omega_N \rangle$ and $|\Omega_{-N} \rangle$
\[
R|\Omega_N \rangle =(-1)^N |\Omega_{-N} \rangle \;.
\]
By applying $R$ to the $\theta$-vacuum we obtain
\[
R|\theta\rangle = \sum_{N=-\infty}^{\infty}e^{-iN\theta}R|\Omega_{N} \rangle
= \sum_{N=-\infty}^{\infty}e^{-iN\theta}e^{iN\pi}|\Omega_{-N} \rangle
= \sum_{N=-\infty}^{\infty}e^{-iN(\pi-\theta)}|\Omega_{N} \rangle
\]
and we see that only two values of the parameter $\theta$ , namely $\theta=\pm \frac{\pi}{2}$, give rise to states which are invariant under both the $T$
and $R$ residual symmetries. We therefore have two physically acceptable vacua, in agreement
with  refs. \cite{witten}\cite{smilga}\cite{lenz}\cite{paper}. 
\chapter{THE CONDENSATE}
We want to use our results for the vacuum to obtain a perturbative evaluation of the gauge-invariant  fermion condensate, defined as
\[
\frac{\langle\theta | {\rm Tr}\bar{\Psi}\Psi | \theta \rangle}{\langle\theta  | \theta \rangle} \;.
\]
We have
\[
{\rm Tr}\bar{\Psi}\Psi =i {\rm Tr }(\Psi_L^\dag \Psi_R-\Psi_R^\dag\Psi_L) =
i\left(\phi_L \phi_R + \psi_L^\dag \psi_R -  \psi_R^\dag \psi_L \right)
\]
and
\[
\langle\theta | {\rm Tr}\bar{\Psi}\Psi | \theta \rangle
=i\! \sum_{N,M=-\infty}^\infty e^{i (M-N)\theta}\, \langle \Omega_M |\left(\phi_L \phi_R + \psi_L^\dag \psi_R -  \psi_R^\dag \psi_L \right) |\Omega_N\rangle \;.
\]
Being a time-independent quantity, the fermion condensate can be evaluated at $t=0$. \\
Writing
\[
\Omega_N =\Omega_N^{(0)} +\Omega_N^{(1)} +\Omega_N^{(2)} +\ldots
\]
we have
\[
\langle\theta | {\rm Tr}\bar{\Psi}\Psi | \theta \rangle
=i \sum_{j,k} \sum_{N,M=-\infty}^\infty e^{i (M-N)\theta}\, \langle \Omega_M^{(j)} |\left(\phi_L \phi_R + \psi_L^\dag \psi_R -  \psi_R^\dag \psi_L \right)\!(0,\x)\;|\Omega_N^{(k)}\rangle
\]
where
\[
|\Omega^{(i)} _{N} \rangle= T^N |\Omega^{(i)} \rangle \quad {\rm and}
 \quad  | \Omega^{(0)} \rangle \equiv \0 \;.
\]
Explicitly
\[
| \Omega^{(0)}_N\rangle = T^N \0  = e^{-\frac{iN\pi\sq}{g\sqrt{L}}b_0^3} \left( \sigma_L^\dag \sigma_R\right)^N \0 \equiv
 e^{-\frac{iN\pi\sq}{g\sqrt{L}}b_0^3}|N\rangle
\]
where
\bea
&&|N\rangle =\beta_{N-\half}^\dag d_{N-\half}^\dag\cdots \beta_{\half}^\dag d_{\half}^\dag \0 \qquad
{\rm for} \ N>0\\
&&|N\rangle =\delta_{N-\half}^\dag b_{N-\half}^\dag\cdots \delta_{\half}^\dag b_{\half}^\dag \0 \qquad
\;{\rm for} \ N<0 \, ,
\eea
\bea
&&|\Omega^{(1)}_N\rangle =
-\frac{g \sq}{\sqrt{L}}\sum_{N=1}^{\infty}\left(\frac{{a_N^3}^\dag}{2k_N}+\frac{i{b_N^3}^\dag}{(2k_N)^2}
\right){C_N^3}^\dag  | \Omega^{(0)}_N\rangle \\
&&\hspace{25pt}-\frac{g}{\sqrt{mL}}
 \sum_{n=\half}^{\infty}\frac{1}{2k_n+m}\bigg( (A_{n+N}^-)^\dag (C_{n+N}^+)^\dag | \Omega^{(0)}_N\rangle  +(A_{n-N}^+)^\dag (C_{n-N}^-)^\dag | \Omega^{(0)}_N\rangle  \bigg)\, ,
\eea
\vskip0pt

\baselineskip15pt
\bea
|\Omega^{(2)}_N\rangle  \!\!\!&=&\!\!\!
\frac{g^2}{L}\, \sum_{M=1}^{\infty}\sum_{J=1}^{\infty}
\left(\frac{{a_M^3}^\dag}{2k_M}+\frac{i{b_M^3}^\dag}{(2k_M)^2}
\right)\left(\frac{{a_J^3}^\dag}{2k_J}+\frac{i{b_J^3}^\dag}{(2k_J)^2}
\right){C_M^3}^\dag {C_J^3}^\dag  | \Omega^{(0)}_N\rangle \\ \\
&\;+&\!\!\!\! \frac{g^2 \sq}{\sqrt{m}L}\sum_{M=1}^{\infty}\sum_{n=\half}^{\infty}\Bigg(\frac{{a_M^3}^\dag}{2k_{M}+2k_n +m}\\
&&\hspace{30pt}+\frac{i{b_M^3}^\dag}{(2k_{n}+2k_M +m)^2} \Bigg){C_M^3}^\dag \, \frac{ (A_{n+N}^-)^\dag  (C_{n+N}^+)^\dag +(A_{n-N}^+)^\dag (C_{n-N}^-)^\dag }{2k_n+m}| \Omega^{(0)}_N\rangle \\ \\
 &\;+&\!\!\!\! \frac{g^2 \sq}{\sqrt{m}L}\sum_{J=1}^{\infty}\sum_{p=-\infty}^{\infty}
\Bigg\{ \frac{1}{2k_{p}+2k_J +m} \left(\frac{{a_J^3}^\dag}{2k_J}+\frac{i{b_J^3}^\dag}{(2k_J)^2}
\right) \\
&&\hspace{15pt}+ \frac{i{b_J^3}^\dag}{2k_J (2k_{p}+2k_J +m)^2}\Bigg\}
\bigg((A_{p+N}^-)^\dag   (C_{p+N}^+)^\dag +(A_{p-N}^+)^\dag (C_{p-N}^-)^\dag  \bigg){C_J^3}^\dag
| \Omega^{(0)}_N\rangle \\ \\
&\;+&\!\!\!\! \frac{g^2}{mL}\Bigg\{ \sum_{n=\half}^{\infty} \frac{\left( (A_0^3)^\dag -\frac{N\pi \sqrt{m}}{g\sqrt{L}}\right)
 \bigg({ -(A_{n+N}^-)^\dag  (C_{n+N}^+)^\dag +(A_{n-N}^+)^\dag (C_{n-N}^-)^\dag }\bigg)}
{2(k_{n}+m) (2k_n+m)}| \Omega^{(0)}_N\rangle\\ \\
&& \hspace{20pt}+\sum_{p=\half}^{\infty}\sum_{n=\half}^{\infty} \frac{(A_{p-N}^+)^\dag (A_{n-N}^+)^\dag(C_{p-N}^-)^\dag   (C_{n-N}^-)^\dag  }{2(2k_p +m)(2k_n+m)}| \Omega^{(0)}_N\rangle\\ \\
&& \hspace{20pt}+ \sum_{p=\half}^{\infty}\sum_{n=\half}^{\infty}+\frac{(A_{p+N}^-)^\dag (A_{n+N}^-)^\dag(C_{p+N}^+)^\dag   (C_{n+N}^+)^\dag  }{2(2k_p +m)(2k_n+m)}|\Omega^{(0)}_N\rangle\\ \\
&& \hspace{20pt}+\sum_{p=-\infty}^{\infty}\sum_{n=\half}^{\infty} \frac{(A_{p+N}^-)^\dag (A_{n-N}^+)^\dag(C_{p+N}^+)^\dag   (C_{n-N}^-)^\dag  }{2(k_{n}+k_p +m)(2k_n+m)}
| \Omega^{(0)}_N\rangle\\ \\
&& \hspace{20pt}+\sum_{p=-\infty}^{\infty}\sum_{n=\half}^{\infty}
\frac{(A_{p-N}^+)^\dag (A_{n+N}^-)^\dag(C_{p-N}^-)^\dag   (C_{n+N}^+)^\dag  }{2(k_{n}+k_p +m)(2k_n+m)}|\Omega^{(0)}_N\rangle
 \Bigg\}.
\eea
\lines
We define
\[
|\theta ^{(i)}\rangle = \sum_{N=-\infty}^{\infty}e^{-iN\theta}|\Omega^{(i)} _{N} \rangle
\]
so that we can write
\[
|\theta\rangle= |\theta ^{(0)}\rangle +|\theta ^{(1)}\rangle + |\theta ^{(2)}\rangle +\ldots
\]
and
\[
 \langle\theta | {\rm Tr}\bar{\Psi}\Psi | \theta \rangle=\sum_{i,j}
\langle\theta^{(i)} | {\rm Tr}\bar{\Psi}\Psi | \theta^{(j)} \rangle
\]

\section{\textbf{The complex field contribution.}}

Let us consider
\[
\langle \theta^{(0)} |\psi_L^\dag (0,\x) \psi_R (0,\x) | \theta^{(0)}\rangle = \sum_{N,M=-\infty}^\infty e^{i (M-N)\theta}\, \langle \Omega_M^{(0)}| \psi_L^\dag (0,\x) \psi_R (0,\x) | \Omega^{(0)}_N\rangle \;.
\]
From the bosonized expressions (\ref{psiRbos}--\ref{psiLbos}) we get
\bea
\langle \Omega_M^{(0)}| \psi_L^\dag (0,\x) \psi_R (0,\x) | \Omega^{(0)}_N\rangle
 &=& \frac{1}{2L}\langle \Omega_M^{(0)}| \sigma_L^\dag \sigma_R | \Omega^{(0)}_N\rangle\\
&=& \frac{1}{2L}\langle 0|\left( \factor\right) ^{N-M}\left( \sigma_L^\dag \sigma_R\right)^{N+1-M} | 0\rangle\\
&=&  \frac{1}{2L}\langle 0|e^{\frac{i\pi\sq}{g\sqrt{L}}b_0^3} | 0\rangle \delta_{M,N+1} \;.
\eea
Using the relation
\[
e^{A+B}=e^A e^B e^{-\half [A,B]} \qquad {\rm if } \quad  [A,B]= c-{\rm number}
\]
we can write
\[
 \langle 0|e^{\frac{i\pi\sq}{g\sqrt{L}}b_0^3} | 0\rangle =
 \langle 0|e^{-\frac{\sqrt{\pi}}{\sqrt{mL}}(A_0^3)^\dag } e^{\frac{\sqrt{\pi}}{\sqrt{mL}}A_0^3}
e^{-\frac{\pi}{2mL}}| 0\rangle =e^{-\frac{\pi}{2mL}}
\]
so that
\be\label{zero}
\langle \Omega_M^{(0)}| \psi_L^\dag (0,\x) \psi_R (0,\x) | \Omega^{(0)}_N\rangle
= \frac{1}{2L}e^{-\frac{\pi}{2mL}} \delta_{M,N+1}
\ee
and
\[
\langle \theta^{(0)}|\psi_L^\dag (0,\x) \psi_R (0,\x) | \theta^{(0)}\rangle= \frac{1}{2L} \sum_{N=-\infty}^\infty e^{i\theta}\, e^{-\frac{\pi}{2mL}}\;.
\]
Analogously, from
\[
\langle \Omega_M^{(0)}| \psi_R^\dag (0,\x) \psi_L (0,\x) | \Omega^{(0)}_N\rangle
= \frac{1}{2L}e^{-\frac{\pi}{2mL}} \delta_{N,M+1}
\]
we get
\[
\langle \theta^{(0)}|\psi_R^\dag (0,\x) \psi_L (0,\x) | \theta^{(0)}\rangle = \frac{1}{2L} \sum_{N=-\infty}^\infty e^{-i\theta}\, e^{-\frac{\pi}{2mL}}\;.
\]
and finally
\be  \label{condzero}
\langle \theta^{(0)}|i\left(\psi_L^\dag (0,\x) \psi_R (0,\x) -\psi_R^\dag (0,\x) \psi_L (0,\x)\right)| \theta^{(0)}\rangle=
 -\frac{1}{L}
\sum_{N=-\infty}^\infty \sin \theta e^{-\frac{\pi}{2mL}} \ee Note
that \be \label{normzero} \langle \theta ^{(0)}|
\theta^{(0)}\rangle= \sum_{N,M=-\infty}^\infty e^{i (M-N)\theta}\,
\langle \Omega_M^{(0)}| \Omega^{(0)}_N\rangle =
\sum_{N,M=-\infty}^\infty e^{i (M-N)\theta}\delta_{M,N}=
\sum_{N=-\infty}^\infty 1 \ee so that the infinite sum over $N$
disappears when we divide by the norm of the \\ $\theta$-vacuum:
\[
\frac{ \langle \theta^{(0)}|\, i\left(\psi_L^\dag (0,\x) \psi_R (0,\x) -\psi_R^\dag (0,\x) \psi_L (0,\x)\right)| \theta^{(0)}\rangle}{\!\langle\theta^{(0)}  | \theta ^{(0)}\rangle}
=  -\frac{1}{L}\sin \theta e^{-\frac{\pi}{2mL}}\;.
\]
One immediately sees that
\[
 \langle \theta^{(0)}|\, i\left(\psi_L^\dag (0,\x) \psi_R (0,\x) -\psi_R^\dag (0,\x) \psi_L (0,\x)\right)| \theta^{(1)}\rangle=0 \;.
\]
Let us consider now
\[
 \langle \theta^{(1)}|\, \psi_L^\dag (0,\x) \psi_R (0,\x) | \theta^{(1)}\rangle=
\sum_{N,M=-\infty}^\infty e^{i (M-N)\theta}\, \langle \Omega_M^{(1)}| \psi_L^\dag (0,\x) \psi_R (0,\x) | \Omega^{(1)}_N\rangle \;.
\]
We have
\baselineskip20pt
\bea
\langle \Omega_M^{(1)}|\, \psi_L^\dag (0,\x) \psi_R (0,\x) | \Omega_N^{(1)}\rangle &\!\!\!=&\!\!\!
 \frac{g^2}{mL}\, \Bigg\{-m \sum_{J=1}^{\infty}\frac{\langle \Omega^{(0)}_M | C_J^3 \psi_L^\dag (0,\x) \psi_R (0,\x) (C_J^3)^\dag | \Omega^{(0)}_N\rangle}{2(k_J )^3} \\ \\
&&\hspace{-18pt}+\sum_{n,p=\half}^{\infty}\frac{\langle \Omega^{(0)}_M |C_{p+M}^+ \psi_L^\dag (0,\x) \psi_R (0,\x)  (C_{n+N}^+)^\dag  | \Omega^{(0)}_N\rangle}{(2k_n+m)(2k_p+m)}\delta_{p+M, n+N} \\ \\
&&\hspace{-18pt}+\sum_{n,p=\half}^{\infty}\frac{\langle \Omega^{(0)}_M |C_{p-M}^- \psi_L^\dag (0,\x) \psi_R (0,\x)  (C_{n-N}^-)^\dag  | \Omega^{(0)}_N\rangle}{(2k_n+m)(2k_p+m)}\delta_{p-M, n-N}
\Bigg\}
\eea
\lines
Using the bosonized form for $\psi$ and the fact that
\[
e^A B=Be^A +[A,B]e ^A \qquad {\rm if} \ [A,B]=c-{\rm number}
\]
it is easy to prove that
\bean
&&\psi_R  (0,\x) (C_N^3)^\dag = \left( (C_N^3)^\dag + e^{ik_N\x}\right)\psi_R (0,\x) \label{psic}\\
&&C_N^3 \psi_R (0,\x)= \psi_R (0,\x) \left(C_N^3 -e^{-ik_N \x}\right)\label{cpsi}
\eean
so that, considering that $[\psi_L , C_N^3]=0$ , we have
\bea
\langle\Omega^{(0)}_M | C_J^3 \psi_L^\dag (0,\x) \psi_R (0,\x) (C_J^3)^\dag | \Omega^{(0)}_N\rangle
=\langle \Omega^{(0)}_M |  \psi_L^\dag (0,\x) \psi_R (0,\x)C_J^3 (C_J^3)^\dag | \Omega^{(0)}_N\rangle \nn \\
-e^{-ik_J \x}\langle \Omega^{(0)}_M |  \psi_L^\dag (0,\x) \psi_R (0,\x) (C_J^3)^\dag | \Omega^{(0)}_N\rangle
\hspace{-30pt}
\nn \\
=(J-1) \langle\Omega^{(0)}_M |\psi_L^\dag (0,\x) \psi_R (0,\x) | \Omega^{(0)}_N\rangle
\eea
and from (\ref{zero})
\be\label{useful}
\langle\Omega^{(0)}_M | C_J^3 \psi_L^\dag (0,\x) \psi_R (0,\x) (C_J^3)^\dag | \Omega^{(0)}_N\rangle
=(J-1) \frac{1}{2L}e^{-\frac{\pi}{2mL}} \delta_{M,N+1}\;.
\ee
Let us consider now the $+$ and $-$ terms.\\
The following commutation relations can easily be verified:
\bean
&&[b_m , C_n^+ ]=-r_{n+m} \label{commbc}\\
&&[b_m^\dag , C_n^+ ]=[b_n , C_m^- ]=0\\
&&[b_m^\dag , C_n^- ]= r_{m-n}^\dag \theta (m-n+\half)+ r_{n-m}\theta (n-m+\half)\\
&&[d_m , C_n^- ]=r_{n+m}\\
&&[d_m^\dag , C_n^- ]=[d_n , C_m^+ ]=0 \\
&&[d_m^\dag , C_n^+ ]= -r_{m-n}^\dag \theta (m-n+\half)- r_{n-m}\theta (n-m+\half)
 \label{commddagc}
\eean
As a consequence
\bea
&&[\psi_R , C_n^-]=[\psi_R , (C_n^+)^\dag]=0\\
&&[\psi_R^\dag, C_n^+ ]=[\psi_R^\dag, (C_n^-)^\dag]=0\;.
\eea
Therefore

\baselineskip10pt
\bea
&&\sum_{n,p=\half}^{\infty}\frac{\langle \Omega^{(0)}_M |C_{p+M}^+  \psi_L^\dag (0,\x) \psi_R (0,\x)  (C_{n+N}^+)^\dag  | \Omega^{(0)}_N\rangle}{(2k_n+m)(2k_p+m)}\, \delta_{p+M, n+N}=\\ \\
&&= \sum_{p=\half}^{\infty}\frac{\langle \Omega^{(0)}_M |C_{p+M}^+   (C_{p+M}^+)^\dag \psi_L^\dag (0,\x) \psi_R (0,\x) | \Omega^{(0)}_N\rangle}{(2k_{p+M-N }+m)(2k_p+m)}\,\theta (p+M-N) 
\eea
and
\bea
&&\sum_{n,p=\half}^{\infty}\frac{\langle \Omega^{(0)}_M |C^-_{p-M} \psi_L^\dag (0,\x) \psi_R (0,\x)  (C_{n-N}^-)^\dag  | \Omega^{(0)}_N\rangle}{(2k_n+m)(2k_p+m)}\, \delta_{p-M, n-N}=\\ \\
&&= \sum_{n=\half}^{\infty}\frac{\langle \Omega^{(0)}_M | \psi_L^\dag (0,\x) \psi_R (0,\x) C_{n-N}^-   (C_{n-N}^-)^\dag| \Omega^{(0)}_N\rangle}{(2k_{n }+m)(2k_{n+M-N}+m)}\,\theta (n+M-N)
\;. \\
\eea
\lines
From the fact that
$
C_n^\pm \0=0
$, by applying the unitary operator $T^N$
we can see that
\[
0=T^N C_n^\pm \0 = C_{n\pm N}^\pm|\Omega_N^{(0)} \rangle \;.
\]
We then have
\bea
C_{n\pm N}^\pm (C_{n\pm N}^\pm)^\dag |\Omega_N^{(0)} \rangle &=&\left( (C_{n\pm N}^\pm)^\dag C_{n \pm N}^\pm
+\left[C_{n\pm N}^\pm , (C_{n\pm N}^\pm)^\dag \right]\right) |\Omega_N^{(0)} \rangle \\ 
&=&\left( \pm C_0^3 + n\pm N \right) |\Omega_N^{(0)} \rangle
\eea
\lines
and, since $C_0^3 |\Omega_N^{(0)} \rangle= -N |\Omega_N^{(0)} \rangle$,
\be\label{cc+}
C_{n\pm N}^\pm (C_{n\pm N}^\pm)^\dag |\Omega_N^{(0)} \rangle =n |\Omega_N^{(0)} \rangle \;.
\ee
Therefore
\baselineskip20pt
\bea
&&\sum_{n,p=\half}^{\infty}\frac{\langle \Omega^{(0)}_M |C_{p+M}^+  \psi_L^\dag (0,\x) \psi_R (0,\x)  (C_{n+N}^+)^\dag  | \Omega^{(0)}_N\rangle}{(2k_p+m)(2k_n+m)}\, \delta_{p+M, n+N}=\\ \\
&&= \sum_{p=\half}^{\infty}\frac{p \langle \Omega^{(0)}_M | \psi_L^\dag (0,\x) \psi_R (0,\x) | \Omega^{(0)}_N\rangle}{(2k_p+m)(2k_{p+M-N }+m)}\,\theta (p+M-N) =\\ \\
&&= \frac{1}{2L}e^{-\frac{\pi}{2mL}}\sum_{p=\half}^{\infty}\frac{p }{(2k_p+m)(2k_{p+M-N }+m)}\,\theta (p+M-N) \delta_{M,N+1} =\\ \\
&&= \frac{1}{2L}e^{-\frac{\pi}{2mL}}\sum_{p=\half}^{\infty}\frac{p }{(2k_p+m)(2k_{p+1}+m)}  \delta_{M,N+1}
\eea
\lines
and
\baselineskip20pt
\bea
&&\sum_{n,p=\half}^{\infty}\frac{\langle \Omega^{(0)}_M |C^-_{p-M} \psi_L^\dag (0,\x) \psi_R (0,\x)  (C_{n-N}^-)^\dag  | \Omega^{(0)}_N\rangle}{(2k_p+m)(2k_n+m)}\, \delta_{p-M, n-N}=\\ \\
&&= \frac{1}{2L}e^{-\frac{\pi}{2mL}}
 \sum_{n=\half}^{\infty}\frac{ n}{(2k_{n+M-N}+m)(2k_{n }+m)}\,\theta (n+M-N) \delta_{M,N+1}=\\ \\
&&= \frac{1}{2L}e^{-\frac{\pi}{2mL}}
 \sum_{n=\half}^{\infty}\frac{ n}{(2k_{n+1}+m)(2k_{n }+m)}\,\theta (n+1) \delta_{M,N+1}=\\ \\
&&= \frac{1}{2L}e^{-\frac{\pi}{2mL}}
 \sum_{n=\half}^{\infty}\frac{ n}{(2k_{n+1}+m)(2k_{n}+m)}\, \delta_{M,N+1}\;.
\eea
\lines
Finally we get
\baselineskip18pt
\bea
 \!\langle \theta^{(1)}|\,i \psi_L^\dag (0,\x) \psi_R (0,\x) | \theta^{(1)}\rangle =
\frac{ig^2}{2mL^2}e^{-\frac{\pi}{2mL}}\sum_{N=-\infty}^{\infty}e^{i\theta}\Bigg\{
 \sum_{n=\half}^{\infty}\frac{2 n}{(2k_{n}+m)(2k_{n+1}+m)}\\ \\
\hspace{45pt}-m \sum_{J=1}^{\infty}\frac{(J-1)}{2k_J^3}\Bigg\}
\eea
Taking the complex conjugate gives
\baselineskip18pt
\bea
 -\langle \theta^{(1)}|\,i \psi_R^\dag (0,\x) \psi_L (0,\x) | \theta^{(1)}\rangle =
-\frac{ig^2}{2mL^2}e^{-\frac{\pi}{2mL}}\sum_{N=-\infty}^{\infty}e^{-i\theta}\Bigg\{
 \sum_{n=\half}^{\infty}\frac{2 n}{(2k_{n}+m)(2k_{n+1}+m)}\\ \\
\hspace{45pt}-m\sum_{J=1}^{\infty}\frac{(J-1)}{2k_J^3}\Bigg\}
\eea
and therefore
\bean
&&\hspace{-50pt} \langle \theta^{(1)}|\,i\left( \psi_L^\dag (0,\x) \psi_R (0,\x)- \psi_L^\dag (0,\x) \psi_R (0,\x)\right) | \theta^{(1)}\rangle = \nn \\ \nn  \\
&&\hspace{-30pt}-\frac{g^2}{mL^2}e^{-\frac{\pi}{2mL}}\sum_{N=-\infty}^{\infty}\sin \theta\Bigg\{
 \sum_{n=\half}^{\infty}\frac{2 n}{(2k_{n}+m)(2k_{n+1}+m)}-m\sum_{J=1}^{\infty}\frac{(J-1)}{2k_J^3}\Bigg\}\,. \label{condone} \\ \nn
\eean
\lines
\vspace{12pt}
Let us calculate the contribution of $| \theta^{(1)}\rangle$ to the norm of $|\theta\rangle$.
We have
\baselineskip20pt
\bea
\langle \Omega_M^{(1)}| \Omega_N^{(1)}\rangle &\!\!\!=&\!\!\!
 \frac{g^2}{mL}\, \Bigg\{ -m\sum_{J=1}^{\infty}\frac{\langle \Omega^{(0)}_M | C_J^3  (C_J^3)^\dag | \Omega^{(0)}_N\rangle}{2k_J ^3}
 \\ \\
&&\hspace{20pt}+\sum_{n,p=\half}^{\infty}\frac{\langle \Omega^{(0)}_M |C_{p+M}^+  (C_{n+N}^+)^\dag  | \Omega^{(0)}_N\rangle}{(2k_n+m)(2k_p+m)}\delta_{p+M, n+N} \\ \\
&&\hspace{20pt}+\sum_{n,p=\half}^{\infty}\frac{\langle \Omega^{(0)}_M |C_{p-M}^-   (C_{n-N}^-)^\dag  | \Omega^{(0)}_N\rangle}{(2k_n+m)(2k_p+m)}\delta_{p-M, n-N}
\Bigg\} \\ \\
&\!\!\!=&\!\!\! \frac{g^2}{mL}\, \Bigg\{\sum_{J=1}^{\infty}\left( \frac{-mJ}{2k_J^3}\right)\delta_{M,N}
+\sum_{n=\half}^{\infty}\frac{2 n}{(2k_{n}+m)^2}\delta_{M,N}\Bigg\}
\eea
\lines
and
\be \label{normone}
 \langle \theta^{(1)}| \theta^{(1)}\rangle=\frac{g^2}{mL}\sum_{N=-\infty}^{\infty}\Bigg\{-\sum_{J=1}^{\infty} \frac{mJ}{2k_J^3}
+\sum_{n=\half}^{\infty}\frac{2 n}{(2k_{n}+m)^2}\Bigg\}
\ee
Let us consider now
$
 \langle \theta^{(1)}|\, \psi_L^\dag (0,\x) \psi_R (0,\x) | \theta^{(2)}\rangle \;.
$
The only term in  $| \Omega_N^{(2)}\rangle$ that contributes to
$ \langle \Omega_M^{(1)}|\, \psi_L^\dag (0,\x) \psi_R (0,\x) | \Omega_N^{(2)}\rangle$
is
\[
 \frac{g^2}{mL} \sum_{n=\half}^{\infty} \frac{\left( (A_0^3)^\dag -\frac{N\pi \sqrt{m}}{g\sqrt{L}}\right)
 \bigg({ -(A_{n+N}^-)^\dag  (C_{n+N}^+)^\dag +(A_{n-N}^+)^\dag (C_{n-N}^-)^\dag }\bigg)}
{2(k_{n}+m) (2k_n+m)}| \Omega^{(0)}_N\rangle
\]
and so
\baselineskip20pt
\bea
 &&\hspace{-20pt}\langle \Omega_M^{(1)}|\, \psi_L^\dag (0,\x) \psi_R (0,\x) | \Omega_N^{(2)}\rangle =\\ \\
&&\hspace{-20pt}-\frac{g^3}{(mL)^{\frac{3}{2}}}\Bigg\{\sum_{n,p=\half}^{\infty}\frac{\langle \Omega^{(0)}_M |C_{p-M}^- \left( (A_0^3)^\dag -\frac{N\pi \sqrt{m}}{g\sqrt{L}}\right)
 \psi_L^\dag (0,\x) \psi_R (0,\x)  (C_{n-N}^- )^\dag  | \Omega^{(0)}_N\rangle}{2(2k_p+m)(k_{n}+m)(2k_n+m)}\, \delta_{p-M, n-N}\\ \\
&&\hspace{15pt}-\sum_{n,p=\half}^{\infty}\frac{\langle \Omega^{(0)}_M |C_{p+M}^+ \left( (A_0^3)^\dag -\frac{N\pi \sqrt{m}}{g\sqrt{L}}\right)
 \psi_L^\dag (0,\x) \psi_R (0,\x)  (C_{n+N}^+)^\dag  | \Omega^{(0)}_N\rangle}{2(2k_p+m)(k_{n}+m)(2k_n+m)}\, \delta_{p+M, n+N}\Bigg\}\;.\\
\eea
\lines
We can write
\bean
\langle \Omega^{(0)}_M |C_{p+M}^\pm \left( (A_0^3)^\dag -\frac{N\pi \sqrt{m}}{g\sqrt{L}}\right)
&=&\langle 0| C_p^\pm T^{-M}\left( (A_0^3)^\dag -\frac{N\pi \sqrt{m}}{g\sqrt{L}}\right) \nn \\
&=&\langle 0| C_p^\pm \left( (A_0^3)^\dag +\frac{(M-N)\pi \sqrt{m}}{g\sqrt{L}}\right)T^{-M}\nn \\
&=& \frac{(M-N)\pi \sqrt{m}}{g\sqrt{L}}\langle  \Omega^{(0)}_M | C_{p\pm M}^\pm \label{azc}
\eean
and
\baselineskip20pt
\bea
 &&\langle \Omega_M^{(1)}|\, \psi_L^\dag (0,\x) \psi_R (0,\x) | \Omega_N^{(2)}\rangle = \\ \\
&&\hspace{20pt}- \frac{(M-N)g^2\pi }{m L^2}\Bigg\{ \sum_{n,p=\half}^{\infty}\frac{\langle \Omega^{(0)}_M |C_{p-M}^-
 \psi_L^\dag (0,\x) \psi_R (0,\x)  (C_{n-N}^- )^\dag  | \Omega^{(0)}_N\rangle}{2(2k_p+m)(k_{n}+m)(2k_n+m)}\, \delta_{p-M, n-N}\\ \\
&&\hspace{100pt}- \sum_{n,p=\half}^{\infty}\frac{\langle \Omega^{(0)}_M |C_{p+M}^+
 \psi_L^\dag (0,\x) \psi_R (0,\x)  (C_{n+N}^+)^\dag  | \Omega^{(0)}_N\rangle}{2(2k_p+m)(k_{n}+m)(2k_n+m)}\, \delta_{p+M, n+N}\Bigg\} \\ \\
&&\hspace{20pt}= - \frac{g^2\pi }{2mL^3}e^{-\frac{\pi}{2mL}}\Bigg\{
\sum_{n=\half}^{\infty}\frac{ n}{2(2k_{n+1}+m)(k_{n}+m)(2k_{n}+m)}\, \delta_{M,N+1}\\ \\
&&\hspace{100pt}-\sum_{p=\half}^{\infty}\frac{p }{2(2k_p+m)(k_{p+1}+m)(2k_{p+1}+m)}  \delta_{M,N+1}\Bigg\}\;.
\eea
We also need
\pagebreak
\baselineskip15pt
\bea
 &&\langle \Omega_M^{(1)}|\, \psi_R^\dag (0,\x) \psi_L (0,\x) | \Omega_N^{(2)}\rangle =\\ \\
&&=- \frac{(M-N)g^2\pi }{m L^2}\Bigg\{ \sum_{n,p=\half}^{\infty}\frac{\langle \Omega^{(0)}_M |C_{p-M}^-
 \psi_R^\dag (0,\x) \psi_L (0,\x)  (C_{n-N}^- )^\dag  | \Omega^{(0)}_N\rangle}{2(2k_p+m)(k_{n}+m)(2k_n+m)}\, \delta_{p-M, n-N}\\ \\
&&\hspace{95pt}- \sum_{n,p=\half}^{\infty}\frac{\langle \Omega^{(0)}_M |C_{p+M}^+
 \psi_R^\dag (0,\x) \psi_L (0,\x)  (C_{n+N}^+)^\dag  | \Omega^{(0)}_N\rangle}{2(2k_p+m)(k_{n}+m)(2k_n+m)}\, \delta_{p+M, n+N}\Bigg\} \\ \\
&& =-\frac{(M-N)g^2\pi }{m L^2}\Bigg\{ \sum_{n,p=\half}^{\infty}\frac{\langle \Omega^{(0)}_M |C_{p-M}^- (C_{n-N}^- )^\dag
 \psi_R^\dag (0,\x) \psi_L (0,\x)    | \Omega^{(0)}_N\rangle}{2(2k_p+m)(k_{n}+m)(2k_n+m)}\, \delta_{p-M, n-N}\\ \\
&&\hspace{95pt}- \sum_{n,p=\half}^{\infty}\frac{\langle \Omega^{(0)}_M |
 \psi_R^\dag (0,\x) \psi_L (0,\x) C_{p+M}^+ (C_{n+N}^+)^\dag  | \Omega^{(0)}_N\rangle}{2(2k_p+m)(k_{n}+m)(2k_n+m)}\, \delta_{p+M, n+N}\Bigg\}\\ \\
&& =-\frac{(M-N)g^2\pi }{m L^2}\Bigg\{ \sum_{p=\half}^{\infty}\frac{p\langle \Omega^{(0)}_M |
 \psi_R^\dag (0,\x) \psi_L (0,\x)    | \Omega^{(0)}_N\rangle}{2(2k_p+m)(k_{p+N-M}+m)(2k_{p+N-M}+m)}\, \theta  (p+N-M) \\ \\
&&\hspace{95pt}- \sum_{n=\half}^{\infty}\frac{n\langle \Omega^{(0)}_M |
 \psi_R^\dag (0,\x) \psi_L (0,\x)  | \Omega^{(0)}_N\rangle}{2(2k_{n+N-M}+m)(k_{n}+m)(2k_n+m)}\, \theta (n+N-M)\Bigg\}\\ \\
&&=  \frac{g^2\pi }{2mL^3}e^{-\frac{\pi}{2mL}}\delta_{N,M+1}\Bigg\{\sum_{p=\half}^{\infty}\frac{p }{2(2k_{p+1}+m)(k_{p+1}+m)(2k_{p}+m)}
\\ \\
&&\hspace{125pt}-\sum_{n=\half}^{\infty}\frac{ n}{2(2k_{n}+m)(k_{n}+m)(2k_{n+1}+m)}\,
 \Bigg\}
\eea
so that
\bean
&&\langle \theta^{(1)}|\,i\left( \psi_L^\dag (0,\x) \psi_R (0,\x)- \psi_R^\dag (0,\x) \psi_L (0,\x)\right) | \theta^{(2)}\rangle=\nn \\ \nn  \\
&&= \frac{g^2\pi }{mL^3}e^{-\frac{\pi}{2mL}}\sum_{N=-\infty}^{\infty}\sin \theta
\Bigg\{\sum_{n=\half}^{\infty}\frac{n}{2(2k_{n+1}+m)(k_{n}+m)(2k_{n}+m)}
\nn \\ \nn \\
&&\hspace{125pt}-\sum_{n=\half}^{\infty}\frac{ n}{2(2k_{n}+m)(k_{n+1}+m)(2k_{n+1}+m)}\,
 \Bigg\}\nn \\ \nn \\
&&= \frac{g^2\pi^2 }{mL^4}e^{-\frac{\pi}{2mL}}\sum_{N=-\infty}^{\infty}\sin \theta
\sum_{n=\half}^{\infty}\frac{n}{2(2k_{n}+m)(2k_{n+1}+m)(k_{n}+m)(k_{n+1}+m)} \;. \label{condonetwo} \nn \\
\eean
\lines
Moreover, since
$\langle \theta^{(2)}|\,i\left( \psi_L^\dag (0,\x) \psi_R (0,\x)- \psi_R^\dag (0,\x) \psi_L (0,\x)\right) | \theta^{(1)}\rangle
$
can be obtained from this quantity by complex conjugation, we see that we must have
\bea
&&\langle \theta^{(2)}|\,i\left( \psi_L^\dag (0,\x) \psi_R (0,\x)- \psi_R^\dag (0,\x) \psi_L (0,\x)\right) | \theta^{(1)}\rangle \\
&&=\langle \theta^{(1)}|\,i\left( \psi_L^\dag (0,\x) \psi_R (0,\x)- \psi_R^\dag (0,\x) \psi_L (0,\x)\right) | \theta^{(2)}\rangle \;.
\eea
\vskip12pt
\lines
\noindent
\vspace{12pt}
No contribution to the norm is given by  $\langle \theta^{(1)}|\theta^{(2)}\rangle$ and
$\langle \theta^{(2)}|\theta^{(1)}\rangle$. As a matter of fact
\baselineskip20pt
\bea
 \langle \Omega_M^{(1)}| \Omega_N^{(2)}\rangle &=&
- \frac{(M-N)g^2\pi }{m L^2}\Bigg\{ \sum_{n,p=\half}^{\infty}\frac{\langle \Omega^{(0)}_M |C_{p-M}^-
 (C_{n-N}^- )^\dag  | \Omega^{(0)}_N\rangle}{2(2k_p+m)(k_{n}+m)(2k_n+m)}\, \delta_{p-M, n-N}\\ \\
&&\hspace{80pt}- \sum_{n,p=\half}^{\infty}\frac{\langle \Omega^{(0)}_M |C_{p+M}^+
  (C_{n+N}^+)^\dag  | \Omega^{(0)}_N\rangle}{2(2k_p+m)(k_{n}+m)(2k_n+m)}\, \delta_{p+M, n+N}\Bigg\} \\ \\
&=& - \frac{(M-N)g^2\pi }{m L^2}\Bigg\{
\sum_{n=\half}^{\infty}\frac{ n}{2(2k_{n+1}+m)(k_{n}+m)(2k_{n}+m)}\, \delta_{M,N}\\ \\
&&\hspace{80pt}-\sum_{n=\half}^{\infty}\frac{n }{2(2k_n+m)(k_{n+1}+m)(2k_{n+1}+m)}  \delta_{M,N}\Bigg\}=0
\eea
\lines
and therefore
\[
\langle \theta^{(1)}|\theta^{(2)}\rangle=\langle \theta^{(2)}|\theta^{(1)}\rangle=0 \;.
\]
In order to calculate $\langle \theta^{(2)}|\,i\left( \psi_L^\dag (0,\x) \psi_R (0,\x)- \psi_R^\dag (0,\x) \psi_L (o,\x)\right) | \theta^{(2)}\rangle$ it is convenient to write $\langle \Omega_M^{(2)}|\, \psi_L^\dag (0,\x) \psi_R (0,\x) | \Omega_N^{(2)}\rangle$ as
\bea
\langle \Omega_M^{(2)}|\, \psi_L^\dag  \psi_R  | \Omega_N^{(2)}\rangle&=&\langle \Omega_M^{(2)}|\, \psi_L^\dag  \psi_R  | \Omega_N^{(2)}\rangle_{33}
+\,\langle\Omega_M^{(2)}|\, \psi_L^\dag \psi_R  | \Omega_N^{(2)}\rangle_{++}\\
&&+\,\langle\Omega_M^{(2)}|\, \psi_L^\dag  \psi_R  | \Omega_N^{(2)}\rangle_{-\, -}+\langle\Omega_M^{(2)}|\, \psi_L^\dag  \psi_R  | \Omega_N^{(2)}\rangle_{3+}\\
&&+\langle\Omega_M^{(2)}|\, \psi_L^\dag  \psi_R  | \Omega_N^{(2)}\rangle_{3-}
+\langle\Omega_M^{(2)}|\, \psi_L^\dag  \psi_R  | \Omega_N^{(2)}\rangle_{+\, -}\;.
\eea
It is understood that $\psi$ is to be evaluated at $(0,\x)$. \\
One can verify that
\bea
\langle \Omega_M^{(2)}|\, \psi_L^\dag  \psi_R  | \Omega_N^{(2)}\rangle_{33}=
\frac{g^4}{L^2}\sum_{I,J=1}^{\infty}\frac{8}{(2k_I)^3 (2k_J)^3}
\langle \Omega_M^{(0)}|\,C_J^3 C_I^3 \psi_L^\dag  \psi_R (C_J^3)^\dag (C_I^3)^\dag | \Omega_N^{(0)}\rangle
\eea

\lines
\bea
&&\hspace{-20pt}\langle\Omega_M^{(2)}|\, \psi_L^\dag \psi_R  | \Omega_N^{(2)}\rangle_{\pm \pm}
=\\ \\
&&\hspace{-5pt}\frac{g^4}{m^2 L^2}\sum_{l,p,n,j=\half}^{\infty}\frac{\langle \Omega_M^{(0)}|\,C_{n\pm M}^\pm C_{p \pm M}^\pm \psi_L^\dag  \psi_R (C_{l\pm N}^\pm)^\dag (C_{j\pm N}^\pm)^\dag | \Omega_N^{(0)}\rangle}
{4(2k_p  +m) (2k_n +m) (2k_l  +m) (2k_j +m)}
(\delta_{n\pm M, l\pm N}\delta_{p \pm M, j \pm N}\\
&&\hspace{312pt}+\delta_{n\pm M, j \pm N}\delta_{p \pm M, l \pm N})\\
\eea
\bea
&&\hspace{-32pt}
\langle\Omega_M^{(2)}|\, \psi_L^\dag  \psi_R  | \Omega_N^{(2)}\rangle_{3\pm}
=\frac{2g^4}{m L^2}
\sum_{J=1}^{\infty}
\Bigg\{\sum_{p,n=\half}^{\infty}\frac{\langle \Omega_M^{(0)}|
 C^\pm_{p\pm M }C_J^3 \psi_L^\dag  \psi_R (C_J^3)^\dag (C^\pm_{n\pm N })^\dag
| \Omega_N^{(0)}\rangle }{(2k_p+m)(2k_n+m)}\\
&&\hspace{160pt}\Bigg(-\frac{1}{(2k_p +2k_J +m)^2 (2k_n +2k_J +m)}\\
&&\hspace{170pt}-\frac{1}{(2k_p +2k_J +m) (2k_n +2k_J +m)^2}\Bigg)
\delta_{p\pm M , n\pm N}\\ \\
&&\hspace{80pt}-\sum_{n=-\infty}^{\infty}\sum_{p=\half}^{\infty}\frac{\langle \Omega_M^{(0)}|
 C^\pm_{p\pm M }C_J^3 \psi_L^\dag  \psi_R (C^\pm_{n\pm N })^\dag (C_J^3)^\dag
| \Omega_N^{(0)}\rangle }
{(2k_p+m) }\\
&&\hspace{140pt}\Bigg(\frac{1}{(2k_J)^2 (2k_p +2k_J +m)(2k_n +2k_J +m)}\\
&&\hspace{140pt}+\frac{1}{(2k_J)(2k_p +2k_J +m)^2 (2k_n +2k_J +m)}\\
&&\hspace{140pt}+\frac{1}{(2k_J)(2k_p +2k_J +m) (2k_n +2k_J +m)^2}\Bigg)
\delta_{p\pm M , n\pm N}\\ \\
&&\hspace{80pt}-\sum_{p=-\infty}^{\infty}\sum_{n=\half}^{\infty}\frac{\langle \Omega_M^{(0)}|
C_J^3  C^\pm_{p\pm M }\psi_L^\dag  \psi_R (C_J^3)^\dag (C^\pm_{n\pm N })^\dag
| \Omega_N^{(0)}\rangle }
{(2k_n+m)}\\
&&\hspace{140pt}\Bigg(\frac{1}{(2k_J)^2 (2k_p +2k_J +m)(2k_n +2k_J +m)}\\
&&\hspace{140pt}+\frac{1}{(2k_J)(2k_p +2k_J +m)^2 (2k_n +2k_J +m)}\\
&&\hspace{140pt}+\frac{1}{(2k_J)(2k_p +2k_J +m) (2k_n +2k_J +m)^2}\Bigg)
\delta_{p\pm M , n\pm N}\\ \\
&&\hspace{10pt}+\sum_{p,n=-\infty}^{\infty}\langle \Omega_M^{(0)}|
C_J^3  C^\pm_{p\pm M }\psi_L^\dag  \psi_R  (C^\pm_{n\pm N })^\dag (C_J^3)^\dag
| \Omega_N^{(0)}\rangle
\Bigg(-\frac{2}{(2k_J)^3 (2k_p +2k_J +m)^2}\\
&&\hspace{197pt}-\frac{1}{(2k_J)^2 (2k_p +2k_J +m)^2
 (2k_n +2k_J +m)}\\
&&\hspace{137pt}-\frac{1}{(2k_J)^2 (2k_p +2k_J +m) (2k_n +2k_J +m)^2}\Bigg)
\delta_{p\pm M , n\pm N}\Bigg\}\\ \\
&& \hspace{10pt}+ \sum_{n,p=\half}^{\infty} \frac{\langle \Omega^{(0)}_M|
C_{p\pm M}^\pm  \left( A_0^3-\frac{M\pi \sqrt{m}}{g\sqrt{L}}\right)\psi_L^\dag \psi_R \left( (A_0^3)^\dag -\frac{N\pi \sqrt{m}}{g\sqrt{L}}\right)
 (C_{n\pm N}^\pm)^\dag | \Omega^{(0)}_N\rangle}{4(k_{p}+m) (2k_p+m)(k_{n}+m) (2k_n+m)}
\delta_{p\pm M, n\pm N}\\
\eea

\baselineskip12pt
\bea
&&\hspace{-20pt}\langle\Omega_M^{(2)}|\, \psi_L^\dag \psi_R  | \Omega_N^{(2)}\rangle_{+\, -}
=\\ \\
&&\hspace{-21pt}\frac{g^4}{m^2 L^2}\!\sum_{l,p=-\infty}^{\infty}\sum_{n,j=\half}^{\infty}\!\Bigg\{ \frac{\langle \Omega_M^{(0)}|\,C_{n-M}^- C_{p + M}^+ \psi_L^\dag  \psi_R (C_{l+N}^+)^\dag (C_{j- N}^-)^\dag | \Omega_N^{(0)}\rangle}
{4(k_p +k_n +m) (2k_n +m) (k_l +k_j +m) (2k_j +m)}\delta_{n- M, j - N}\delta_{p +M, l + N}
\eea
\vspace{-16pt}
\bea
&&\hspace{60pt}+\frac{\langle \Omega_M^{(0)}|\,C_{n+M}^+ C_{p - M}^- \psi_L^\dag  \psi_R (C_{l+N}^+)^\dag (C_{j- N}^-)^\dag | \Omega_N^{(0)}\rangle}
{4(k_p +k_n +m) (2k_n +m) (k_l +k_j +m) (2k_j +m)}\delta_{n+M, l + N}\delta_{p - M, j - N}\\ \\
&&\hspace{60pt}+\frac{\langle \Omega_M^{(0)}|\,C_{n-M}^- C_{p + M}^+ \psi_L^\dag  \psi_R (C_{l-N}^-)^\dag (C_{j+ N}^+)^\dag | \Omega_N^{(0)}\rangle}
{4(k_p +k_n +m) (2k_n +m) (k_l +k_j +m) (2k_j +m)}\delta_{n- M, l - N}\delta_{p + M, j+ N}\\ \\
&&\hspace{60pt}+\frac{\langle \Omega_M^{(0)}|\,C_{n+M}^+ C_{p - M}^- \psi_L^\dag  \psi_R (C_{l-N}^-)^\dag (C_{j+ N}^+)^\dag | \Omega_N^{(0)}\rangle}
{4(k_p +k_n +m) (2k_n +m) (k_l +k_j +m) (2k_j +m)}\delta_{n+M, j+ N}\delta_{p - M, l- N}\Bigg\}\\
\eea
\lines
The calculation of $\langle \Omega_M^{(2)}|\, \psi_L^\dag (0,\x) \psi_R (0,\x) | \Omega_N^{(2)}\rangle$ can be simplified by noting that
\bea
&&
\langle \Omega_M^{(2)}|\, \psi_L^\dag (0,\x) \psi_R (0,\x) | \Omega_N^{(2)}\rangle=\\
&&\langle \Omega_1^{(2)}|\, \psi_L^\dag (0,\x) \psi_R (0,\x) | \Omega^{(2)}\rangle\delta_{M,N+1}\;.
\eea
As a matter of fact we can write
\bea
&&\langle \Omega_M^{(2)}|\, \psi_L^\dag  \psi_R  | \Omega_N^{(2)}\rangle =\langle \Omega_M^{(2)} |T^N T^{-N}\psi_L^\dag  \psi_R T^N| \Omega^{(2)}\rangle
\eea
and, since the product $\psi_L^\dag  \psi_R $ is gauge invariant,
\bea
\langle \Omega_M^{(2)}|\, \psi_L^\dag  \psi_R  | \Omega_N^{(2)}\rangle =\langle \Omega_M^{(2)}| T^N \psi_L^\dag  \psi_R | \Omega^{(2)}\rangle  =
\langle \Omega_{M-N}^{(2)}|\, \psi_L^\dag  \psi_R  | \Omega^{(2)}\rangle
\eea
Moreover, from
\[
[D_0^3, \psi_L^\dag  \psi_R ]=\psi_L^\dag  \psi_R\ ,  \quad [D_0^3, C^{3,\pm}]=0 \quad {\rm and}
 \quad  D_0^3 | \Omega_N^{(0)}\rangle= N | \Omega_N^{(0)}\rangle
\]
we get
\bea
&&  D_0^3 | \Omega_N^{(2)}\rangle= N | \Omega_N^{(2)}\rangle \\
&&  D_0^3 \psi_L^\dag  \psi_R| \Omega_N^{(2)}\rangle= (N+1) | \Omega_N^{(0)}\rangle \;.
\eea
Therefore, the states   $| \Omega_{M-N}^{(2)}\rangle$ and $\psi_L^\dag  \psi_R| \Omega^{(2)}\rangle$, being eigenstates of the self-adjoint operator $D_0^3$ with eigenvalues $M-N$ and $1$ respectively, must be orthogonal unless $M-N=1$.\\
We therefore have
\[
\langle \Omega_M^{(0)}|\,C_J^3 C_I^3  \psi_L^\dag \psi_R (C_I^3)^\dag  (C_J^3)^\dag
  | \Omega_N^{(0)}\rangle=\langle \Omega_1^{(0)}|\,C_J^3 C_I^3  \psi_L^\dag \psi_R (C_I^3)^\dag  (C_J^3)^\dag \0 \delta_{M,N+1}
\]
and using (\ref{psic}), (\ref{cpsi})  we can write
\baselineskip12pt
\bea
&&\langle \Omega_1^{(0)}|\,C_J^3 C_I^3  \psi_L^\dag \psi_R (C_I^3)^\dag  (C_J^3)^\dag
 \0
=\langle \Omega_1^{(0)}|\,C_J^3  \psi_L^\dag \psi_R \left(C_I^3 -e^{-ik_I \x}\right) (C_I^3)^\dag  (C_J^3)^\dag
 \0\\ \\
&&\hspace{100pt}=I\langle \Omega_1^{(0)}|\,C_J^3   \psi_L^\dag \psi_R  (C_J^3)^\dag
 \0 + I\delta_{IJ}\langle \Omega_1^{(0)}|\,C_J^3  \psi_L^\dag \psi_R  (C_I^3)^\dag
 \0\\ \\
&&\hspace{165pt}-e^{-ik_I \x}\langle \Omega_1^{(0)}|\,C_J^3  \left((C_I^3)^\dag +e^{ik_I \x}\right)  \psi_L^\dag \psi_R  (C_J^3)^\dag
 \0\\ \\
&&\hspace{100pt}=(I+I\delta_{IJ}-1)\langle \Omega_1^{(0)}|\,C_J^3   \psi_L^\dag \psi_R  (C_J^3)^\dag
 \0\\ \\
&&\hspace{165pt}-e^{-ik_I \x}I\delta_{IJ}\langle \Omega_1^{(0)}|\,  \left((C_J^3)^\dag +e^{ik_I \x}\right)  \psi_L^\dag \psi_R  \0
\eea
\lines
and from (\ref{zero}), (\ref{useful})
\bea
\langle \Omega_1^{(0)}|\,C_J^3 C_I^3  \psi_L^\dag \psi_R (C_I^3)^\dag  (C_J^3)^\dag
 \0=
\left[(I+I\delta_{IJ}-1)(J-1) -I\delta_{IJ} \right] \frac{1}{2L}e^{-\frac{\pi}{2mL}} \;.
\eea \\
Therefore
\baselineskip15pt
\bea
\langle \Omega_M^{(2)}|\, \psi_L^\dag  \psi_R  | \Omega_N^{(2)}\rangle_{33}=
\frac{g^4}{ L^2} \frac{1}{2L}e^{-\frac{\pi}{2mL}} \delta_{M,N+1}
\Bigg\{ \sum_{I,J=1}^{\infty}\frac{8}{(2k_I)^3 (2k_J)^3}
 (IJ-I-J+1)\\ \\
+ \sum_{J=1}^{\infty}\frac{8}{ (2k_J)^6}(J^2-2J)\Bigg\} \; .\hspace{-20pt}\\
\eea
\lines
Let us consider now $\langle\Omega_1^{(2)}|\, \psi_L^\dag \psi_R | \Omega^{(2)}\rangle_{++}$ .\\
We have

\baselineskip15pt
\bea
\hspace{-20pt}\langle\Omega_1^{(2)}|\, \psi_L^\dag \psi_R  | \Omega^{(2)}\rangle_{++}&=&
\hspace{-5pt}\frac{g^4}{m^2 L^2}\sum_{p,n=\half}^{\infty}\frac{\langle \Omega_1^{(0)}|\,C_{n+ 1}^+ C_{p + 1}^+ \psi_L^\dag  \psi_R (C_{p + 1}^+)^\dag (C_{n+ 1}^+)^\dag \0}
{2(2k_p  +m) (2k_n +m) (2k_{p+1} +m) (2k_{n+1} +m)}\\ \\
&&\hspace{-5pt}\frac{g^4}{m^2 L^2}\sum_{p,n=\half}^{\infty}\frac{\langle \Omega_1^{(0)}|\,C_{n+ 1}^+ C_{p + 1}^+  (C_{p + 1}^+)^\dag (C_{n+ 1}^+)^\dag\psi_L^\dag  \psi_R \0}
{2(2k_p  +m) (2k_n +m) (2k_{p+1} +m) (2k_{n+1} +m)}\;. \\
\eea
\lines
Using the commutators  (\ref{commc3}-\ref{commcpcm}),  relation (\ref{cc+}) and the fact that \[
C_0^3|\Omega_N^{(0)}\rangle=-N|\Omega_N^{(0)}\rangle
\]
 we can write
\bea
&&C_{n+ 1}^+C_{p + 1}^+(C_{p + 1}^+)^\dag (C_{n+ 1}^+)^\dag |\Omega_1^{(0)}\rangle =
C_{p + 1}^+(C_{p + 1}^+)^\dag C_{n+ 1}^+ (C_{n+ 1}^+)^\dag|\Omega_1^{(0)}\rangle  \\
&&\quad \hspace{150pt}+C^+ _{p + 1}\Big(C^3_{n-p}+(n+1)\delta_{pn}\Big)(C_{n+ 1}^+)^\dag |\Omega_1^{(0)}\rangle \\
&&\qquad =\Big(np+n(n+1)\delta_{pn}\Big)|\Omega_1^{(0)}\rangle +\Big(C^3_{n-p}C^+ _{p + 1}-C^+ _{n+1}\Big)(C_{n+ 1}^+)^\dag
|\Omega_1^{(0)}\rangle \\
&&\qquad =\Big(np+n(n+1)\delta_{pn}-n\Big)|\Omega_1^{(0)}\rangle +C^3_{n-p}\Big(C^3_{p-n}+(n+1)\delta_{np}\Big)
|\Omega_1^{(0)}\rangle\\
&&\qquad =\Big(np+n(n+1)\delta_{pn}-n+(n-p)\theta(n-p)+\delta_{np}-(n+1)\delta_{np}\Big)|\Omega_1^{(0)}\rangle\\
&&\qquad =\Big(np+n^2\delta_{pn}-n+(n-p)\theta(n-p)\Big)|\Omega_1^{(0)}\rangle
\eea
and
\baselineskip15pt
\bea
\langle\Omega_1^{(2)}|\, \psi_L^\dag \psi_R  | \Omega^{(2)}\rangle_{++}&\hspace{-9pt}=&
\hspace{-8pt}\frac{g^4e^{-\frac{\pi}{2mL}}}{4m^2 L^3}\sum_{p,n=\half}^{\infty}\frac{np+n^2\delta_{pn}-n+(n-p)\theta(n-p)}
{(2k_p  +m) (2k_n +m) (2k_{p+1} +m) (2k_{n+1} +m)}\\ \\
&=&\hspace{-5pt}\frac{g^4e^{-\frac{\pi}{2mL}}}{4m^2 L^3}\Bigg\{ \sum_{p,n=\half}^{\infty}\frac{np-n}{(2k_p  +m) (2k_n +m) (2k_{p+1} +m) (2k_{n+1} +m)}\\ \\
&&\hspace{25pt}+\sum_{n=\half}^{\infty}
\sum_{p=\half}^{n-1}\frac{n-p}{(2k_p  +m) (2k_n +m) (2k_{p+1} +m) (2k_{n+1} +m)}\\ \\
&&\hspace{25pt}+\sum_{n=\half}^{\infty}\frac{n^2}{(2k_n  +m)^2 (2k_{n+1} +m)^2 }\Bigg\}\;. \\
\eea
\lines
The same result holds for $ \langle\Omega_1^{(2)}|\, \psi_L^\dag \psi_R  | \Omega^{(2)}\rangle_{- -}$ .\\
We then have
\[
\langle\Omega_M^{(2)}|\, \psi_L^\dag \psi_R  |\Omega_N^{(2)}\rangle_{\pm \pm}=
\langle\Omega_1^{(2)}|\, \psi_L^\dag \psi_R  | \Omega^{(2)}\rangle_{++}\delta_{M, N+1}
\]
\\ \\
$\langle\Omega_1^{(2)}|\, \psi_L^\dag \psi_R  | \Omega^{(2)}\rangle_{3+}$ is given by
\baselineskip15pt
\bea
&&\hspace{-32pt}
\langle\Omega_1^{(2)}|\, \psi_L^\dag  \psi_R  | \Omega^{(2)}\rangle_{3+}=\\ \\
&&\hspace{-32pt}\frac{2g^4}{m L^2}
\sum_{J=1}^{\infty}
\Bigg\{\!\!-\sum_{p=\half}^{\infty}\frac{\langle \Omega_1^{(0)}|
 C^+_{p+ 1 }C_J^3 \psi_L^\dag  \psi_R (C_J^3)^\dag (C^+_{p})^\dag
 \0 }{(2k_p+m)(2k_{p+1}+m)}
\Bigg(\frac{1}{(2k_{p+J}+m)^2 (2k_{p+J+1} +m)}\\
&&\hspace{210pt}+\frac{1}{(2k_{p+J} +m) (2k_{p+J+1}  +m)^2}\Bigg)
\eea
\lines
\bea
&&\hspace{-10pt}-\sum_{p=\half}^{\infty}\frac{\langle \Omega_1^{(0)}|
 C^+_{p+ 1 }C_J^3 \psi_L^\dag  \psi_R (C^+_{{p+1} })^\dag (C_J^3)^\dag
\0 }
{ (2k_p+m) }
\Bigg(\frac{1}{(2k_J)^2 (2k_{p+J}  +m)(2k_{p+J+1} +m)}\\
&&\hspace{188pt}+\frac{1}{(2k_J)(2k_{p+J}+m)^2 (2k_{p+J+1} +m)}\\
&&\hspace{188pt}+\frac{1}{(2k_J)(2k_{p+J} +m) (2k_{p+J+1} +m)^2}\Bigg)\\ \\
&&\hspace{-10pt}-\sum_{p=\half}^{\infty}\frac{\langle \Omega_1^{(0)}|
C_J^3  C^+_{p+ 1 }\psi_L^\dag  \psi_R (C_J^3)^\dag (C^+_{{p+1} })^\dag \0 }
{(2k_{p+1}+m)}
\Bigg(\frac{1}{(2k_J)^2 (2k_{p+J}+m)(2k_{p+J+1} +m)}\\
&&\hspace{188pt}+\frac{1}{(2k_J)(2k_{p+J} +m)^2 (2k_{p+J+1}+m)}\\
&&\hspace{188pt}+\frac{1}{(2k_J)(2k_{p+J}+m) (2k_{p+J+1} +m)^2}\Bigg)\\ \\
&&\hspace{-10pt}-\frac{\langle \Omega_1^{(0)}|
C_J^3  C^+_{\half }\psi_L^\dag  \psi_R (C_J^3)^\dag (C^+_{\half })^\dag \0 }
{(2k_{\half}+m)}
\Bigg(\frac{1}{(2k_J)^2 (2k_{J-\half} +m)(2k_{J+\half} +m)}\\
&&\hspace{158pt}+\frac{1}{(2k_J)(2k_{J-\half} +m)^2 (2k_{J+\half} +m)}\\
&&\hspace{158pt}+\frac{1}{(2k_J)(2k_{J-\half} +m) (2k_{J+\half}+m)^2}\Bigg)\\ \\
&&\hspace{-10pt}-\sum_{p=\half}^{\infty}\langle \Omega_1^{(0)}|
C_J^3  C^+_{p+ 1 }\psi_L^\dag  \psi_R  (C^+_{{p+1} })^\dag (C_J^3)^\dag
\0
\Bigg(\frac{2}{(2k_J)^3 (2k_{p+J} +m)(2k_{p+J+1} +m)}\\
&&\hspace{183pt}+\frac{1}{(2k_J)^2 (2k_{p+J}+m)^2 (2k_{p+J+1} +m)}\\
&&\hspace{183pt}+\frac{1}{(2k_J)^2 (2k_{p+J}+m) (2k_{p+J+1}  +m)^2}
\Bigg)\\ \\
&&\hspace{-10pt}-\sum_{p=\half}^{\infty}\langle \Omega_1^{(0)}|
C_J^3  (C^-_{p- 1 })^\dag \psi_L^\dag  \psi_R  C^-_{p-1 } (C_J^3)^\dag\0
\Bigg(\frac{2}{(2k_J)^3 (2k_{J-p} +m)(2k_{J-p+1} +m)}\\
&&\hspace{183pt}+\frac{1}{(2k_J)^2 (2k_{J-p} +m)^2 (2k_{J-p+1} +m)}\\
&&\hspace{183pt}+\frac{1}{(2k_J)^2 (2k_{J-p} +m) (2k_{J-p+1} +m)^2}
\Bigg)
\Bigg\}\\ \\
&& \hspace{-10pt}+\frac{g^4}{m^2 L^2} \sum_{p=\half}^{\infty} \frac{\langle \Omega^{(0)}_1|
C_{p+ 1}^+  \left( A_0^3-\frac{\pi \sqrt{m}}{g\sqrt{L}}\right)\psi_L^\dag \psi_R  (A_0^3)^\dag
 (C_{{p+1}}^+)^\dag | \0 \rangle}{4(k_{p}+m) (2k_p+m)(k_{{p+1}}+m) (2k_{p+1}+m)}
\\
\eea
\lines
We can write
\baselineskip12pt
\bea
&&\langle\Omega_1^{(0)}|C_{p+1}^+ C^3_J \psi_L^\dag \psi_R (C^3_J )^\dag (C_{p+1}^+)^\dag \0 = \\ \\
 &&= \langle\Omega_1^{(0)}|C_{p+1}^+ C^3_J \psi_L^\dag \psi_R (C_{p+1}^+)^\dag  (C^3_J )^\dag  \0
 - \langle\Omega_1^{(0)}|C_{p+1}^+ C^3_J \psi_L^\dag \psi_R (C_{p+1+J}^+)^\dag  \0
\eea

\baselineskip12pt
\bea
&& \langle\Omega_1^{(0)}| C^3_JC_{p+1}^+ \psi_L^\dag \psi_R (C^3_J )^\dag (C_{p+1}^+)^\dag    \0 =\\ \\
&&\langle\Omega_1^{(0)}|C_{p+1}^+ C^3_J \psi_L^\dag \psi_R (C^3_J )^\dag (C_{p+1}^+)^\dag \0
+\langle\Omega_1^{(0)}|C_{p+1+J}^+  \psi_L^\dag \psi_R(C^3_J )^\dag (C_{p+1}^+)^\dag \0 \\ \\
&&=\langle\Omega_1^{(0)}|C_{p+1}^+ C^3_J \psi_L^\dag \psi_R (C^3_J )^\dag (C_{p+1}^+)^\dag \0
+\langle\Omega_1^{(0)}|C_{p+1+J}^+  \psi_L^\dag \psi_R (C_{p+1}^+)^\dag (C^3_J )^\dag  \0 \\ \\
&&-\langle\Omega_1^{(0)}|C_{p+1+J}^+  \psi_L^\dag \psi_R(C_{p+1+J}^+)^\dag \0
\eea

\bea
&&\langle\Omega_1^{(0)}| C^3_JC_{p+1}^+ \psi_L^\dag \psi_R (C_{p+1}^+)^\dag  (C^3_J )^\dag  \0 = \\ \\
&& =\langle\Omega_1^{(0)}|C_{p+1}^+  C^3_J \psi_L^\dag \psi_R (C_{p+1}^+)^\dag  (C^3_J )^\dag  \0
+\langle\Omega_1^{(0)}|C_{p+1+J}^+  \psi_L^\dag \psi_R (C_{p+1}^+)^\dag  (C^3_J )^\dag \0 \\
\eea
\vspace{12pt}
and we have
\bea
&&\langle\Omega_1^{(0)}|C_{p+1}^+ C^3_J \psi_L^\dag \psi_R (C_{p+1+J}^+)^\dag  \0 =\\ \\
&&= \langle\Omega_1^{(0)}|C_{p+1}^+ (C_{p+1+J}^+)^\dag C^3_J \psi_L^\dag \psi_R  \0
- \langle\Omega_1^{(0)}|C_{p+1}^+ (C_{p+1}^+)^\dag  \psi_L^\dag \psi_R  \0 \\ \\
&&=\langle\Omega_1^{(0)}| (C^3_J )^\dag C^3_J \psi_L^\dag \psi_R \0
-p \langle\Omega_1^{(0)}|  \psi_L^\dag \psi_R \0  \\ \\
&&=-p\langle\Omega_1^{(0)}|  \psi_L^\dag \psi_R \0 \\ \\ \\
&&\langle\Omega_1^{(0)}|C_{p+1+J}^+  \psi_L^\dag \psi_R (C_{p+1}^+)^\dag  (C^3_J )^\dag  \0 = \\ \\
&&=\langle\Omega_1^{(0)}| C^3_J  \psi_L^\dag \psi_R (C^3_J )^\dag  \0 =(J-1)\langle\Omega_1^{(0)}| \psi_L^\dag \psi_R  \0 \\ \\ \\
&& \langle\Omega_1^{(0)}|C_{p+1}^+ C^3_J \psi_L^\dag \psi_R (C_{p+1}^+)^\dag  (C^3_J )^\dag  \0=
\\ \\
&&=\langle\Omega_1^{(0)}|C_{p+1}^+ (C_{p+1}^+)^\dag C^3_J \psi_L^\dag \psi_R  (C^3_J )^\dag  \0
-\langle\Omega_1^{(0)}|C_{p+1}^+ (C_{p+1-J}^+)^\dag \psi_L^\dag \psi_R  (C^3_J )^\dag  \0\\ \\
&&=p \langle\Omega_1^{(0)}|C^3_J \psi_L^\dag \psi_R (C^3_J )^\dag \0
-\langle\Omega_1^{(0)}|(C_{p+1-J}^+)^\dag C_{p+1}^+  \psi_L^\dag \psi_R  (C^3_J )^\dag  \0 \\ \\
&&\hspace{10pt}- \langle\Omega_1^{(0)}|C^3_J \psi_L^\dag \psi_R (C^3_J )^\dag \0 \\ \\
&&=\Big[(p-1)(J-1)\Big] \langle\Omega_1^{(0)}|  \psi_L^\dag \psi_R \0
-\theta(J-p) \langle\Omega_1^{(0)}|C_{p+1}^+ C_{J-p-1}^-  \psi_L^\dag \psi_R (C^3_J )^\dag \0 \\ \\
&&\hspace{10pt}+\theta(J-p) \langle\Omega_1^{(0)}|  C^3_J \psi_L^\dag \psi_R (C^3_J )^\dag \0 \\ \\
&&  = \Big[(p-1)(J-1) +\theta(J-p) (J-1)\Big]  \langle\Omega_1^{(0)}|  \psi_L^\dag \psi_R \0  \\ \\
&&\hspace{10pt} -\theta(J-p) \langle\Omega_1^{(0)}|C_{p+1}^+  \psi_L^\dag \psi_R (C^3_J )^\dag C_{J-p-1}^-  \0
 -\theta(J-p) \langle\Omega_1^{(0)}|C_{p+1}^+  \psi_L^\dag \psi_R (C_{p+1}^+)^\dag \0\\ \\
&&  = \Big[ (p-1)(J-1)+\theta(J-p) (J-p-1)\Big]  \langle\Omega_1^{(0)}|  \psi_L^\dag \psi_R \0  \\ \\
&&\hspace{10pt}-\delta_{J,p+\half} \langle\Omega_1^{(0)}|C_{p+1}^+ \Big((C^3_J)^\dag + e^{ik_J \x}\Big) \psi_L^\dag \psi_R (C_{\half}^+)^\dag \0 \\ \\
&&  = \Big[(p-1)(J-1) +\theta(J-p) (J-p-1)\Big]  \langle\Omega_1^{(0)}|  \psi_L^\dag \psi_R \0  \\ \\
&&\hspace{10pt}+\delta_{J,p+\half} \langle\Omega_1^{(0)}|(C_{-\half}^-)^\dag \psi_L^\dag \psi_R (C_{\half}^+)^\dag \0-\delta_{J,p+\half} e^{ik_J \x}\langle\Omega_1^{(0)}|C_{p+1}^+ \psi_L^\dag \psi_R (C_{\half}^+)^\dag \0 \\ \\
&&  = \Big[(p-1)(J-1) +\theta(J-p) (J-p-1)\Big]  \langle\Omega_1^{(0)}|  \psi_L^\dag \psi_R \0  \\ \\
&&\hspace{10pt}-\delta_{J,p+\half} e^{ik_{p+\half} \x}\langle\Omega_1^{(0)}|C_{p+1}^+ \psi_L^\dag \psi_R (C_{\half}^+)^\dag \0\;.\\
\eea
\lines
Let us evaluate
$\langle\Omega_1^{(0)}|C_{p+1}^+ \psi_L^\dag \psi_R (C_{\half}^+)^\dag \0$. Using
(\ref{commbc}) and (\ref{commddagc}) we get
\bea
[ C_{n+1}^+, \psi_R] &=& \nor \Bigg \{ \sum_{n=\half}^{\infty}r_{p+n+1}e^{ik_n \x}+
\sum_{n=\half}^p r_{p+1-n}e^{-ik_n \x}\\
&&\hspace{20pt}+\sum_{n=p+2}^{\infty} r_{n-p-1}^\dag e^{-ik_n \x}
+(r_{0}+r_{0}^\dag)e^{-ik_{p+1} \x}\Bigg \}
\eea
and since
\[
(C_{\half}^+)^\dag \0 = d_{\half}^\dag (r_{0}+r_{0}^\dag) \0
\]
we can write
\pagebreak
\baselineskip12pt
\bea
&&\langle\Omega_1^{(0)}|C_{p+1}^+ \psi_L^\dag \psi_R (C_{\half}^+)^\dag \0 = \\ \\
&&=\langle\Omega_1^{(0)}|\psi_L^\dag \psi_R C_{p+1}^+ (C_{\half}^+)^\dag \0 +\langle\Omega_1^{(0)}|\psi_L^\dag [ C_{n+1}^+, \psi_R]d_{\half}^\dag (r_{0}+r_{0}^\dag) \0 \\ \\
&&=\langle\Omega_1^{(0)}|\psi_L^\dag \psi_R C^3_{p+\half}\0
-\nor e^{-ik_{p+1} \x}\langle\Omega_1^{(0)}|\psi_L^\dag(r_{0}+r_{0}^\dag)^2 d_{\half}^\dag \0 \\ \\
&&=-\sqrt{2L}e^{-ik_{p+1} \x}\langle\Omega_1^{(0)}|\psi_L^\dag  \,({\stackrel {\;o} {\phi}}_R  )^2 d_{\half}^\dag \0
\eea
\lines
and from (\ref{zmccrs}) we can see that
\[({\stackrel {\;o} {\phi}}_R  )^2=\frac{1}{4L}\;.
\]
Note now that since
\[
|\Omega_1^{(0)}\rangle = e^{-\frac{i\pi\sq}{g\sqrt{L}}b_0^3} \beta_{\half}^\dag d_{\half}^\dag \0
\]
we can write
\[
\langle\Omega_1^{(0)}|\psi_L^\dag \psi_R \0
=e^{-ik_{\half}\x}\nor \langle\Omega_1^{(0)}|\psi_L^\dag  d_{\half}^\dag \0 \;.
\]
Therefore
\be
\langle\Omega_1^{(0)}|C_{p+1}^+ \psi_L^\dag \psi_R (C_{\half}^+)^\dag \0 =
-\half e^{-ik_{p+\half}\x} \langle\Omega_1^{(0)}|\psi_L^\dag \psi_R \0 \label{chalf}
\ee
and finally
\baselineskip12pt
\bea
&&\langle\Omega_1^{(0)}|C_{p+1}^+ C^3_J \psi_L^\dag \psi_R (C_{p+1}^+)^\dag (C^3_J )^\dag \0 =\\ \\
&&=  \frac{1}{2L}e^{-\frac{\pi}{2mL}}\Big[pJ-J+1-p +\theta(J-p) (J-p-1)+\half \delta_{J,p+\half}\Big] \\ \\
&&= \frac{1}{2L}e^{-\frac{\pi}{2mL}}\Big[pJ-J+1-p +\theta(J-p-1) (J-p-1)\Big] \;. \\
\eea
\lines
Moreover we have
\bean
&&\langle \Omega_1^{(0)}|
C_J^3  C^+_{\half }\psi_L^\dag  \psi_R (C_J^3)^\dag (C^+_{\half })^\dag \0=\langle \Omega_1^{(0)}|C^+ _{J+\half}  \psi_L^\dag  \psi_R (C_J^3)^\dag (C^+_{\half })^\dag \0 \nn\\
&&\hspace{-20pt}=\langle \Omega_1^{(0)}|C^+ _{J+\half}\Big( (C_J^3)^\dag +e^{ik_{J}\x}\Big) (C^+_{\half })^\dag \0 \nn\\
&&\hspace{-20pt}=-\langle \Omega_1^{(0)}|C^+ _\half  \psi_L^\dag  \psi_R
(C^+_{\half })^\dag \0  +e^{ik_{J}\x}\langle \Omega_1^{(0)}|C^+ _{J+\half}
\psi_L^\dag  \psi_R
(C^+_{\half })^\dag \0 \nn
\eean
and using  (\ref{chalf}) with $p+1\equiv J+\half$
\be\label{-half}
\langle \Omega_1^{(0)}|
C_J^3  C^+_{\half }\psi_L^\dag  \psi_R (C_J^3)^\dag (C^+_{\half })^\dag \0=
-\half  \langle\Omega_1^{(0)}|\psi_L^\dag \psi_R \0 \;.
\ee
We also need
\bea
&&\hspace{-20pt}\langle \Omega_1^{(0)}|
C_J^3  (C^-_{p- 1 })^\dag \psi_L^\dag  \psi_R  C^-_{p-1 } (C_J^3)^\dag
\0=\\
 &&\hspace{-20pt}=\langle \Omega_1^{(0)}|(C^- _{p-J-1})^\dag  \psi_L^\dag  \psi_R (C_J^3)^\dag C^-_{p-1 }\0
+ \langle \Omega_1^{(0)}|(C^- _{p-J-1})^\dag  \psi_L^\dag  \psi_R C^- _{p-J-1}\0 \\
&&\hspace{-20pt}=\langle \Omega_1^{(0)}|C^+ _{J+\half}  \psi_L^\dag  \psi_R (C_J^3)^\dag (C^+_{\half })^\dag \0 \delta_{p,\half} +\theta(J-p)(J-p)
\eea
and from (\ref{-half})
\bea
\langle \Omega_1^{(0)}|
C_J^3  (C^-_{p- 1 })^\dag \psi_L^\dag  \psi_R  C^-_{p-1 } (C_J^3)^\dag
\0=
\left(-\half \delta_{p,\half} +\theta(J-p)(J-p) \right)\langle\Omega_1^{(0)}|\psi_L^\dag \psi_R \0 \;.
\eea
\lines
Finally, using (\ref{azc}) we get
\baselineskip12pt
\bea
&&\langle \Omega^{(0)}_1|
C_{p+1}^+  \left( A_0^3-\frac{\pi \sqrt{m}}{g\sqrt{L}}\right) \psi_L^\dag \psi_R (A_0^3)^\dag
 (C_{n}^+)^\dag  \0= \\ \\
&&=\langle \Omega^{(0)}_1|C_{p+1}^+  (A_0^3)^\dag \psi_L^\dag \psi_R
 \left( A_0^3-\frac{\pi \sqrt{m}}{g\sqrt{L}}\right)(C_{p+1}^+)^\dag \0  +
\langle \Omega^{(0)}_1|C_{p+1}^+   \psi_L^\dag \psi_R  (C_{p+1}^+)^\dag  \0 \\ \\
&&=\Bigg(\frac{1-\pi^2 m}{g^2 L}\Bigg)\langle \Omega^{(0)}_1|C_{p+1}^+   \psi_L^\dag \psi_R  (C_{p+1}^+)^\dag  \0 = \frac{1}{2L}e^{-\frac{\pi}{2mL}}\Bigg(1-\frac{\pi^2 m}{g^2 L}\Bigg)p
\eea
\lines
Substituting these results into $\langle\Omega_1^{(2)}|\, \psi_L^\dag  \psi_R  | \Omega^{(2)}\rangle_{3+}$ and after some rearrangements
\baselineskip15pt
\bea
&&\hspace{-10pt}
\langle\Omega_1^{(2)}|\, \psi_L^\dag  \psi_R  | \Omega^{(2)}\rangle_{3+}=\\ \\
&&\hspace{-10pt}\frac{g^4e^{-\frac{\pi}{2mL}}}{2m^2 L^3}
\sum_{J=1}^{\infty}
\Bigg\{ -4m\sum_{p=\half}^{\infty}\frac{pJ-J+1-p +\theta(J-p-1) (J-p-1)}
{(2k_p +m) (2k_{p+1} +m)(2k_J)^3}\\ \\
&&\hspace{50pt} -2m\sum_{p=\half}^{\infty} \frac{p}{(2k_p+m)(2k_{p+1}+m)}
\Bigg(\frac{1}{(2k_p +2k_J +m)^2 (2k_{p+1} +2k_J +m)}\\
&&\hspace{210pt}+\frac{1}{(2k_p +2k_J +m) (2k_{p+1} +2k_J +m)^2}\Bigg)\\ \\
&&\hspace{50pt}\!+m \sum_{p=\half}^{\infty}\! \frac{1 }{(2k_{p+1}+m)}
\Bigg(\! \frac{2}{(2k_J)^3 (2k_p +2k_J +m)}+\frac{1}{(2k_J)^2 (2k_p +2k_J +m)^2}\!\Bigg) \\ \\
&&\hspace{50pt}-2m\sum_{p=\half}^{\infty}
\Bigg(\frac{2J}{(2k_J)^3 (2k_p +2k_J +m)(2k_{p+1} +2k_J +m)}\\
&&\hspace{100pt}+\frac{J}{(2k_J)^2 (2k_p +2k_J +m)^2
 (2k_{p+1} +2k_J +m)}\\
&&\hspace{100pt}+ \frac{J}{(2k_J)^2 (2k_p +2k_J +m) (2k_{p+1} +2k_J +m)^2}\Bigg)\\ \\
&&\hspace{50pt}+\frac{m }
{2(2k_{\half}+m)}\Bigg(\frac{2}{(2k_J)^3 (2k_{J-\half} +m)}+\frac{1}{(2k_J)^2 (2k_{J+\half} +m)^2}\Bigg) \\ \\
&&\hspace{50pt}-2m \sum_{p=\half}^{\infty}
\Bigg(\frac{2\theta(J-p)(J-p)}{(2k_J)^3 ( 2k_{J-p} +m)(2k_{J-p+1} +m)}\\
&&\hspace{100pt}+\frac{\theta(J-p)(J-p)}{(2k_J)^2 (2k_{J-p} +m)^2
 (2k_{J-p+1} +m)}\\
&&\hspace{100pt}+ \frac{\theta(J-p)(J-p)}{(2k_J)^2 (2k_{J-p} +m) (2k_{J-p+1} +m)^2}\Bigg)
 \Bigg\}
\\ \\
&& +\frac{g^4e^{-\frac{\pi}{2mL}}}{2m^2 L^3} \sum_{p=\half}^{\infty} \frac{\left(1-\frac{\pi^2 m}{g^2 L}\right)p
}{4(k_{p}+m) (2k_p+m)(k_{{p+1}}+m) (2k_{p+1}+m)}\\
\eea
Let us consider now the term $\langle\Omega_1^{(2)}|\, \psi_L^\dag \psi_R  | \Omega^{(2)}\rangle_{+\, -}$ . We have

\baselineskip15pt
\bea
&&\hspace{-20pt}\langle\Omega_1^{(2)}|\, \psi_L^\dag \psi_R  | \Omega^{(2)}\rangle_{+\, -}
=
\frac{g^4}{m^2 L^2}\Bigg\{ \sum_{n=-\infty}^{\infty}\sum_{p=\half}^{\infty} \frac{\langle \Omega_1^{(0)}|\,C_{p}^- C_{n + 1}^+ \psi_L^\dag  \psi_R (C_{n+1}^+)^\dag (C_{p}^-)^\dag \0}
{4(2k_p +m)(2k_{p+1} +m)(k_{n+p+1} +m)^2 }\\ \\
&&\hspace{128pt}+\sum_{n=\half}^{\infty}\sum_{p=\half}^{\infty}\frac{\langle \Omega_1^{(0)}|\,C_{n+1}^+ C_{p}^- \psi_L^\dag  \psi_R (C_{n+1}^+)^\dag (C_{p}^-)^\dag \0 }
{4(2k_n +m) (2k_p +m) (k_{n+p+1} +m)^2}\\ \\
&&\hspace{128pt}+\sum_{p=-\half}^{\infty}\sum_{n=-\half}^{\infty}\frac{\langle \Omega_1^{(0)}|\,C_{p}^- C_{n + 1}^+ \psi_L^\dag  \psi_R (C_{p}^-)^\dag (C_{n+1}^+)^\dag \0}
{4 (2k_{p+1} +m)  (2k_{n+1} +m)(k_{n+p+1}  +m)^2}\\ \\
&&\hspace{128pt}+\sum_{p=-\infty}^{\infty}\sum_{n=\half}^{\infty}\frac{\langle \Omega_1^{(0)}|\,C_{n+1}^+ C_{p}^- \psi_L^\dag  \psi_R (C_{p}^-)^\dag (C_{n+1}^+)^\dag \0}
{4 (2k_n +m)  (2k_{n+1} +m)(k_{n+p+1} +m)^2}\: \Bigg\}\;.
\eea
\lines
Considering that
\[
(C_{p}^-)^\dag (C_{n+1}^+)^\dag =
(C_{n+1}^+)^\dag  (C_{p}^-)^\dag +(C^3_{n+p+1})^\dag +(n+1)\delta_{n+1, -p}
\]
and
\[
C_{n+1}^+ C_{p}^- = C_{p}^- C_{n+1}^+ + C^3_{n+p+1}+ (n+1) \delta_{n+1 , -p}
\]
one gets

\baselineskip15pt
\bea
&&\hspace{-20pt}\langle\Omega_1^{(2)}|\, \psi_L^\dag \psi_R  | \Omega^{(2)}\rangle_{+\, -}
=
\frac{g^4}{m^2 L^2}\Bigg\{ \sum_{n=\half}^{\infty}\sum_{p=\half}^{\infty} \frac{\langle \Omega_1^{(0)}|\,C_{p}^- C_{n + 1}^+ \psi_L^\dag  \psi_R (C_{n+1}^+)^\dag (C_{p}^-)^\dag \0}
{(2k_n +m)(2k_{n+1} +m)(2k_p +m)(2k_{p+1} +m)}\\ \\
&&\hspace{85pt}+\sum_{n=\half}^{\infty}\sum_{p=\half}^{\infty} \frac{\langle \Omega_1^{(0)}|\,C_{p}^- C_{-n + 1}^+ \psi_L^\dag  \psi_R (C_{-n+1}^+)^\dag (C_{p}^-)^\dag \0}
{4(2k_p +m)(2k_{p+1} +m)(k_{p-n+1} +m)^2 }\\ \\
&&\hspace{85pt}-\sum_{p=\half}^{\infty}\frac{\langle \Omega_1^{(0)}|\,C^3_{p+\half} \psi_L^\dag  \psi_R (C_{\half}^+)^\dag (C_{p}^-)^\dag\0}
{4 (2k_{p+1} +m)  (2k_{\half} +m)(k_{p+\half}  +m)^2}\\ \\
&&\hspace{85pt}+ \sum_{n=\half}^{\infty}\sum_{p=\half}^{\infty}
\frac{\langle \Omega_1^{(0)}|\,C^3_{n+p+1}\psi_L^\dag  \psi_R
(C^+ _{n+1})^\dag (C^- _{p})^\dag \0}{2(2k_n +m)(2k_{n+1} +m)(2k_p +m)(k_{n+p+1} +m)}\\ \\
&&\hspace{85pt}+\sum_{p=-\half}^{\infty}\sum_{n=\half}^{\infty}\frac{\langle \Omega_1^{(0)}|\,C^- _{p}C^+ _{n+1}\psi_L^\dag  \psi_R
(C^3_{n+p+1})^\dag \0}{2(2k_n +m)(2k_{n+1} +m)(2k_{p+1} +m)(k_{n+p+1} +m)}\\ \\
&&\hspace{85pt}-\sum_{p=\half}^{\infty}\frac{\langle \Omega_1^{(0)}|\,C^3_{p+\half}\psi_L^\dag  \psi_R (C^3_{p+\half})^\dag \0}
{4 (2k_{p+1} +m)  (2k_{\half} +m)(k_{p+\half}  +m)^2}\\ \\
&&\hspace{85pt}+ \sum_{n=\half}^{\infty}\sum_{p=\half}^{\infty}
\frac{\langle \Omega_1^{(0)}|\, C^3_{n+p+1}\psi_L^\dag  \psi_R (C^3_{n+p+1})^\dag \0}
{4 (2k_n +m)  (2k_{n+1} +m)(k_{n+p+1} +m)^2}
\\ \\
&&\hspace{85pt}+\sum_{p=\half}^{\infty}\sum_{n=\half}^{\infty}\frac{\langle \Omega_1^{(0)}|\,C^3_{n-p+1} \psi_L^\dag  \psi_R (C^3_{n-p+1})^\dag \0}
{4 (2k_n +m)  (2k_{n+1} +m)(k_{n-p+1} +m)^2}\\ \\
&&\hspace{85pt}+ \frac{\langle \Omega_1^{(0)}|\, \psi_L^\dag  \psi_R  \0}
{16 (2k_{\half} +m)  (2k_{\half} +m)  m^2}+\sum_{n=\half}^{\infty}\frac{n(n+1) \langle \Omega_1^{(0)}|\psi_L^\dag  \psi_R \0}
{4 (2k_n +m)  (2k_{n+1} +m) m^2}\: \Bigg\}\:.
\eea
Since
\bea
&&\sum_{n=\half}^{\infty}\sum_{p=\half}^{\infty} \frac{\langle \Omega_1^{(0)}|\,C_{p}^- C_{-n + 1}^+ \psi_L^\dag  \psi_R (C_{-n+1}^+)^\dag (C_{p}^-)^\dag \0}
{4(2k_p +m)(2k_{p+1} +m)(k_{p-n+1} +m)^2 }= \\ \\
&& \qquad=-\sum_{p=\half}^{\infty}
\frac{\langle \Omega_1^{(0)}|\,C^3_{p+\half} \psi_L^\dag  \psi_R
(C_{p}^-  )^\dag (C_\half ^ + )^\dag \0}
{4(2k_p +m)(2k_{p+1} +m)(k_{p+\half} +m)^2 }
\\ \\
&&\qquad \quad+\sum_{n=\half}^{\infty}\sum_{p=\half}^{\infty} \frac{\langle \Omega_1^{(0)}|\,C_{p-n + 1}^3 \psi_L^\dag  \psi_R (C_{p-n+1}^3)^\dag  \0}
{4(2k_p +m)(2k_{p+1} +m)(k_{p-n+1} +m)^2 }\\ \\
&&\qquad \quad+\sum_{p=\half}^{\infty} \frac{p(p+1) \langle \Omega_1^{(0)}|\, \psi_L^\dag  \psi_R  \0}
{4(2k_p +m)(2k_{p+1} +m) m^2 }\\ \\
&&\qquad = -\sum_{p=\half}^{\infty}
\frac{\langle \Omega_1^{(0)}|\,C^3_{p+\half} \psi_L^\dag  \psi_R
 (C_\half ^ + )^\dag (C_{p}^-  )^\dag \0}
{4(2k_p +m)(2k_{p+1} +m)(k_{p+\half} +m)^2 }
\\ \\
&&\qquad \quad+\sum_{n=\half}^{\infty}\sum_{p=\half}^{\infty} \frac{\langle \Omega_1^{(0)}|\,C_{p-n }^3 \psi_L^\dag  \psi_R (C_{p-n}^3)^\dag  \0}
{4(2k_p +m)(2k_{p+1} +m)(k_{p-n} +m)^2 }\\ \\
&&\qquad \quad + \sum_{p=\half}^{\infty} \frac{p(p+1) \langle \Omega_1^{(0)}|\, \psi_L^\dag  \psi_R  \0}
{4(2k_p +m)(2k_{p+1} +m) m^2 }
\eea
we can write

\bea
&&\hspace{-20pt}\langle\Omega_1^{(2)}|\, \psi_L^\dag \psi_R  | \Omega^{(2)}\rangle_{+\, -}
=
\frac{g^4}{m^2 L^2}\Bigg\{ \sum_{n=\half}^{\infty}\sum_{p=\half}^{\infty} \frac{\langle \Omega_1^{(0)}|\,C_{p}^- C_{n + 1}^+ \psi_L^\dag  \psi_R (C_{n+1}^+)^\dag (C_{p}^-)^\dag \0}
{(2k_n +m)(2k_{n+1} +m)(2k_p +m)(2k_{p+1} +m)}\\ \\
&&\hspace{85pt}-\sum_{n=\half}^{\infty}\frac{\langle \Omega_1^{(0)}|\,C^3_{n+\half} \psi_L^\dag  \psi_R (C_{\half}^+)^\dag (C_{n}^-)^\dag\0}
 {2  (2k_{\half} +m)(2k_{n} +m)(2k_{n+1} +m)(k_{n+\half}  +m)}\\ \\
&&\hspace{85pt}+ \sum_{n=\half}^{\infty}\sum_{p=\half}^{\infty}
\frac{\langle \Omega_1^{(0)}|\,C^3_{n+p+1}\psi_L^\dag  \psi_R
(C^+ _{n+1})^\dag (C^- _{p})^\dag \0}{2(2k_n +m)(2k_{n+1} +m)(2k_p +m)(k_{n+p+1} +m)}\\ \\
&&\hspace{85pt}+\sum_{p=\half}^{\infty}\sum_{n=\half}^{\infty}\frac{\langle \Omega_1^{(0)}|\,C^- _{p}C^+ _{n+1}\psi_L^\dag  \psi_R
(C^3_{n+p+1})^\dag \0}{2(2k_n +m)(2k_{n+1} +m)(2k_{p+1} +m)(k_{n+p+1} +m)}\\ \\
&&\hspace{85pt}+\sum_{n=\half}^{\infty}\frac{\langle \Omega_1^{(0)}|\,C^- _{-\half}C^+ _{n+1}\psi_L^\dag  \psi_R
(C^3_{n+\half})^\dag \0}{2(2k_n +m)(2k_{n+1} +m)(2k_{\half} +m)(k_{n+\half} +m)}\\ \\
&&\hspace{85pt}-\sum_{p=\half}^{\infty}\frac{\langle \Omega_1^{(0)}|\,C^3_{p+\half}\psi_L^\dag  \psi_R (C^3_{p+\half})^\dag \0}
{4 (2k_{p+1} +m)  (2k_{\half} +m)(k_{p+\half}  +m)^2}\\ \\
&&\hspace{85pt}+ \sum_{n=\half}^{\infty}\sum_{p=\half}^{\infty}
\frac{\langle \Omega_1^{(0)}|\, C^3_{n+p+1}\psi_L^\dag  \psi_R (C^3_{n+p+1})^\dag \0}
{4 (2k_n +m)  (2k_{n+1} +m)(k_{n+p+1} +m)^2}
\\ \\
&&\hspace{85pt}+\sum_{p=\half}^{\infty}\sum_{n=\half}^{\infty}\frac{\langle \Omega_1^{(0)}|\,C^3_{n-p} \psi_L^\dag  \psi_R (C^3_{n-p})^\dag \0}
{2(2k_n +m)  (2k_{n+1} +m)(k_{n-p} +m)^2}\\ \\
&&\hspace{85pt}+\sum_{n=\half}^{\infty}\frac{\langle \Omega_1^{(0)}|\,C^3_{n+\half} \psi_L^\dag  \psi_R (C^3_{n+\half})^\dag \0}
{2(2k_n +m)  (2k_{n+1} +m)(k_{n+\half} +m)^2}\\ \\
&&\hspace{85pt}+ \frac{\langle \Omega_1^{(0)}|\, \psi_L^\dag  \psi_R  \0}
{16 (2k_{\half} +m)  (2k_{\half} +m)  m^2}\\ \\
&&\hspace{85pt}+\sum_{n=\half}^{\infty}\frac{n(n+1) \langle \Omega_1^{(0)}|\psi_L^\dag  \psi_R \0}
{2 (2k_n +m)  (2k_{n+1} +m) m^2}\: \Bigg\}\:.
\eea
\lines
We have
\bea
&&\langle \Omega_1^{(0)}|\,C_{p}^- C_{n + 1}^+ \psi_L^\dag  \psi_R (C_{n+1}^+)^\dag (C_{p}^-)^\dag \0 = \\
&&\qquad=\langle \Omega_1^{(0)}|\,C_{n + 1}^+C_{p}^-  \psi_L^\dag  \psi_R (C_{n+1}^+)^\dag (C_{p}^-)^\dag \0
- \langle \Omega_1^{(0)}|\,C_{n+p+1}^3 \psi_L^\dag  \psi_R (C_{n+1}^+)^\dag (C_{p}^-)^\dag \0 \\
&&\qquad =\langle \Omega_1^{(0)}|\,C_{n + 1}^+ (C_{n+1}^+)^\dag \psi_L^\dag  \psi_R  C_{p}^- (C_{p}^-)^\dag \0 + \langle \Omega_1^{(0)}| C_p^- \psi_L^\dag  \psi_R  (C_p^-)^\dag \0 \\
&&\qquad = (np+p) \langle \Omega_1^{(0)}| \psi_L^\dag  \psi_R \0
\eea
\bea
&&\langle \Omega_1^{(0)}|\,C^3_{n+\half} \psi_L^\dag  \psi_R (C_{\half}^+)^\dag (C_{n}^-)^\dag\0 = \langle \Omega_1^{(0)}|\psi_L^\dag  \psi_R (C^3_{n+\half}-e^{-ik_{n+\half} \x} )
 (C_{\half}^+)^\dag (C_{n}^-)^\dag\0 \\
&&\qquad = \langle \Omega_1^{(0)}|\psi_L^\dag  \psi_R (C_{\half}^+)^\dag C^3_{n+\half}(C_{n}^-)^\dag\0 -  \langle \Omega_1^{(0)}|\psi_L^\dag  \psi_R C_{n}^- (C_{n}^-)^\dag\0 \\
&&\qquad \quad -e^{-ik_{n+\half} \x}\langle \Omega_1^{(0)}|\psi_L^\dag  \psi_R   (C_{\half}^+)^\dag  (C_{n}^-)^\dag\0   \\
&&\qquad = \langle \Omega_1^{(0)}|\psi_L^\dag  \psi_R  (C_{\half}^+)^\dag C_{\half}^+ \0
-n  \langle \Omega_1^{(0)}| \psi_L^\dag  \psi_R \0 -e^{-ik_{n+\half} \x}\langle \Omega_1^{(0)}|(C_{\half}^+)^\dag \psi_L^\dag  \psi_R    (C_{n}^-)^\dag\0
\eea
In order to calculate
$\langle \Omega_1^{(0)}|(C_{\half}^+)^\dag\psi_L^\dag  \psi_R     (C_{n}^-)^\dag\0$
we use
\bea
&&[b_m , (C_n^-)^\dag ] =-r_{m-n}^\dag  \theta (m-n +\half) - r_{m-n}  \theta (n-m +\half)\\
&&[d_m ^\dag , (C_n^-)^\dag ] = -r_{m+n}
\eea
\bea
[ \psi_R , (C_n^-)^\dag ] = - \nor \left( \sum_{m=n+1}^\infty r_{m-n}^\dag e^{-ik_m \x} +\sum_{m=\half} ^{n-1}r_{n-m}e^{-ik_m \x}
+(r_0 + r_0^\dag ) e^{ik_n \x}\right)
\eea
and
\bea
&&C_{\half}^+ | \Omega_1^{(0)}\rangle = (r_0 + r_0^\dag ) d_\half  | \Omega_1^{(0)}\rangle\\
&& C_{n}^- (r_0 + r_0^\dag ) d_\half | \Omega_1^{(0)}\rangle =0  \;.
\eea
We then have
\bea
&&\langle \Omega_1^{(0)}|(C_{\half}^+)^\dag\psi_L^\dag  \psi_R     (C_{n}^-)^\dag\0 =
\langle \Omega_1^{(0)}|d_\half ^\dag (r_0 + r_0^\dag )(C_{n}^-)^\dag \psi_L^\dag  \psi_R \0 \\
&&\qquad \quad \hspace{110pt}- \nor e^{ik_n \x}\langle \Omega_1^{(0)}|d_\half ^\dag (r_0 + r_0^\dag ) \psi_L^\dag(r_0 + r_0^\dag )\0  \\
&&\qquad = \nor e^{ik_n \x}\langle \Omega_1^{(0)}|d_\half ^\dag \psi_L^\dag (r_0 + r_0^\dag )^2 \0
= \sqrt{2L} e^{ik_n \x}\langle \Omega_1^{(0)}|
d_\half ^\dag \psi_L^\dag ({\stackrel {\;o} {\psi}}_R  )^2 \0 \\
&&\qquad = -\half \nor e^{ik_n \x}\langle \Omega_1^{(0)}|\psi_L^\dag d_\half ^\dag \0 =
-\half  e^{ik_{n+\half} \x}\langle \Omega_1^{(0)}|\psi_L^\dag  \psi_R \0
\eea
and
\[
\langle \Omega_1^{(0)}|\,C^3_{n+\half} \psi_L^\dag  \psi_R (C_{\half}^+)^\dag (C_{n}^-)^\dag\0
=\left(\half -n \right) \langle \Omega_1^{(0)}|\psi_L^\dag  \psi_R \0 \; .
\]
Moreover
\bea
\langle \Omega_1^{(0)}|\,C^3_{n+p+1}\psi_L^\dag  \psi_R
(C^+ _{n+1})^\dag (C^- _{p})^\dag \0 =-\langle \Omega_1^{(0)}|\,C^- _{p}\psi_L^\dag  \psi_R
(C^- _{p})^\dag \0 =-p \langle \Omega_1^{(0)}|\psi_L^\dag  \psi_R \0
\eea
\bea
&&\langle \Omega_1^{(0)}|\,C^- _{p}C^+ _{n+1}\psi_L^\dag  \psi_R
(C^3_{n+p+1})^\dag \0 =
\\
&&\qquad = \langle \Omega_1^{(0)}|\,C^+ _{n+1}C^- _{p}\psi_L^\dag  \psi_R
(C^3_{n+p+1})^\dag \0
- \langle \Omega_1^{(0)}|\,(C^3_{n+p+1})\psi_L^\dag  \psi_R
(C^3_{n+p+1})^\dag \0 \\
&&\qquad=\langle \Omega_1^{(0)}|\,C^+ _{n+1}\psi_L^\dag  \psi_R (C^+ _{n+1})^\dag \0
-(n+p) \langle \Omega_1^{(0)}|\psi_L^\dag  \psi_R \0\\
&&\qquad = -p\langle \Omega_1^{(0)}|\psi_L^\dag  \psi_R \0
\eea
\bea
&&\langle \Omega_1^{(0)}|\,C^- _{-\half}C^+ _{n+1}\psi_L^\dag  \psi_R
(C^3_{n+\half})^\dag \0 = \\
&&\qquad = \langle \Omega_1^{(0)}|\,C^+ _{n+1}C^- _{-\half}\left((C^3_{n+\half})^\dag +
e^{ik_{n+\half} \x}\right) \psi_L^\dag  \psi_R \0
- \langle \Omega_1^{(0)}|\, C^3_{n+\half}\psi_L^\dag  \psi_R(C^3_{n+\half})^\dag \0 \\
&&\qquad =  \langle \Omega_1^{(0)}|\,C^+ _{n+1}(C^3_{n+\half})^\dag C^- _{-\half}
 \psi_L^\dag  \psi_R \0
+ \langle \Omega_1^{(0)}|\,C^+ _{n+1}(C^+ _{n+1})^\dag  \psi_L^\dag  \psi_R \0 \\
&&\qquad \quad +e^{ik_{n+\half} \x} \langle \Omega_1^{(0)}|\,C^+ _{n+1} C^- _{-\half}
 \psi_L^\dag  \psi_R \0 -  \left( n-\half \right)  \langle \Omega_1^{(0)}|\psi_L^\dag  \psi_R \0 \\
&&\qquad = -\langle \Omega_1^{(0)}|\,(C^- _{-\half})^\dag C^- _{-\half} \psi_L^\dag  \psi_R \0 =0
\eea
\lines
\vspace{4pt}
 where (\ref{chalf}) has been used. \\
Therefore we have
\baselineskip18pt
\bea
&&\hspace{-20pt}\langle\Omega_1^{(2)}|\psi_L^\dag \psi_R  | \Omega^{(2)}\rangle_{+\, -}
=\frac{g^4e^{-\frac{\pi}{2mL}}}{2m^2 L^3}
\Bigg\{ \sum_{n=\half}^{\infty}\sum_{p=\half}^{\infty} \frac{
 p(n+1)}
{(2k_n +m)(2k_{n+1} +m)(2k_p +m)(2k_{p+1} +m)}\\ \\
&&\hspace{85pt}+\sum_{n=\half}^{\infty}\frac{
\left(n- \half  \right)}
 {2  (2k_{\half} +m)(2k_{n} +m)(2k_{n+1} +m)(k_{n+\half}  +m)}\\ \\
&&\hspace{85pt}- \sum_{n=\half}^{\infty}\sum_{p=\half}^{\infty}
\frac{p}{2(2k_n +m)(2k_{n+1} +m)(2k_p +m)(k_{n+p+1} +m)}\\ \\
&&\hspace{85pt}-\sum_{p=\half}^{\infty}\sum_{n=\half}^{\infty}\frac{p}{2(2k_n +m)(2k_{n+1} +m)(2k_{p+1} +m)(k_{n+p+1} +m)}\\ \\
&&\hspace{85pt}-\sum_{p=\half}^{\infty}\frac{p-\half}
{4 (2k_{p+1} +m)  (2k_{\half} +m)(k_{p+\half}  +m)^2}\\ \\
&&\hspace{85pt}+ \sum_{n=\half}^{\infty}\sum_{p=\half}^{\infty}
\frac{n+p}
{4 (2k_n +m)  (2k_{n+1} +m)(k_{n+p+1} +m)^2}
\\ \\
&&\hspace{85pt}+\sum_{p=\half}^{\infty}\sum_{n=\half}^{\infty}\frac{(n-p-1)\theta(n-p-\half)}
{2(2k_n +m)  (2k_{n+1} +m)(k_{n-p} +m)^2}\\ \\
&&\hspace{85pt}+\sum_{n=\half}^{\infty}\frac{n-\half}
{4(2k_n +m)  (2k_{n+1} +m)(k_{n+\half} +m)^2}\\ \\
&&\hspace{85pt}+ \frac{1}
{16 (2k_{\half} +m)  (2k_{\half} +m)  m^2}
+\sum_{n=\half}^{\infty}\frac{n(n+1) }
{2 (2k_n +m)  (2k_{n+1} +m) m^2}\: \Bigg\}\\ \\
&&\hspace{75pt}=\; \frac{g^4e^{-\frac{\pi}{2mL}}}{2m^2 L^3}
\Bigg\{ \sum_{n=\half}^{\infty}\sum_{p=\half}^{\infty} \frac{
 pn}
{(2k_n +m)(2k_{n+1} +m)(2k_p +m)(2k_{p+1} +m)}\\ \\
&&\hspace{85pt}+ \sum_{n=\half}^{\infty}\sum_{p=\half}^{\infty}
\frac{n}
{2 (2k_n +m)  (2k_{n+1} +m)(k_{n+p+1} +m)^2}
\\ \\
&&\hspace{85pt}+\sum_{p=\half}^{\infty}\sum_{n=\half}^{\infty}\frac{(n-p-1)\theta(n-p-\half)}
{2(2k_n +m)  (2k_{n+1} +m)(k_{n-p} +m)^2}\\ \\
&&\hspace{85pt}+\sum_{n=\half}^{\infty}\frac{n-\half}
{2(2k_n +m)  (2k_{n+1} +m)(k_{n+\half} +m)^2}\\ \\
&&\hspace{85pt}+ \frac{1}
{16  (2k_{\half} +m)^2  m^2}
+\sum_{n=\half}^{\infty}\frac{n(n+1) }
{2 (2k_n +m)  (2k_{n+1} +m) m^2}\, \Bigg\}\\ \\
\:.
\eea
\lines

Collecting all the terms that contribute to $\langle \theta^{(2)}| i\psi_L^\dag(0,\x) \psir |  \theta^{(2)}\rangle$ we obtain

\baselineskip18pt
\bea
&&\langle \theta^{(2)}| i\psi_L^\dag  \psir-i\psi_R^\dag \psi_L   | \theta^{(2)}\rangle= \\ \\
&&\quad=-\frac{g^4e^{-\frac{\pi}{2mL}}}{m^2 L^3}\sum_{N=-\infty}^{\infty}\sin \theta
\Bigg\{
\sum_{I,J=1}^{\infty}
\frac{8m^2}{(2k_I)^3 ((2k_J)^3}(IJ-I-J+1)\\
&&\hspace{35pt}+ \sum_{J=1}^{\infty}
\frac{8m^2}{(2k_J)^6}(J^2-2J) \\ \\
&&\hspace{35pt}-8m \sum_{J=1}^{\infty}
\sum_{p=\half}^{\infty}\frac{pJ-J+1-p +\theta(J-p-1) (J-p-1)}
{(2k_p +m) (2k_{p+1} +m)(2k_J)^3}\\ \\
&&\hspace{35pt} -4m\sum_{J=1}^{\infty}\sum_{p=\half}^{\infty} \frac{p}{(2k_p+m)(2k_{p+1}+m)}
\Bigg(\frac{1}{(2k_p +2k_J +m)^2 (2k_{p+1} +2k_J +m)}\\
&&\hspace{205pt}+\frac{1}{(2k_p +2k_J +m) (2k_{p+1} +2k_J +m)^2}\Bigg)\\ \\
&&\hspace{35pt}\!+2m \sum_{J=1}^{\infty}\sum_{p=\half}^{\infty}\! \frac{1 }{(2k_{p+1}+m)}
\Bigg(\! \frac{2}{(2k_J)^3 (2k_p +2k_J +m)}\\
&&\hspace{165pt}+\frac{1}{(2k_J)^2 (2k_p +2k_J +m)^2}\!\Bigg) \\ \\
&&\hspace{35pt}-4m\sum_{J=1}^{\infty}\sum_{p=\half}^{\infty}
\Bigg(\frac{2J}{(2k_J)^3 (2k_p +2k_J +m)(2k_{p+1} +2k_J +m)}\\
&&\hspace{85pt}+\frac{J}{(2k_J)^2 (2k_p +2k_J +m)^2
 (2k_{p+1} +2k_J +m)}\\
&&\hspace{85pt}+ \frac{J}{(2k_J)^2 (2k_p +2k_J +m) (2k_{p+1} +2k_J +m)^2}\Bigg)\\ \\
&&\hspace{35pt}+\sum_{J=1}^{\infty}\frac{m }
{(2k_{\half}+m)}\Bigg(\frac{2}{(2k_J)^3 (2k_{J-\half} +m)}+\frac{1}{(2k_J)^2 (2k_{J+\half} +m)^2}\Bigg) \\ \\
&&\hspace{35pt}-4m \sum_{J=1}^{\infty}\sum_{p=\half}^{\infty}
\Bigg(\frac{2\theta(J-p)(J-p)}{(2k_J)^3 ( 2k_{J-p} +m)(2k_{J-p+1} +m)}\\
&&\hspace{85pt}+\frac{\theta(J-p)(J-p)}{(2k_J)^2 (2k_{J-p} +m)^2
 (2k_{J-p+1} +m)}\\
&&\hspace{85pt}+ \frac{\theta(J-p)(J-p)}{(2k_J)^2 (2k_{J-p} +m) (2k_{J-p+1} +m)^2}\Bigg)
\\ \\
&& \hspace{35pt}+ \sum_{p=\half}^{\infty} \frac{\left(1-\frac{\pi^2 m}{g^2 L}\right)p
}{2(k_{p}+m) (2k_p+m)(k_{{p+1}}+m) (2k_{p+1}+m)}\\ \\
&&\hspace{35pt}+\sum_{p,n=\half}^{\infty}\frac{2np-n+(n-p)\theta(n-p -\half)}
{(2k_p  +m) (2k_n +m) (2k_{p+1} +m) (2k_{n+1} +m)}\\ \\
&&\hspace{35pt}+\sum_{n=\half}^{\infty}\frac{n^2}{(2k_n  +m)^2 (2k_{n+1} +m)^2 }\\ \\
&&\hspace{40pt}+ \sum_{n=\half}^{\infty}\sum_{p=\half}^{\infty}
\frac{n}
{2 (2k_n +m)  (2k_{n+1} +m)(k_{n+p+1} +m)^2}
\\ \\
&&\hspace{40pt}+\sum_{p=\half}^{\infty}\sum_{n=\half}^{\infty}\frac{(n-p-1)\theta(n-p-\half)}
{2(2k_n +m)  (2k_{n+1} +m)(k_{n-p} +m)^2}\\ \\
&&\hspace{40pt}+\sum_{n=\half}^{\infty}\frac{n-\half}
{2(2k_n +m)  (2k_{n+1} +m)(k_{n+\half} +m)^2}\\ \\
&&\hspace{40pt}+ \frac{1}
{16  (2k_{\half} +m)^2  m^2}
+\sum_{n=\half}^{\infty}\frac{n(n+1) }
{2 (2k_n +m)  (2k_{n+1} +m) m^2}\, \Bigg\}
\:.
\\
\eea
\lines
Finally we have
\bea
\langle \theta |i( \psi_L^\dag \psi_R  - \psi_R^\dag \psi_L )  | \theta \rangle & \!\!\! \simeq &
\!\!\! \langle \theta^{(0)}| i(\psi_L^\dag  \psi_R -\psi_R^\dag \psi_L )   | \theta^{(0)}\rangle
+\langle \theta^{(1)}| i(\psi_L^\dag  \psi_R-\psi_R^\dag \psi_L )   | \theta^{(1)}\rangle \\
&&\!\!+2 \langle \theta^{(1)}| i(\psi_L^\dag  \psi_R-\psi_R^\dag \psi_L )   | \theta^{(2)}\rangle
+\langle \theta^{(2)}| i( \psi_L^\dag  \psi_R-\psi_R^\dag \psi_L  ) | \theta^{(2)}\rangle
\eea
where  $ \langle \theta^{(0)}| i(\psi_L^\dag  \psi_R -\psi_R^\dag \psi_L )   | \theta^{(0)}\rangle$,
$\langle \theta^{(1)}| i(\psi_L^\dag  \psi_R-\psi_R^\dag \psi_L )   | \theta^{(1)}\rangle $ and \\
$ \langle \theta^{(1)}| i(\psi_L^\dag  \psi_R-\psi_R^\dag \psi_L )   | \theta^{(2)}\rangle$
are given by  (\ref{condzero}) , (\ref{condone}) (\ref{condonetwo}).\\ \\
Let us calculate $\langle \theta^{(2)}| \theta^{(2)}\rangle$. We have

\baselineskip15pt
\bea
\langle \Omega_M^{(2)}|\, \psi_L^\dag  \psi_R  | \Omega_N^{(2)}\rangle_{33}=
\frac{g^4}{ L^2}\sum_{I,J=1}^{\infty}\frac{8}{(2k_I)^3 (2k_J)^3}
\langle \Omega_M^{(0)}|\,C_J^3 C_I^3  (C_J^3)^\dag (C_I^3)^\dag | \Omega_N^{(0)}\rangle\\ \\
=\frac{g^4}{ L^2}\delta_{MN}\sum_{I,J=1}^{\infty}\frac{8}{(2k_I)^3 (2k_J)^3}
\Big(IJ+J^2\delta_{IJ}\Big)\\
=\frac{g^4}{ L^2}\Bigg\{\sum_{I,J=1}^{\infty}\frac{8IJ}{(2k_I)^3 (2k_J)^3}+\sum_{J=1}^{\infty}
\frac{8J^2}{ (2k_J)^6}\Bigg\}
\eea

\bea
\langle\Omega_M^{(2)}|\Omega_N^{(2)}\rangle_{\pm \pm}&=&\frac{g^4}{m^2 L^2}\delta_{MN }\sum_{p,n,j,l=\half}^{\infty}\frac{\langle 0|\,C_{n}^\pm C_{p \pm}^+(C_{j }^\pm)^\dag (C_{l}^\pm)^\dag \0 (\delta_{nj}\delta_{pl}+\delta_{nl}\delta_{pj})}
{4(2k_n  +m) (2k_p +m)(2k_j  +m) (2k_l +m)}\\ \\
&=&\frac{g^4}{m^2 L^2}\delta_{MN }\sum_{p,n=\half}^{\infty}\frac{\langle 0|\,C_{n}^\pm C_{p }^\pm(C_{p }^\pm)^\dag (C_{n}^\pm)^\dag \0}
{2(2k_p  +m)^2 (2k_n +m)^2}\\ \\
&=&\frac{g^4}{m^2 L^2}\delta_{MN }\sum_{p,n=\half}^{\infty}\frac{\langle 0|\,C_{p}^\pm
\Big(
(C_{p }^\pm)^\dag C_{n }^\pm \pm C^3_{n-p}+n\delta_{np}\Big)(C_{n}^\pm)^\dag \0 }{2(2k_p  +m)^2 (2k_n +m)^2}\\ \\
&=&\frac{g^4}{m^2 L^2}\delta_{MN }\sum_{p,n=\half}^{\infty}\frac{np+n^2\delta_{np}\pm
\langle 0|\Big(C^3_{n-p}C_{p}^\pm \mp C^\pm_{n}\Big)(C_{n}^\pm)^\dag \0}{2(2k_p  +m)^2 (2k_n +m)^2}\\ \\
&=&\frac{g^4}{m^2 L^2}\delta_{MN }\sum_{p,n=\half}^{\infty}\frac{np+n^2\delta_{np}-n \pm
\langle 0|C^3_{n-p}\Big(\pm C^3_{p-n}+n\delta_{pn}\Big)\0}{2(2k_p  +m)^2 (2k_n +m)^2}\\ \\
&=&\frac{g^4}{m^2 L^2}\delta_{MN }\sum_{p,n=\half}^{\infty}\frac{np+n^2\delta_{np}-n
+(n-p)\theta(n-p)}{2(2k_p  +m)^2 (2k_n +m)^2}\\ \\
\eea

$\langle\Omega_M^{(2)}  | \Omega^{(2)}_N \rangle_{3+}
=\delta_{MN} \langle\Omega^{(2)}  | \Omega^{(2)} \rangle_{3+}$
and
\baselineskip15pt
\bea
&&\hspace{-20pt}\langle\Omega^{(2)}  | \Omega^{(2)}\rangle_{3+}
=\frac{g^4}{m^2 L^2}
\sum_{p=\half}^{\infty}\Bigg\{\,-4m \sum_{J=1}^{\infty}\frac{\langle 0|
 C^+_{p }C_J^3(C_J^3)^\dag (C^+_{p})^\dag
 \0 }{(2k_p+m)^2 (2k_p +2k_J + m )^3}\\
&&\hspace{-10pt}-2m\sum_{J=1}^{\infty}\frac{\langle 0 |
 C^+_{p }C_J^3(C^+_{{p} })^\dag (C_J^3)^\dag
\0 }
{ (2k_p+m) }\Bigg(\frac{2}{2k_J (2k_p +2k_J+m )^3 }+ \frac{1}{(2k_J)^2 (2k_p +2k_J +m)^2}
\Bigg)\\ \\
&&\hspace{-10pt}-2m \sum_{J=1}^{\infty}\frac{\langle 0 |
C_J^3  C^+_{p } (C_J^3)^\dag (C^+_{{p} })^\dag \0 }
{(2k_{p}+m)}\Bigg(\frac{2}{2k_J (2k_p +2k_J+m )^3 }+ \frac{1}{(2k_J)^2 (2k_p +2k_J +m)^2}
\Bigg)\\ \\
&&\hspace{-10pt}-4m\sum_{J=1}^{\infty}\langle 0 |
C_J^3  C^+_{p }(C^+_{{p} })^\dag (C_J^3 )^\dag
\0 \Bigg( \frac{1}
{(2k_p +2k_J+m )^2 (2k_J )^3 }\\
&&\hspace{137pt}+\frac{1}{(2k_p +2k_J +m)^3 (2k_J)^2 }
\Bigg)\\ \\
&&\hspace{-10pt}-4m\sum_{J=1}^{\infty}\langle 0 |
C_J^3  (C^-_{p })^\dag  C^-_{p } (C_J^3 )^\dag
\0 \Bigg(\!\frac{1}
{(-2k_p +2k_J +m )^2 (2k_J)^3 }\\
&&\hspace{137pt}+\frac{1}{(-2k_p +2k_J +m)^3 (2k_J )^2 }
\Bigg)\\ \\
&& \hspace{-10pt}+  \frac{\langle 0 |
C_{p}^+  A_0^3 (A_0^3)^\dag
 (C_{{p}}^+)^\dag | \0 \rangle}{4(k_{p}+m)^2 (2k_p+m)^2}\; \Bigg\}
\\
\eea

\lines
where
\bea
\langle 0 | C^+_{p }C_J^3(C^+_{p} )^\dag (C_J^3)^\dag \0 & =&
\langle 0 | C^+_{p }(C^+_{p} )^\dag C_J^3(C_J^3)^\dag \0 -
\langle 0 | C^+_{p }(C^+ _{p-J})^\dag(C_J^3)^\dag \0 \\
&=& pJ - \langle 0 | C^+_{p }(C_J^3)^\dag (C^+ _{p-J})^\dag \0
 - \langle 0 | C^+_{p }(C^+_{p })^\dag \0 \\
&=& pJ-p +\theta (p-J) (p-J)\\ \\
\langle 0| C^+_{p }C_J^3(C_J^3)^\dag (C^+_{p})^\dag \0 &=&
 \langle 0 | C^+_{p }C_J^3(C^+_{p} )^\dag (C_J^3)^\dag \0 -
 \langle 0 | C^+_{p }C_J^3 (C^+_{p+J})^\dag \0 \\
&& \langle 0 | C^+_{p }C_J^3(C^+_{p} )^\dag (C_J^3)^\dag \0 +p \\ \\
\langle 0 |
C_J^3  C^+_{p } (C_J^3)^\dag (C^+_{{p} })^\dag \0 &=&
\langle 0| C^+_{p }C_J^3(C_J^3)^\dag (C^+_{p})^\dag \0 +\langle 0| C^+ _{P+J}(C^+_{p} )^\dag (C_J^3)^\dag \0\\
&& -\langle 0| C^+ _{P+J}(C^+ _{P+J})^\dag \0 \\
&=& \langle 0 | C^+_{p }C_J^3(C^+_{p} )^\dag (C_J^3)^\dag \0 +p +J -(p+J)\\ \\
\langle 0 |C_J^3  C^+_{p }(C^+_{{p} })^\dag (C_J^3)^\dag \0
&=& \langle 0 | C^+_{p }C_J^3(C^+_{p} )^\dag (C_J^3)^\dag \0 +
\langle 0| C^+ _{P+J}(C^+_{p} )^\dag (C_J^3)^\dag \0\\
&=& \langle 0 | C^+_{p }C_J^3(C^+_{p} )^\dag (C_J^3)^\dag \0 +J \\ \\
 \langle 0 |C_J^3  (C^-_{p })^\dag  C^-_{p } (C_J^3)^\dag \0 &=&
\langle 0 |(C^-_{p-J })^\dag  C^-_{p-J }  \0 = \theta(J-p) (J-p)\\
\eea
so that

\baselineskip18pt
\bea
\langle\Omega^{(2)}  | \Omega^{(2)}\rangle_{3+}\!\!\!&=\!\!\!&\frac{g^4}{m^2 L^2}
\sum_{p=\half}^{\infty}\Bigg\{ \,
-4m \sum_{J=1}^{\infty}
\frac{p}{(2k_p+m)^2  (2k_p +2k_J + m )^3}\\ \\
&&\hspace{10pt}-4m\sum_{J=1}^{\infty}
\Bigg( \frac{J}
{(2k_p +2k_J+m )^2 (2k_J )^3 }+\frac{J}{(2k_p +2k_J +m)^3 (2k_J)^2 }
\Bigg)\\ \\
&&\hspace{10pt}-4m\sum_{J=1}^{\infty}
\Bigg( \frac{\theta(J-p) (J-p)}
{(-2k_p +2k_J+m )^2 (2k_J )^3 }+\frac{\theta(J-p) (J-p)}{(-2k_p +2k_J +m)^3 (2k_J)^2 }
\Bigg)\\ \\
&&\hspace{10pt} -4m\sum_{J=1}^{\infty} \frac{pJ-p +\theta (p-J) (p-J)}{(2k_p+m)^2 (2k_J)^3}\\ \\
&& \hspace{10pt}+  \frac{p}{4(k_{p}+m)^2 (2k_p+m)^2}\; \Bigg\}
\\
\eea

\baselineskip18pt
\bea
\hspace{-20pt}\langle\Omega_M^{(2)}  | \Omega_N^{(2)}\rangle_{+\, -}
&=&\frac{g^4}{m^2 L^2}\! \sum_{n, p=\half}^{\infty}\!\Bigg\{\frac{\langle 0|\,C_{n}^- C_{p}^+ (C_{p}^+)^\dag (C_{n}^-)^\dag | \0}
{4(k_{p +n} +m)^2 (2k_n +m)^2 }+ \!\frac{\langle 0|\,C_{n}^- (C_{p}^-)^\dag C_{p}^-  (C_{n}^-)^\dag | \0}
{4(k_{n-p} +m)^2 (2k_n +m)^2 }\\ \\
&&\hspace{60pt}+\frac{\langle 0 |\,C_{n}^- C_{p}^+  (C_{p}^+)^\dag (C_{n}^-)^\dag \0 +
\langle 0 |\,C^3_{p+n}  (C_{p}^+)^\dag (C_{n}^-)^\dag \0}
{4(k_{p +n} +m)^2 (2k_n +m)(2k_p+m)}\\ \\
&&\hspace{60pt}+\frac{\langle 0 |\,C_{n}^- C_{p}^+ (C_{p}^+)^\dag (C_{n}^-)^\dag  \0
+\langle 0 |\,C_{n}^- C_{p}^+ (C^3_{p+n})^\dag \0}
{4(k_{p +n} +m)^2 (2k_n +m)  (2k_p +m)}\\ \\
&&\hspace{60pt}+\frac{\langle 0 |\,C_{n}^-  C_{p}^+ (C_{p}^+)^\dag  (C_{n}^-)^\dag \0
+\langle 0 |\,C_{n}^-  C_{p}^+  (C^3_{p+n})^\dag \0  }{4(k_{p +n} +m)^2 (2k_p +m)^2  }\\ \\
&&\hspace{60pt}+\frac{\langle 0 |\,C^3_{p+n}(C_{p}^+)^\dag  (C_{n}^-)^\dag \0 +\langle 0 |\,C^3_{p+n} (C^3_{p+n})^\dag \0}
{4(k_{p +n} +m)^2 (2k_p +m)^2  }\\ \\
&&\hspace{60pt}+\frac{\langle 0 |\,C_{p}^+ (C_{n}^+)^\dag  C_{n}^+ (C_{p}^+)^\dag \0}
{4(k_{p-n} +m)^2 (2k_p +m)^2  }\Bigg\}\\ \\
&=&\frac{g^4}{m^2 L^2}\sum_{n, p=\half}^{\infty}\Bigg\{\! \!\frac{n(p+1)}
{(2k_p +m)^2 (2k_n +m)^2 }
+ \!\frac{(n-p)\theta(n-p-\half)}
{2(k_{n-p} +m)^2 (2k_n +m)^2 }\\ \\
&&\hspace{60pt}+ \!\frac{ n^2}
{2m^2 (2k_n +m)^2 }-\frac{n}
{(k_{p +n} +m) (2k_n +m)(2k_p+m)^2}\\ \\
&&\hspace{60pt}+\frac{n+p}
{4(k_{p +n} +m)^2 (2k_p +m)^2  }
\Bigg\}\\ \\
&=&\frac{g^4}{m^2 L^2}\sum_{n, p=\half}^{\infty}\Bigg\{\! \!\frac{np}
{(2k_p +m)^2 (2k_n +m)^2 }
+ \!\frac{(n-p)\theta(n-p-\half)}
{2(k_{n-p} +m)^2 (2k_n +m)^2 }\\ \\
&&\hspace{60pt}+ \!\frac{ n^2}
{2m^2 (2k_n +m)^2 }+\frac{n}
{2(k_{p +n} +m)^2 (2k_n +m)^2  }
\Bigg\}
\eea
\lines


Finally we get

\baselineskip18pt
\bea
&&\langle \theta^{(2)}|  \theta^{(2)}\rangle =
\frac{g^4}{m^2 L^2}\sum_{n=-\infty }^{\infty}\Bigg\{
\sum_{I,J=1}^{\infty}IJ \frac{8m^2}{(2k_I)^3 ((2k_J)^3}
+\sum_{J=1}^{\infty}\frac{8m^2}{(2k_J)^6}J^2
 \\ \\
&&\hspace{50pt} -8m\sum_{p=\half}^{\infty}\sum_{J=1}^{\infty} \frac{pJ-p +\theta (p-J) (p-J)}{(2k_p+m)^2 (2k_J)^3}\\ \\
&& \hspace{50pt}
-8m \sum_{p=\half}^{\infty}\sum_{J=1}^{\infty}
\frac{p}{(2k_p+m)^2  (2k_p +2k_J + m )^3}\\ \\
&&\hspace{50pt}-8m\sum_{p=\half}^{\infty}\sum_{J=1}^{\infty}
\Bigg( \frac{J}
{(2k_p +2k_J+m )^2 (2k_J )^3 }+\frac{J}{(2k_p +2k_J +m)^3 (2k_J)^2 }
\Bigg)\\ \\
&&\hspace{50pt}-8m\sum_{p=\half}^{\infty}\sum_{J=1}^{\infty}
\Bigg( \frac{\theta(J-p) (J-p)}
{(-2k_p +2k_J+m )^2 (2k_J )^3 }+\frac{\theta(J-p) (J-p)}{(-2k_p +2k_J +m)^3 (2k_J)^2 }
\Bigg)\\ \\
&& \hspace{50pt}+ \sum_{p=\half}^{\infty} \frac{p}{2(k_{p}+m)^2 (2k_p+m)^2}
 +\sum_{n=\half}^{\infty} \frac{n^2}{(2k_n +m)^4}\\ \\
&& \hspace{50pt}+\sum_{p,n=\half}^{\infty}\frac{2np-n
+(n-p)\theta(n-p-\half)}{(2k_p  +m)^2 (2k_n +m)^2}
+ \sum_{n=\half}^{\infty} \!\frac{ n^2}
{2m^2 (2k_n +m)^2 }\\ \\
&& \hspace{50pt}+ \sum_{n, p=\half}^{\infty}\Bigg[
 \!\frac{(n-p)\theta(n-p-\half)}
{2(k_{n-p} +m)^2 (2k_n +m)^2 }
+\frac{n}
{2(k_{p +n} +m)^2 (2k_n +m)^2  }\Bigg]
\Bigg\}\;.
\eea
\lines
and
\bea
\langle \theta  | \theta \rangle &\simeq &
\langle \theta^{(0)}|  \theta^{(0)}\rangle
+\langle \theta^{(1)}|  \theta^{(1)}\rangle
+\langle \theta^{(2)}|  \theta^{(2)}\rangle
\eea
where $\langle \theta^{(0)}|  \theta^{(0)}\rangle$ and  $\langle \theta^{(1)}|  \theta^{(1)}\rangle$
are given by (\ref{normzero}) and (\ref{normone}) .\\ \\
We now need to evaluate  $\frac{\langle \theta | \psi_L^\dag \psi_R | \theta \rangle}
{\langle \theta  | \theta \rangle }$.  We expand $\frac{1}{\langle \theta  | \theta \rangle }$
obtaining

\baselineskip18pt
\bea
\frac{1}{\langle \theta  | \theta \rangle }&\!\!\! \simeq & \!\!\!\Bigg(\sum_{N=-\infty}^{\infty}1\Bigg) ^{-1}\Bigg\{
1-\frac{g^2}{mL}\Bigg[ \sum_{n=\half}^{\infty}\frac{2n}{(2k_n +m)^2}
- \sum_{J=1}^{\infty}\frac{mJ}{ 2k_J^3}\Bigg] \\ \\
&&+\frac{g^4}{m^2 L^2}
\Bigg[ \sum_{n=\half}^{\infty}\frac{2n}{(2k_n +m)^2}
- \sum_{J=1}^{\infty}\frac{mJ}{ 2k_J^3}\Bigg]^2 -\Bigg(\sum_{N=-\infty}^{\infty}1\Bigg) ^{-1} \langle \theta^{(2)}|  \theta^{(2)}\rangle  \Bigg\}\\ \\
&=&\Bigg(\sum_{N=-\infty}^{\infty}1\Bigg) ^{-1}\Bigg\{ 1- \Bigg(\sum_{N=-\infty}^{\infty}1\Bigg) ^{-1}\langle \theta^{(1)}|  \theta^{(1)}\rangle \\ \\
&& + \Bigg(\sum_{N=-\infty}^{\infty}1\Bigg) ^{-2}\langle \theta^{(1)}|  \theta^{(1)}\rangle^2 - \Bigg(\sum_{N=-\infty}^{\infty}1\Bigg) ^{-1} \langle \theta^{(2)}|  \theta^{(2)}\rangle  \Bigg\}
\eea
\lines
so that we can write

\baselineskip12pt
\bea
&&\hspace{-20pt}\frac{ \langle \theta | i(\psi_L^\dag \psi_R - \psi_R^\dag \psi_L )| \theta \rangle}
{ \langle \theta  | \theta \rangle }\simeq \\ \\
&&-\frac{\sin \theta e^{-\frac{\pi}{2mL}}}{ L}+
\left(\sum_{N=-\infty}^{\infty}1 \right)^{-1} \Bigg\{
\langle \theta^{(1)}| i\psi_L^\dag  \psi_R-i\psi_R^\dag \psi_L   | \theta^{(1)}\rangle
+\frac{\sin \theta e^{-\frac{\pi}{2mL}}}{ L} \langle \theta^{(1)}| \theta^{(1)}\rangle\Bigg\}\\ \\
&&-\left(\sum_{N=-\infty}^{\infty}1 \right)^{-2} \langle \theta^{(1)}| \theta^{(1)}\rangle
\Bigg[ \langle \theta^{(1)}| i\psi_L^\dag  \psi_R -i\psi_R^\dag \psi_L   | \theta^{(1)}\rangle
+\frac{\sin \theta e^{-\frac{\pi}{2mL}}}{ L} \langle \theta^{(1)}| \theta^{(1)}\rangle
\Bigg] \\ \\
&&+\left(\sum_{N=-\infty}^{\infty}1 \right)^{-1}  \Bigg\{
2  \langle \theta^{(1)}| i\psi_L^\dag  \psi_R -i\psi_R^\dag \psi_L   | \theta^{(2)}\rangle
+ \langle \theta^{(2)}| i\psi_L^\dag  \psi_R -i\psi_R^\dag \psi_L   | \theta^{(2)}\rangle\\ \\
&&
+\frac{\sin \theta e^{-\frac{\pi}{2mL}}}{ L} \langle \theta^{(2)}| \theta^{(2)}\rangle\Bigg\}
\eea
and

\baselineskip18pt
\bea
&&\hspace{-20pt}\frac{L e^{\frac{\pi}{2mL}}}{\sin\theta}\frac{ \langle \theta | i(\psi_L^\dag \psi_R - \psi_R^\dag \psi_L )| \theta \rangle}
{ \langle \theta  | \theta \rangle }\simeq \\ \\
&&\hspace{-20pt}\simeq -1+\frac{g^2}{mL}\Bigg\{  \sum_{n=\half}^{\infty}\frac{2n}{(2k_n +m)}\Bigg(\frac{1}{2k_n +m} -\frac{1}{(2k_{n+1} +m)}\Bigg)
-\sum_{J=1}^{\infty}\frac{4m}{(2k_J )^3}\Bigg\} \\ \\
&&\hspace{-20pt}+\frac{g^4}{m^2 L^2}\Bigg\{ - \sum_{n,p=\half}^{\infty}\frac{8 k_n p}{(2k_n +m)^2 (2k_p +m)^2 (2k_{p+1} +m)}\\ \\
&&\hspace{10pt}+\sum_{n=\half}^{\infty}\frac{2n}{(2k_n +m)^2}\sum_{J=1}^{\infty}
\frac{4m}{(2k_J )^3}+\sum_{J=1}^{\infty}
\frac{4mJ }{(2k_J )^3}
\sum_{n=\half}^{\infty}\frac{4k_n}{(2k_n +m)^2 (2k_{n+1} +m)}\\ \\
&&\hspace{10pt}- \sum_{I,J=1}^{\infty}
\frac{4mJ }{(2k_J )^3}
\frac{4m}{(2k_I )^3}+\sum_{I,J=1}^{\infty}
\frac{8m^2}{(2k_I )^3 (2k_J)^3}(I+J-1)
+\sum_{J=1}^{\infty}\frac{16m^2 J}{ (2k_J)^6}\\ \\
&&\hspace{10pt}+\sum_{p=\half}^{\infty}
\sum_{J=1}^{\infty}
\frac{4mp}{(2k_p+m) (2k_{p} +2k_J+m )}\Bigg(\frac{1}{(2k_{p+1} +m )(2k_{p+1} +2k_J+m )^2 }\\
&&\hspace{10pt}
+\frac{1}{(2k_{p+1} +m )(2k_{p+1} +2k_J+m ) (2k_{p} +2k_J+m )}\\
&&\hspace{10pt}-\frac{2}{(2k_p+m) (2k_{p} +2k_J+m )^2} \Bigg)\\ \\
&&\hspace{10pt}-\sum_{J=1}^{\infty}\sum_{p=\half}^{\infty}\frac{ 2m}
{(2k_{p+1}+m)(2k_p +2k_J +m)(2k_J)^2}\Bigg( \frac{1}{(2k_p +2k_J +m)}+ \frac{1}{k_J}
\Bigg)\\ \\
&&\hspace{10pt}+\sum_{p=\half}^{\infty}\sum_{J=1}^{\infty} \frac{4mJ}
{(2k_p +2k_J +m) (2k_J)^2 }\Bigg(\frac{2}{2k_J (2k_{p+1} +2k_J +m )}\\
&&\hspace{10pt}+\frac{1}{(2k_{p+1} +2k_J +m )^2}+\frac{1}{(2k_{p+1} +2k_J +m )(2k_p +2k_J +m)}\\
&&\hspace{10pt}-\frac{2}{2k_J (2k_{p} +2k_J +m)}-\frac{2}{ (2k_{p} +2k_J +m)^2}
\Bigg)\\ \\
&&\hspace{10pt}+\sum_{J=1}^{\infty}\frac{ m}
{(2k_{\half}+m)(2k_{J-\half} +m)(2k_J)^2}\Bigg( \frac{1}{(2k_{J-\half} +m)}+ \frac{1}{k_J}
\Bigg)\\ \\
&&\hspace{10pt}+\sum_{p=\half}^{\infty}\sum_{J=1}^{\infty} \frac{4m\theta(J-p) (J-p)}
{(-2k_p +2k_J +m) (2k_J)^2 }\Bigg(\frac{2}{2k_J (-2k_{p-1} +2k_J +m )}\\
&&\hspace{10pt}+\frac{1}{(-2k_{p-1} +2k_J +m )^2}+\frac{1}{(-2k_{p-1} +2k_J +m )(-2k_p +2k_J +m)}\\
&&\hspace{10pt}-\frac{2}{2k_J (-2k_{p} +2k_J +m)}-\frac{2}{ (-2k_{p} +2k_J +m)^2}
\Bigg)\\ \\
&&\hspace{10pt}- \sum_{p=\half}^{\infty}\sum_{J=1}^{\infty}\frac{8m}{(2k_J)^3}
 \frac{pJ-J +\theta (J-p) (J-p)}{(2k_p+m)}\Bigg( \frac{1}{2k_p+m}-\frac{1}{2k_{p+1}+m}
\Bigg)\\ \\
&&\hspace{10pt}+\sum_{p=\half}^{\infty}\sum_{J=1}^{\infty}\frac{8m}{(2k_J)^3}
\frac{-p+1 -\theta(J-p-1) -\half \delta_{J,p+\half}}{(2k_p +m) (2k_{p+1} +m)}\\ \\
&&\hspace{10pt}+\sum_{p=\half}^{\infty} \frac{p}{2(k_{p}+m) (2k_p+m)}
\Bigg( \frac{1}{(k_{p}+m) (2k_p+m)}-\frac{1}{(k_{p+1}+m) (2k_{p+1}+m)}\Bigg)\\ \\
&&\hspace{10pt}+\frac{3\pi^2 m}{2g^2 L}\sum_{p=\half}^{\infty} \frac{p
}{(k_{p}+m) (2k_p+m)(k_{{p+1}}+m) (2k_{p+1}+m)}\\ \\
&&\hspace{10pt}+\sum_{p,n=\half}^{\infty}\frac{2np-n+(n-p)\theta(n-p-\half)}
{(2k_p  +m) (2k_n +m) }\Bigg( \frac{1}{(2k_p  +m) (2k_n +m)}\\
&&\hspace{160pt}-\frac{1}{(2k_{p+1} +m) (2k_{n+1} +m)}\Bigg)\\ \\
&&\hspace{10pt}+\sum_{n=\half}^{\infty} \frac{n^2}{(2k_n +m)^2}\Bigg(\frac{1}{(2k_n  +m)^2 }
-\frac{1}{(2k_{n+1} +m)^2 }\Bigg) \\ \\
&&\hspace{10pt}+ \sum_{n=\half}^{\infty} \!\frac{ n^2}
{2m^2 (2k_n +m)}\Bigg(\frac{1}{2k_n +m}- \frac{1}{2k_{n+1} +m}\Bigg)
\\ \\
&&\hspace{10pt}-\sum_{n=\half}^{\infty}\frac{n }
{2 (2k_n +m)  (2k_{n+1} +m) m^2}\\ \\
&& \hspace{10pt}+\sum_{n, p=\half}^{\infty}
 \!\frac{(n-p)\theta(n-p-\half)}
{2(k_{n-p} +m)^2 (2k_n +m)}\Bigg(\frac{1}{2k_n +m}- \frac{1}{2k_{n+1} +m}\Bigg)\\ \\
&&\hspace{10pt}+\sum_{p=\half}^{\infty}\sum_{n=\half}^{\infty}\frac{\theta(n-p-\half)}
{2(2k_n +m)  (2k_{n+1} +m)(k_{n-p} +m)^2}\\ \\
&&\hspace{10pt}+\sum_{n=\half}^{\infty}\frac{n-\half}
{2(2k_n +m)  (2k_{n+1} +m)(k_{n+\half} +m)^2}
+ \frac{1}{16  (2k_{\half} +m)^2  m^2}\\ \\
&&\hspace{10pt}+ \sum_{n=\half}^{\infty}\sum_{p=\half}^{\infty}
\frac{n}
{2 (2k_n +m ) }\Bigg(\frac{1}{(k_{p +n} +m)^2(2k_n +m)}- \frac{1}{(k_{p +n+1} +m)^2(2k_{n+1} +m)}\Bigg)\Bigg\} \\ \\
&=&-1+\frac{g^2}{mL}\Bigg\{  \sum_{n=\half}^{\infty}\frac{4k_n}{(2k_n +m)^2 (2k_{n+1} +m)}
-\sum_{J=1}^{\infty}\frac{4m}{(2k_J )^3}\Bigg\} \\ \\
&&\hspace{-20pt}+\frac{g^4}{m^2 L^2}\Bigg\{ - \sum_{n,p=\half}^{\infty}\frac{8 k_n p}{(2k_n +m)^2 (2k_p +m)^2 (2k_{p+1} +m)}\\ \\
&&\hspace{10pt}+\sum_{n=\half}^{\infty}\frac{2n}{(2k_n +m)^2}\sum_{J=1}^{\infty}
\frac{4m}{(2k_J )^3}
+\sum_{J=1}^{\infty}
\frac{4mJ }{(2k_J )^3}
\sum_{n=\half}^{\infty}\frac{4k_n}{(2k_n +m)^2 (2k_{n+1} +m)}\\ \\
&&\hspace{10pt}- \sum_{I,J=1}^{\infty}
\frac{4mJ }{(2k_J )^3}
\frac{4m}{(2k_I )^3}
+\sum_{I,J=1}^{\infty}
\frac{8m^2}{(2k_I )^3 (2k_J)^3}(I+J-1)
+\sum_{J=1}^{\infty}\frac{16m^2 J}{ (2k_J )^6}\\ \\
&&\hspace{10pt}+\sum_{p=\half}^{\infty}
\sum_{J=1}^{\infty}
\frac{4mp}{(2k_p+m) (2k_{p} +2k_J+m )}\Bigg(\frac{1}{(2k_{p+1} +m )(2k_{p+1} +2k_J+m )^2 }\\
&&\hspace{10pt}
+\frac{1}{(2k_{p+1} +m )(2k_{p+1} +2k_J+m ) (2k_{p} +2k_J+m )}\\
&&\hspace{10pt}-\frac{2}{(2k_p+m) (2k_{p} +2k_J+m )^2} \Bigg)\\ \\
&&\hspace{10pt}-\sum_{J=1}^{\infty}\sum_{p=\half}^{\infty}\frac{ 2m}
{(2k_{p+1}+m)(2k_p +2k_J +m)(2k_J)^2}\Bigg( \frac{1}{(2k_p +2k_J +m)}+ \frac{1}{k_J}
\Bigg)\\ \\
&&\hspace{10pt}-\sum_{p=\half}^{\infty}\sum_{J=1}^{\infty} \frac{8mk_J}
{(2k_p +2k_J +m) (2k_J)^2 }\Bigg(\frac{1}{k_J (2k_p +2k_J +m)(2k_{p+1} +2k_J +m)}\\
&&\hspace{10pt}\!+\frac{1}{(2k_p +2k_J +m)(2k_{p+1} +2k_J +m)^2}
+\! \frac{2}{(2k_p +2k_J +m)^2(2k_{p+1} +2k_J +m)}\Bigg)\\ \\
&&\hspace{10pt}+\sum_{J=1}^{\infty}\frac{ m}
{(2k_{\half}+m)(2k_{J-\half} +m)(2k_J)^2}\Bigg( \frac{1}{(2k_{J-\half} +m)}+ \frac{1}{k_J}
\Bigg)\\ \\
&&\hspace{10pt}-\pl \sum_{p=\half}^{\infty}\sum_{J=1}^{\infty} \frac{8m\theta(J-p) (J-p)}
{(-2k_p +2k_J +m) (2k_J)^2 }\Bigg(\frac{1}{(-2k_p +2k_J +m)(-2k_{p-1} +2k_J +m)^2}\\
&&\hspace{10pt}+\frac{1}{k_J (-2k_p +2k_J +m)(-2k_{p-1} +2k_J +m)}\\
&&\hspace{10pt}+\! \frac{2}{(-2k_p +2k_J +m)^2(-2k_{p-1} +2k_J +m)}\Bigg)\\ \\
&&\hspace{10pt}-  \sum_{p=\half}^{\infty}\sum_{J=1}^{\infty}\frac{16m}{(2k_J)^3}
 \frac{k_p J -k_J +\theta (J-p) (k_J -k_p )}{(2k_p+m)^2 (2k_{p+1} +m)}\\ \\
&&\hspace{10pt}+\sum_{p=\half}^{\infty}\sum_{J=1}^{\infty}\frac{8m}{(2k_J)^3}
\frac{-p+1 -\theta(J-p-1) }{(2k_p +m) (2k_{p+1} +m)}\\ \\
&&\hspace{10pt}-\sum_{p=\half}^{\infty}\frac{4m}{(2k_{p+\half})^3(2k_p +m) (2k_{p+1} +m)}\\ \\
&&\hspace{10pt}+\sum_{p=\half}^{\infty} \frac{k_p}{2(k_{p}+m)^2 (2k_p+m) (2k_{p+1}+m)}
\Bigg( \frac{2}{(2k_p+m)}+\frac{1}{(k_{p+1}+m)}\Bigg)\\ \\
&&\hspace{10pt}+\frac{3\pi^2 m}{2g^2 L}\sum_{p=\half}^{\infty} \frac{p
}{(k_{p}+m) (2k_p+m)(k_{{p+1}}+m) (2k_{p+1}+m)}\\ \\
&&\hspace{10pt}+\sum_{p,n=\half}^{\infty}\frac{4k_n p-2k_n+2(k_n -k_p )\theta(n-p-\half)}
{(2k_p  +m) (2k_n +m)^2 (2k_{p+1} +m)}\Bigg( \frac{ 1}{(2k_p  +m) }+\frac{1}{  (2k_{n+1} +m)} \Bigg)\\ \\
&&\hspace{10pt}+\sum_{n=\half}^{\infty} \frac{n k_n}{(2k_n +m)^3(2k_{n+1} +m)}
\Bigg(\frac{1}{(2k_{n+1} +m) }+
\frac{1}{(2k_n  +m) }
\Bigg) \\ \\
&&\hspace{10pt}-\sum_{n=\half}^{\infty} \!\frac{ n }
{2m (2k_n +m)^2 (2k_{n+1} +m) }\\ \\
&& \hspace{10pt}+\sum_{n, p=\half}^{\infty}
 \!\frac{(k_n-k_p)\theta(n-p-\half)}
{(k_{n-p} +m)^2 (2k_n +m)^2 (2k_{n+1} +m)}\\ \\
&&\hspace{10pt}+\sum_{p=\half}^{\infty}\sum_{n=\half}^{\infty}\frac{\theta(n-p-\half)}
{2(2k_n +m)  (2k_{n+1} +m)(k_{n-p} +m)^2}\\ \\
&&\hspace{10pt}+\sum_{n=\half}^{\infty}\frac{n-\half}
{2(2k_n +m)  (2k_{n+1} +m)(k_{n+\half} +m)^2}
+ \frac{1}{16  (2k_{\half} +m)^2  m^2}\\ \\
&&\hspace{10pt}+ \sum_{n=\half}^{\infty}\sum_{p=\half}^{\infty}
\frac{k_n}
{2 (2k_n +m ) (2k_{n+1} +m)(k_{p +n} +m)}\Bigg(\frac{1}{(k_{p +n} +m) (k_{p +n+1} +m)}\\ \\
&&\hspace{165pt}+\frac{1}{ (k_{p +n+1} +m)^2 }
+ \frac{2}{(k_{p +n} +m) (2k_n +m )}\Bigg) \Bigg\} \\
\eea
Combining the  terms that contain divergent series we get
\bea
&&\frac{g^4}{m^2 L^2}\Bigg\{ - \sum_{n=\half}^{\infty}\frac{k_n }{(2k_n +m)^2 }
\sum_{p=\half}^{\infty}\frac{8  p}{ (2k_p +m)^2 (2k_{p+1} +m)}\\ \\
&&\hspace{20pt}+\sum_{n=\half}^{\infty}\frac{2n}{(2k_n +m)^2}\sum_{J=1}^{\infty}
\frac{4m}{(2k_J )^3}\\ \\
&&\hspace{20pt}+\sum_{p,n=\half}^{\infty}\frac{4k_n p-2k_n+2(k_n -k_p )\theta(n-p-\half) }
{(2k_p  +m) (2k_n +m)^2 (2k_{p+1} +m)}\Bigg( \frac{ 1}{(2k_p  +m) }+\frac{1}{  (2k_{n+1} +m)} \Bigg)\\ \\
&&\hspace{20pt}+\sum_{p=\half}^{\infty}\sum_{J=1}^{\infty}\frac{8m}{(2k_J)^3}
\frac{-p }{(2k_p +m) (2k_{p+1} +m)} \Bigg\}= \\ \\ \\
=&&\frac{g^4}{m^2 L^2}\Bigg\{ - \sum_{n=\half}^{\infty}\frac{k_n }{(2k_n +m)^2 }
\sum_{p=\half}^{\infty}\frac{8  p}{ (2k_p +m)^2 (2k_{p+1} +m)}\\ \\
&&\hspace{20pt}+\sum_{n=\half}^{\infty}\frac{4k_n}{(2k_n +m)^2 (2k_{n+1} +m)}\sum_{J=1}^{\infty}
\frac{4m}{(2k_J)^3}\\ \\
&&\hspace{20pt} +\sum_{n=\half}^{\infty}\frac{k_n }{(2k_n +m)^2 }
\sum_{p=\half}^{\infty}\frac{4  p}{ (2k_p +m)^2 (2k_{p+1} +m)}\\ \\
&&\hspace{20pt}+ \sum_{p=\half}^{\infty}\frac{k_p }{(2k_p +m)(2k_{p+1} +m) }
\sum_{n=\half}^{\infty}\frac{4  n}{ (2k_n +m)^2 (2k_{n+1} +m)}\\ \\
&&\hspace{20pt}+ \sum_{p,n=\half}^{\infty}\frac{-2k_n \theta(p-n+\half)-2 k_p \theta(n-p-\half) }
{(2k_p  +m) (2k_n +m)^2 (2k_{p+1} +m)}\Bigg( \frac{ 1}{(2k_p  +m) }+\frac{1}{  (2k_{n+1} +m)} \Bigg) \Bigg\} = \\ \\ \\
=&&\frac{g^4}{m^2 L^2}\Bigg\{-  \sum_{n=\half}^{\infty}\frac{k_n }{(2k_n +m)^2 (2k_{n+1} +m)}
\sum_{p=\half}^{\infty}\frac{4 k_p}{ (2k_p +m)^2 (2k_{p+1} +m)}\\ \\
&&\hspace{20pt}+\sum_{n=\half}^{\infty}\frac{4k_n}{(2k_n +m)^2 (2k_{n+1} +m)}\sum_{J=1}^{\infty}
\frac{4m}{(2k_J)^3}\\ \\
&&\hspace{20pt}+ \sum_{p,n=\half}^{\infty}\frac{-2k_n \theta(p-n+\half)-2 k_p \theta(n-p-\half) }
{(2k_p  +m) (2k_n +m)^2 (2k_{p+1} +m)}\Bigg( \frac{ 1}{(2k_p  +m) }+\frac{1}{  (2k_{n+1} +m)} \Bigg) \Bigg\} \\
\eea
\lines
which is convergent.\\
We therefore have the following finite  result

\baselineskip18pt
\bea
&&\hspace{-20pt}\frac{ \langle \theta | i(\psi_L^\dag \psi_R - \psi_R^\dag \psi_L )| \theta \rangle}
{ \langle \theta  | \theta \rangle }\simeq \\ \\
&&\hspace{-20pt}\simeq \sin\theta e^{-\frac{\pi}{2mL}}\Bigg\{
-\frac{1}{L}+\frac{g^2}{mL^2}\Bigg[  \sum_{n=\half}^{\infty}\frac{4k_n}{(2k_n +m)^2 (2k_{n+1} +m)}
-\sum_{J=1}^{\infty}\frac{4m}{(2k_J )^3}\Bigg] \\ \\
&&\hspace{-15pt}+\frac{g^4}{m^2 L^3}\Bigg[ - \sum_{n=\half}^{\infty}\frac{k_n }{(2k_n +m)^2 (2k_{n+1} +m)}
\sum_{p=\half}^{\infty}\frac{4 k_p}{ (2k_p +m)^2 (2k_{p+1} +m)}\\ \\
&&\hspace{20pt}+\sum_{n=\half}^{\infty}\frac{4k_n}{(2k_n +m)^2 (2k_{n+1} +m)}\sum_{J=1}^{\infty}
\frac{4m}{(2k_J )^3}\\ \\
&&\hspace{20pt}+\sum_{J=1}^{\infty}
\frac{4mJ}{(2k_J )^3}
\sum_{n=\half}^{\infty}\frac{4k_n}{(2k_n +m)^2 (2k_{n+1} +m)}-\sum_{I,J=1}^{\infty}
\frac{8m^2}{(2k_I )^3 (2k_J)^3}
+\sum_{J=1}^{\infty}\frac{16m^2 J}{ (2k_J)^6}\\ \\
&&\hspace{20pt}+\sum_{p=\half}^{\infty}
\sum_{J=1}^{\infty}
\frac{4mp}{(2k_p+m) (2k_{p} +2k_J+m )}\Bigg(\frac{1}{(2k_{p+1} +m )(2k_{p+1} +2k_J+m )^2 }\\
&&\hspace{20pt}
+\frac{1}{(2k_{p+1} +m )(2k_{p+1} +2k_J+m ) (2k_{p} +2k_J+m )}\\
&&\hspace{20pt}-\frac{2}{(2k_p+m) (2k_{p} +2k_J+m )^2} \Bigg)\\ \\
&&\hspace{20pt}-\sum_{J=1}^{\infty}\sum_{p=\half}^{\infty}\frac{ 2m}
{(2k_{p+1}+m)(2k_p +2k_J +m)(2k_J)^2}\Bigg( \frac{1}{(2k_p +2k_J +m)}+ \frac{1}{k_J}
\Bigg)\\ \\
&&\hspace{20pt}-\sum_{p=\half}^{\infty}\sum_{J=1}^{\infty} \frac{8mk_J}
{(2k_p +2k_J +m) (2k_J)^2 }\Bigg(\frac{1}{k_J (2k_p +2k_J +m)(2k_{p+1} +2k_J +m)}\\
&&\hspace{20pt}\!+\frac{1}{(2k_p +2k_J +m)(2k_{p+1} +2k_J +m)^2}
+\! \frac{2}{(2k_p +2k_J +m)^2(2k_{p+1} +2k_J +m)}\Bigg)\\ \\
&&\hspace{20pt}+\sum_{J=1}^{\infty}\frac{ m}
{(2k_{\half}+m)(2k_{J-\half} +m)(2k_J)^2}\Bigg( \frac{1}{(2k_{J-\half} +m)}+ \frac{1}{k_J}
\Bigg)\\ \\
&&\hspace{20pt}-\pl \sum_{p=\half}^{\infty}\sum_{J=1}^{\infty} \frac{8m\theta(J-p) (J-p)}
{(-2k_p +2k_J +m) (2k_J)^2 }\Bigg(\frac{1}{(-2k_p +2k_J +m)(-2k_{p-1} +2k_J +m)^2}\\
&&\hspace{20pt}+\frac{1}{k_J (-2k_p +2k_J +m)(-2k_{p-1} +2k_J +m)}\\
&&\hspace{20pt}+\! \frac{2}{(-2k_p +2k_J +m)^2(-2k_{p-1} +2k_J +m)}\Bigg)\\ \\
&&\hspace{20pt}-  \sum_{p=\half}^{\infty}\sum_{J=1}^{\infty}\frac{16m}{(2k_J)^3}
 \frac{k_p J -k_J +\theta (J-p) (k_J -k_p )}{(2k_p+m)^2 (2k_{p+1} +m)}\\ \\
&&\hspace{20pt}+\sum_{p=\half}^{\infty}\sum_{J=1}^{\infty}\frac{8m}{(2k_J)^3}
\frac{1 -\theta(J-p-1) }{(2k_p +m) (2k_{p+1} +m)}\\ \\
&&\hspace{20pt}-\sum_{p=\half}^{\infty}\frac{4m}{(2k_{p+\half})^3(2k_p +m) (2k_{p+1} +m)}\\ \
&&\hspace{20pt}+\sum_{p=\half}^{\infty} \frac{k_p}{2(k_{p}+m)^2 (2k_p+m) (2k_{p+1}+m)}
\Bigg( \frac{2}{(2k_p+m)}+\frac{1}{(k_{p+1}+m)}\Bigg)\\ \\
&&\hspace{20pt}+\frac{3\pi^2 m}{2g^2 L}\sum_{p=\half}^{\infty} \frac{p
}{(k_{p}+m) (2k_p+m)(k_{{p+1}}+m) (2k_{p+1}+m)}\\ \\
&&\hspace{20pt}+ \sum_{p,n=\half}^{\infty}\frac{-2k_n \theta(p-n+\half)-2 k_p \theta(n-p-\half) }
{(2k_p  +m) (2k_n +m)^2 (2k_{p+1} +m)}\Bigg( \frac{ 1}{(2k_p  +m) }+\frac{1}{  (2k_{n+1} +m)} \Bigg)  \\ \\
&&\hspace{20pt}+\sum_{n=\half}^{\infty} \frac{n k_n}{(2k_n +m)^3(2k_{n+1} +m)}
\Bigg(\frac{1}{(2k_{n+1} +m) }+
\frac{1}{(2k_n  +m) }
\Bigg) \\ \\
&&\hspace{20pt}-\sum_{n=\half}^{\infty} \!\frac{ n }
{2m (2k_n +m)^2 (2k_{n+1} +m) }\\ \\
&& \hspace{20pt}+\sum_{n, p=\half}^{\infty}
 \!\frac{(k_n-k_p)\theta(n-p-\half)}
{(k_{n-p} +m)^2 (2k_n +m)^2 (2k_{n+1} +m)}\\ \\
&&\hspace{20pt}+\sum_{p=\half}^{\infty}\sum_{n=\half}^{\infty}\frac{\theta(n-p-\half)}
{2(2k_n +m)  (2k_{n+1} +m)(k_{n-p} +m)^2}\\ \\
&&\hspace{20pt}+\sum_{n=\half}^{\infty}\frac{n-\half}
{2(2k_n +m)  (2k_{n+1} +m)(k_{n+\half} +m)^2}
+ \frac{1}{16  (2k_{\half} +m)^2  m^2}\\ \\
&&\hspace{20pt}+ \sum_{n=\half}^{\infty}\sum_{p=\half}^{\infty}
\frac{k_n}
{2 (2k_n +m ) (2k_{n+1} +m)(k_{p +n} +m)}\Bigg(\frac{1}{(k_{p +n} +m) (k_{p +n+1} +m)}\\ \\
&&\hspace{165pt}+\frac{1}{ (k_{p +n+1} +m)^2 }
+ \frac{2}{(k_{p +n} +m) (2k_n +m )}\Bigg)\Bigg] \Bigg\}
\eea\\
\lines
By studying the large-L  behaviour  one can see
that, while several terms go to zero, others  diverge with $L$. The divergent behaviour is expected since it is found also in the expansion of the factor multiplying the exponential $e^{-\frac{\pi}{2mL}}$ in the finite-L condensate for the Schwinger model. In that case, the full nonperturbative result has a finite limit \cite{eliana}.

Setting $g=m\sqrt{\pi}$ we can see that the condensate takes the form of
$m \sin\theta e^{-\frac{\pi}{2mL}}$ multiplied by a function of the product $mL$ that goes
to a pure number, if convergent, in the large-L limit . We therefore find that, as in the case of
the Schwinger model, the condensate is proportional to
the coupling constant. 

Disregarding contributions that vanish as L goes to infinity we get the following estimate for the large-L behaviour of the condensate
\newpage
\begin{eqnarray*}
\frac{ \langle \theta | i(\psi_L^\dag \psi_R - \psi_R^\dag \psi_L )| \theta \rangle}
{ \langle \theta  | \theta \rangle }&\simeq&  m \sin\theta e^{-\frac{\pi}{2mL}}
 \Bigg( \frac{1}{12 \pi}\sum_{J}\frac{1}{J^3} 
 - \frac{7 }{8  \pi}\sum_{J}\frac{1}{J^2}\\ 
&&- \frac{m^3 L^3}{8  \pi^4}\left(\sum_{J}\frac{1}{J^3}\right)^2
+ \frac{m^3 L^3}{4  \pi^4}\sum_{J}\frac{1}{J^5}\Bigg)
\end{eqnarray*}
A standard technique to estimate a series at infinity is that of  Pad\'e approximants\cite{gary}. 
The method of quadratic approximants gives the following function
\[
f(x) = \frac{-1 +\sqrt{1+4 x^3 (a+(a^2 +b)x^3 )}}{2 x ^3}
\]
which has the  power series expansion
\[
f(x) \simeq a + b x^3 \;.
\] 
For our case
 \bea
&&a= \frac{1}{12 \pi}\sum_{J}\frac{1}{J^3} 
 - \frac{7 }{8  \pi}\sum_{J}\frac{1}{J^2}
\simeq -0.426 \\
&& b=- \frac{1}{8  \pi^4}\left(\sum_{J}\frac{1}{J^3}\right)^2
+ \frac{1}{4  \pi^4}\sum_{J}\frac{1}{J^5}\simeq 0.000807\\
&& x=mL\,.
\eea
For $0< x < \infty$  $f$ is between $-0.426$ and $0.427$. It is therefore likely that the number 
multiplying $m\sin\theta$ in the condensate is within this range. But nothing definite can be said about the accuracy of this result. Since we can only form one approximant, we cannot test for convergence, even empirically, and there is no mathematical theorem giving a bound on the error that is made with this approximation. A similar procedure for the Schwinger model gives a correct estimate of the order of magnitude, with an asymptotic value of about 0.45, while the correct value is
about 0.28. In all likelihood our number is of order 1. 


\section{\textbf{The real field contribution}}
It is easy to see that
\bea
\langle \Omega_M^{(0)}| \phi_L (0,\x) \phi_R (0,\x) | \Omega^{(0)}_N\rangle
= \delta_{M,N}\langle 0|{\stackrel {\;o}{\phi}}_L {\stackrel {\;o}{\phi}}_R  \0
\eea
so that
\be
\langle \theta^{(0)}|  {\stackrel {\;o}{\phi}}_L  {\stackrel {\;o}{\phi}}_R | \theta^{(0)}\rangle= \sum_{N=-\infty}^\infty
\langle 0|{\stackrel {\;o}{\phi}}_L {\stackrel {\;o}{\phi}}_R  \0\;.
\ee
Note that the non-vanishing of  $\langle 0|{\stackrel {\;o}{\phi}}_L {\stackrel {\;o}{\phi}}_R  \0$
breaks chiral invariance in free theory. This is an effect of the boundary conditions that
vanishes in the continuum limit.
We also have

\baselineskip15pt
\bea
\langle \Omega_M^{(1)}|\,  \phi_L (0,\x) \phi_R (0,\x) | \Omega_N^{(1)}\rangle &=&
 \frac{g^2}{mL}\delta_{MN}\, \Bigg\{ -2m \sum_{J=1}^{\infty}\frac{\langle 0 | C_J^3  \phi_L (0,\x) \phi_R (0,\x) (C_J^3)^\dag \0}{(2k_J )^3} \\ \\
&&\hspace{50pt}+\sum_{n=\half}^{\infty}\frac{\langle 0 |C_{n}^+
 \phi_L (0,\x) \phi_R (0,\x)   (C_{n}^+)^\dag \0}{(2k_n+m)^2} \\ \\
&&\hspace{50pt}+\sum_{n=\half}^{\infty}\frac{\langle 0 |C_{n}^-
 \phi_L (0,\x) \phi_R (0,\x)   (C_{n}^-)^\dag \0}{(2k_n+m)^2}
\Bigg\}
\eea
\lines

From (\ref{phiR}) , (\ref{zeromode}) and (\ref{Cn+}) we get
\bea
[C_n^+ , \phi_R (0,\x)] &\!\!=&\!\!\! - {1 \over \sqrt {2L}} \sum_{N=1}^\infty
\left[ \left(b_{N-n}^\dag \theta(N-n)+ d_{n-N}\theta(n-N)
\right)e^{-ik_N \x} \right]\\
&&-\frac{1}{ \sqrt {2L}}\sum_{N=0}^\infty d_{n+N}e^{ik _N \x} \\
&=&-e^{-ik_n \x}   \psi_R ^\dag (0,\x)
 \eea
and $[C_{n}^+ , \phi_L (0,\x)] =0 $ , so that
\bea
\langle 0 |C_{n}^+ \phi_L (0,\x) \phi_R (0,\x)   (C_{n}^+)^\dag \0&=& \langle 0 |\phi_L (0,\x) \phi_R (0,\x)  C_{n}^+  (C_{n}^+)^\dag \0\\
&&- e^{-ik_n \x} \langle 0 | {\stackrel {\;o}{\phi}}_L \psi_R ^\dag (0,\x)  (C_{n}^+)^\dag \0
\eea
It is easy to see that
\be\label{phizero+}
e^{-ik_n \x} \langle 0 | {\stackrel {\;o}{\phi}}_L \psi_R ^\dag (0,\x)  (C_{n}^+)^\dag \0
=e^{-ik_n \x} \langle 0 |{\stackrel {\;o}{\phi}}_L  \psi_R ^\dag (0,\x) d_n^\dag (r_0 +r_0^\dag )\0=
\langle 0|{\stackrel {\;o}{\phi}}_L  {\stackrel {\;o}{\phi}}_R  \0
\ee
and
\bea
\langle 0 |C_{n}^+ \phi_L (0,\x) \phi_R (0,\x)   (C_{n}^+)^\dag \0 =
(n-1) \langle 0|{\stackrel {\;o}{\phi}}_L  {\stackrel {\;o}{\phi}}_R  \0 \;.
\eea
From (\ref{Cn-}) we get
\bea
[C_{n}^- ,  \phi _R(0,\x)] &=& {1 \over \sqrt {2L}} \sum_{N=1}^\infty
 \left(d_{N-n}^\dag \theta(N-n) + b_{n-N}\theta(n-N)\right)
 e^{-ik_N \x} \\
&&+ {1 \over \sqrt {2L}} \sum_{N=0}^\infty  b_{n+N} e^{ik _N \x}\\
&=&  e^{-ik _n \x}  \psi_R  (0,\x)
\eea
and
\bea
\langle 0 |C_{n}^-  \phi_L (0,\x) \phi_R (0,\x)   (C_{n}^-)^\dag \0 = n\langle 0|{\stackrel {\;o}{\phi}}_L  {\stackrel {\;o}{\phi}}_R  \0
+e^{-ik _n \x}\langle 0|{\stackrel {\;o}{\phi}}_L  \psi_R  (0,\x) (C_{n}^-)^\dag \0 \;.
\eea
We also have
\be\label{phizero-}
e^{-ik _n \x}\langle 0| {\stackrel {\;o}{\phi}}_L  \psi_R  (0,\x) (C_{n}^-)^\dag \0
=\nor\langle 0| {\stackrel {\;o}{\phi}}_L  b_n (r_0 + r_0^\dag ) b_n^\dag \0 = -\langle 0|{\stackrel {\;o}{\phi}}_L  {\stackrel {\;o}{\phi}}_R  \0\\
\ee
so that
\bea
&&\langle 0 |C_{n}^- {\stackrel {\;o}{\phi}}_L  \phi_R (0,\x)   (C_{n}^-)^\dag \0 =(n-1) \langle 0|{\stackrel {\;o}{\phi}}_L  {\stackrel {\;o}{\phi}}_R  \0 \;.
\eea
Since $[C_N^3, \phi_{R/L}(0,\x)] =0 $ we can finally write
\bea
&&\langle \theta^{(1)}|  \phi_L (0,\x) \phi_R (0,\x)  | \theta^{(1)}\rangle =\\
&=& \langle 0|{\stackrel {\;o}{\phi}}_L  {\stackrel {\;o}{\phi}}_R \0
\sum_{N=-\infty}^\infty
 \frac{g^2}{mL}\Bigg\{  \sum_{J=1}^{\infty}\frac{-4mJ}{(2k_J)^3 }+2\sum_{n=\half}^{\infty}\frac{n-1}{(2k_n+m)^2}
\Bigg\}
\eea
\lines

$\langle\Omega_M^{(2)}|    \phi_L (0,\x) \phi_R (0,\x)  | \Omega^{(2)}_N \rangle_{3+}$ is given by\\
$\langle\Omega_M^{(2)}|    \phi_L (0,\x) \phi_R (0,\x)  | \Omega^{(2)}_N \rangle_{3+}
=\delta_{MN} \langle\Omega^{(2)} |   \phi_L (0,\x) \phi_R (0,\x)   | \Omega^{(2)} \rangle_{3+}$
and

\baselineskip18pt
\bea
&&\hspace{-20pt}\langle\Omega^{(2)} |  \phi_L (0,\x) \phi_R (0,\x)  | \Omega^{(2)}\rangle_{3+}=\\ \\
&&=\frac{g^4}{m^2 L^2}
\sum_{p=\half}^{\infty}\Bigg\{\,
-4m \sum_{J=1}^{\infty}\frac{\langle 0|
 C^+_{p }C_J^3   \phi_L (0,\x) \phi_R (0,\x)  (C_J^3)^\dag (C^+_{p})^\dag
 \0 }{(2k_p+m)^2 (2k_p +2k_J + m )^3}\\
%
&&\hspace{20pt}-2m\sum_{J=1}^{\infty}\frac{\langle 0 |
 C^+_{p }C_J^3   \phi_L (0,\x) \phi_R (0,\x) (C^+_{{p} })^\dag (C_J^3)^\dag
\0 }
{ (2k_p+m) }\Bigg(\frac{2}{2k_J (2k_p +2k_J+m )^3 }\\
&&\hspace{245pt}+ \frac{1}{(2k_J)^2 (2k_p +2k_J +m)^2}
\Bigg)\\ \\
%
&&\hspace{20pt}-2m \sum_{J=1}^{\infty}\frac{\langle 0 |
C_J^3  C^+_{p }  \phi_L (0,\x) \phi_R (0,\x)  (C_J^3)^\dag (C^+_{{p} })^\dag \0 }
{(2k_{p}+m)}\Bigg(\frac{2}{2k_J (2k_p +2k_J+m )^3 }\\
&&\hspace{245pt}+ \frac{1}{(2k_J)^2 (2k_p +2k_J +m)^2}
\Bigg)\\ \\
%
&&\hspace{20pt}-4m\sum_{J=1}^{\infty}\langle 0 |
C_J^3  C^+_{p }  \phi_L (0,\x) \phi_R (0,\x) (C^+_{{p} })^\dag (C_J^3 )^\dag
\0 \Bigg( \frac{1}
{(2k_p +2k_J+m )^2 (2k_J )^3 }\\
&&\hspace{250pt}+\frac{1}{(2k_p +2k_J +m)^3 (2k_J)^2 }
\Bigg)\\ \\
%
&&\hspace{20pt}-4m\sum_{J=1}^{\infty}\langle 0 |
C_J^3  (C^-_{p })^\dag \phi_L (0,\x) \phi_R (0,\x) C^-_{p } (C_J^3 )^\dag
\0 \Bigg(\!\frac{1}
{(-2k_p +2k_J +m )^2 (2k_J)^3 }\\
&&\hspace{250pt}+\frac{1}{(-2k_p +2k_J +m)^3 (2k_J )^2 }
\Bigg)\\ \\
%
&& \hspace{20pt}+  \frac{\langle 0 |
C_{p}^+  \phi_L (0,\x) \phi_R (0,\x)
 (C_{{p}}^+)^\dag | \0 \rangle}{4(k_{p}+m)^2 (2k_p+m)^2}\; \Bigg\}
\\
\eea
\lines

We have
\bea
 \langle 0|C^+_{p }C_J^3    \phi_L \phi_R (C_J^3)^\dag (C^+_{p})^\dag \0 &=&
 \langle 0|   \phi_L \phi_R C^+_{p }C_J^3 (C_J^3)^\dag (C^+_{p})^\dag \0 \\
&&-e^{-ik_p \x} \langle 0|   \phi_L (\psi_R)^\dag C_J^3 (C_J^3)^\dag (C^+_{p})^\dag \0
\eea
The state $C^+_{p }C_J^3 (C_J^3)^\dag (C^+_{p})^\dag \0$ is an eigenstate of $H_0$
with eigenvalue zero. Therefore we must have
\bea
  \langle 0|   \phi_L \phi_R C^+_{p }C_J^3 (C_J^3)^\dag (C^+_{p})^\dag \0
= \langle 0| { \stackrel {\;o}{\phi}}_L  {\stackrel {\;o}{\phi}}_R C^+_{p }C_J^3 (C_J^3)^\dag (C^+_{p})^\dag \0 \;.
\eea
Since $C^+_{p }$ and $C_J^3 $ are bilinear in the fermion operators the term
 $(r_0 + r_0 ^\dag )d_p$
in  $C^+_{p }$ contributes only together with its adjoint in $(C^+_{p })^\dag$, giving a factor
 $(r_0 + r_0 ^\dag )^2 = 2  ({\stackrel {\;o}{\phi}}_R )^2 = \frac{1}{2L}$ .  Therefore ${\stackrel {\;o}{\phi}}_R $ is not present in $C^+_{p }C_J^3 (C_J^3)^\dag (C^+_{p})^\dag \0$ and we can conclude that
\bea
 \langle 0|   \phi_L \phi_R C^+_{p }C_J^3 (C_J^3)^\dag (C^+_{p})^\dag \0 =
 \langle 0| { \stackrel {\;o}{\phi}}_L  {\stackrel {\;o}{\phi}}_R \0   \langle 0| C^+_{p }C_J^3 (C_J^3)^\dag (C^+_{p})^\dag \0 \;.
\eea
Applying the same argument to all the terms in $\langle\Omega^{(2)} |  \phi_L \phi_R  | \Omega^{(2)}\rangle_{3+}$ we can write

\baselineskip18pt
\bea
&&\hspace{-20pt}\langle\Omega^{(2)} |  \phi_L (0,\x) \phi_R (0,\x)  | \Omega^{(2)}\rangle_{3+}=
\langle\Omega^{(2)}  | \Omega^{(2)}\rangle_{3+} \langle 0| { \stackrel {\;o}{\phi}}_L  {\stackrel {\;o}{\phi}}_R \0  \\ \\
&&+ \frac{g^4}{m^2 L^2}
\sum_{p=\half}^{\infty}\Bigg\{\,
-4m \sum_{J=1}^{\infty}\frac{-e^{-ik_p \x} \langle 0|   \phi_L (\psi_R)^\dag C_J^3 (C_J^3)^\dag (C^+_{p})^\dag \0  }{(2k_p+m)^2 (2k_p +2k_J + m )^3}\\ \\
%
&&\hspace{20pt}-2m\sum_{J=1}^{\infty}\frac{-e^{-ik_p \x} \langle 0|   \phi_L (\psi_R)^\dag C_J^3  (C^+_{p})^\dag (C_J^3)^\dag \0 }
{ (2k_p+m) }\Bigg(\frac{2}{2k_J (2k_p +2k_J+m )^3 }\\
&&\hspace{245pt}+ \frac{1}{(2k_J)^2 (2k_p +2k_J +m)^2}
\Bigg)\\ \\
%
&&\hspace{20pt}-2m \sum_{J=1}^{\infty}\frac{-e^{-ik_p \x} \langle 0|   \phi_L C_J^3 (\psi_R)^\dag  (C_J^3)^\dag (C^+_{p})^\dag \0   }
{(2k_{p}+m)}\Bigg(\frac{2}{2k_J (2k_p +2k_J+m )^3 }\\
&&\hspace{245pt}+ \frac{1}{(2k_J)^2 (2k_p +2k_J +m)^2}
\Bigg)\\ \\
%
&&\hspace{20pt}+4m\sum_{J=1}^{\infty}e^{-ik_p \x} \langle 0|   \phi_L C_J^3 (\psi_R)^\dag (C^+_{p})^\dag (C_J^3)^\dag  \0
 \Bigg( \frac{1}
{(2k_p +2k_J+m )^2 (2k_J )^3 }\\
&&\hspace{250pt}+\frac{1}{(2k_p +2k_J +m)^3 (2k_J)^2 }
\Bigg)\\ \\
%
&&\hspace{20pt}+4m\sum_{J=1}^{\infty}e^{-ik_p \x} \langle 0|   \phi_L C_J^3 (\psi_R)^\dag C^-_{p} (C_J^3)^\dag  \0
\Bigg(\!\frac{1}
{(-2k_p +2k_J +m )^2 (2k_J)^3 }\\
&&\hspace{250pt}+\frac{1}{(-2k_p +2k_J +m)^3 (2k_J )^2 }
\Bigg)\\ \\
%
&& \hspace{20pt}- \frac{e^{-ik_p \x} \langle 0|   \phi_L  (\psi_R)^\dag (C^+_{p})^\dag \0 }{4(k_{p}+m)^2 (2k_p+m)^2}\; \Bigg\}
\\
\eea
\lines

We have
\bea
&&\hspace{-20pt}-e^{-ik_p \x} \langle 0|   \phi_L (\psi_R)^\dag C_J^3 (C_J^3)^\dag (C^+_{p})^\dag \0
= \\
&&=-e^{-ik_p \x}\bigg(  \langle 0|   \phi_L (\psi_R)^\dag (C_J^3)^\dag C_J^3  (C^+_{p})^\dag \0
+J  \langle 0|   \phi_L (\psi_R)^\dag  (C^+_{p})^\dag \0 \bigg)\\
&&=e^{-ik_p \x} \langle 0|   \phi_L \left((C_J^3)^\dag -e^{ik_J \x}\right )  (\psi_R)^\dag  (C^+_{p-J})^\dag \0
-J \langle 0| { \stackrel {\;o}{\phi}}_L  {\stackrel {\;o}{\phi}}_R \0 \\
&&= -e^{-ik_{p-J} \x} \langle 0|   \phi_L  (\psi_R)^\dag  (C^+_{p-J})^\dag \0  -J =(-\theta(p-J) -J) \langle 0| { \stackrel {\;o}{\phi}}_L  {\stackrel {\;o}{\phi}}_R \0
\eea
\bea
&&\hspace{-20pt}-e^{-ik_p \x} \langle 0|   \phi_L (\psi_R)^\dag C_J^3 (C^+_{p})^\dag (C_J^3)^\dag  \0=\\
&&= -e^{-ik_p \x} \langle 0|   \phi_L (\psi_R)^\dag C_J^3 (C_J^3)^\dag (C^+_{p})^\dag \0
-e^{-ik_p \x} \langle 0|   \phi_L (\psi_R)^\dag C_J^3 (C^+_{p+J})^\dag \0 \\
&&=(-\theta(p-J) -J+1) \langle 0| { \stackrel {\;o}{\phi}}_L  {\stackrel {\;o}{\phi}}_R \0
\eea
\bea
&&\hspace{-20pt}-e^{-ik_p \x} \langle 0|   \phi_L C_J^3 (\psi_R)^\dag (C_J^3)^\dag (C^+_{p})^\dag \0 \\
&&=-e^{-ik_p \x} \langle 0|   \phi_L  (\psi_R)^\dag C_J^3 (C_J^3)^\dag (C^+_{p})^\dag \0
-e^{-ik_{p+J} \x} \langle 0|   \phi_L  (\psi_R)^\dag  (C_J^3)^\dag (C^+_{p})^\dag \0\\
&&=(-\theta(p-J) -J)  \langle 0| { \stackrel {\;o}{\phi}}_L  {\stackrel {\;o}{\phi}}_R \0 -e^{-ik_{p+J} \x} \langle 0|   \phi_L  \left((C_J^3)^\dag- e^{ik_J \x}\right)   (\psi_R)^\dag (C^+_{p})^\dag \0\\
&&=(-\theta(p-J) -J+1) \langle 0| { \stackrel {\;o}{\phi}}_L  {\stackrel {\;o}{\phi}}_R \0
\eea
\bea
&&\hspace{-20pt}-e^{-ik_p \x} \langle 0|   \phi_L C_J^3 (\psi_R)^\dag (C^+_{p})^\dag  (C_J^3)^\dag \0 =\\
&&-e^{-ik_p \x} \langle 0|   \phi_L C_J^3 (\psi_R)^\dag (C_J^3)^\dag (C^+_{p})^\dag \0
-e^{-ik_p \x} \langle 0|   \phi_L C_J^3 (\psi_R)^\dag (C^+_{p+J})^\dag \0 \\
&&=(-\theta(p-J) -J+1) \langle 0| { \stackrel {\;o}{\phi}}_L  {\stackrel {\;o}{\phi}}_R \0  -e^{-ik_p \x} \langle 0|   \phi_L (\psi_R)^\dag ( C_J^3 +e^{-ik_J \x})(C^+_{p+J})^\dag \0\\
&&=(-\theta(p-J) -J+1) \langle 0| { \stackrel {\;o}{\phi}}_L  {\stackrel {\;o}{\phi}}_R \0
\eea
\bea
&&\hspace{-20pt}- e^{ik_p \x}\langle 0 |
  \phi_L C_J^3  (\psi_R)^\dag C^-_{p } (C_J^3)^\dag\0 =\\
&&= - e^{ik_p \x}\langle 0 | \phi_L (\psi_R)^\dag (C_J^3 +e^{-ik_J \x})C^-_{p-J}\0
=- e^{-ik_{J-p} \x}\langle 0 | \phi_L (\psi_R)^\dag (C^+_{J-p})^\dag \0\\
&&=-\theta(J-p) \langle 0| { \stackrel {\;o}{\phi}}_L  {\stackrel {\;o}{\phi}}_R \0
\eea
where (\ref{phizero+}) and (\ref{phizero-}) have been used.\\
\vspace{-12pt}
We can now write

\baselineskip18pt
\bea
&&\hspace{-20pt}\langle\Omega^{(2)} |  \phi_L (0,\x) \phi_R (0,\x)  | \Omega^{(2)}\rangle_{3+}=
\langle\Omega^{(2)}  | \Omega^{(2)}\rangle_{3+} \langle 0| { \stackrel {\;o}{\phi}}_L  {\stackrel {\;o}{\phi}}_R \0  \\ \\
&&+ \frac{g^4}{m^2 L^2}
\sum_{p=\half}^{\infty}\Bigg\{\,
-4m \sum_{J=1}^{\infty}\frac{-\theta(p-J) -J }{(2k_p+m)^2 (2k_p +2k_J + m )^3}\\ \\
%
&&\hspace{20pt}-4m\sum_{J=1}^{\infty}\frac{-\theta(p-J) -J+1 }
{2k_J  (2k_p+m)(2k_p +2k_J +m)^2 }\Bigg(\frac{2}{2k_p +2k_J+m  }+ \frac{1}{2k_J }
\Bigg)\\ \\
%
%
&&\hspace{20pt}-4m\sum_{J=1}^{\infty}\frac{-\theta(p-J) -J+1}{ (2k_J)^2 (2k_p +2k_J+m )^2 }
 \Bigg( \frac{1}
{2k_J }+\frac{1}{2k_p +2k_J +m }
\Bigg)\\ \\
%
&&\hspace{20pt}+4m\sum_{J=1}^{\infty}\frac{\theta(J-p)}{(2k_J )^2 (-2k_p +2k_J +m )^2 }
\Bigg(\!\frac{1}
{2k_J }+\frac{1}{-2k_p +2k_J +m  }
\Bigg)\\ \\
%
&& \hspace{20pt}- \frac{1}{4(k_{p}+m)^2 (2k_p+m)^2}\; \Bigg\}
\\ \\
&&\hspace{-20pt}=
\langle\Omega^{(2)}  | \Omega^{(2)}\rangle_{3+} \langle 0| { \stackrel {\;o}{\phi}}_L  {\stackrel {\;o}{\phi}}_R \0  \\ \\
&&+ \frac{g^4}{m^2 L^2} \langle 0| { \stackrel {\;o}{\phi}}_L  {\stackrel {\;o}{\phi}}_R \0
\sum_{p=\half}^{\infty}\Bigg\{ -4m  \sum_{J=1}^{\infty}\frac{-\theta(p-J) -J+1}{(2k_J)^3 (2k_p+m)^2}
\\ \\
&&\hspace{20pt}+\sum_{J=1}^{\infty}
4m \sum_{J=1}^{\infty}\frac{1 }{(2k_p+m)^2 (2k_p +2k_J + m )^3}\\ \\
&&\hspace{20pt}+4m\sum_{J=1}^{\infty}\frac{\theta(J-p)}{(2k_J )^2 (-2k_p +2k_J +m )^2 }
\Bigg(\!\frac{1}
{2k_J }+\frac{1}{-2k_p +2k_J +m  }
\Bigg)\\ \\
&& \hspace{20pt}- \frac{1}{4(k_{p}+m)^2 (2k_p+m)^2}\; \Bigg\}\\
\eea
Let us consider now $\langle\Omega_M^{(2)}| \phi_L (0,\x) \phi_R (0,\x) |\Omega_N^{(2)}\rangle_{\pm \pm}$
\lines

\bea
\langle\Omega_M^{(2)}| \phi_L (0,\x) \phi_R (0,\x) |\Omega_N^{(2)}\rangle_{++}=\frac{g^4\delta_{MN }}{m^2 L^2}\!\sum_{p,n=\half}^{\infty}\! \frac{\langle 0|\,C_{n}^+ C_{p }^+
 \phi_L (0,\x) \phi_R (0,\x)
(C_{p }^+)^\dag (C_{n}^+)^\dag \0}
{2(2k_p  +m)^2 (2k_n +m)^2}
\eea
\bea
&&\langle 0|\,C_{n}^+ C_{p }^+ { \stackrel {\;o}{\phi}}_L \phi_R (0,\x)
(C_{p }^+)^\dag (C_{n}^+)^\dag \0 =\\
&&\langle 0|\,C_{n}^+   { \stackrel {\;o}{\phi}}_L  \phi_R (0,\x) C_{p }^+
(C_{p }^+)^\dag (C_{n}^+)^\dag \0
-e^{-ik_p \x}\langle 0|\,C_{n}^+{ \stackrel {\;o}{\phi}}_L  \psi_R^\dag (0,\x)
(C_{p }^+)^\dag (C_{n}^+)^\dag \0\\
&&=\langle 0|\, { \stackrel {\;o}{\phi}}_L  \phi_R (0,\x)C_{n}^+   C_{p }^+
(C_{p }^+)^\dag (C_{n}^+)^\dag \0 -e^{-ik_n \x}\langle 0|\,{ \stackrel {\;o}{\phi}}_L  \psi_R^\dag (0,\x)
C_{p }^+ (C_{p }^+)^\dag (C_{n}^+)^\dag \0\\
&& -e^{-ik_p \x}\langle 0|\,{ \stackrel {\;o}{\phi}}_L  \psi_R^\dag (0,\x)
C_{n }^+ (C_{p }^+)^\dag  (C_{n}^+)^\dag \0\\
\eea
\baselineskip15pt
\bea
&&\langle\Omega_M^{(2)}| \phi_L (0,\x) \phi_R (0,\x) |\Omega_N^{(2)}\rangle_{++}=
\langle 0| { \stackrel {\;o}{\phi}}_L  {\stackrel {\;o}{\phi}}_R \0
\langle\Omega_M^{(2)} |\Omega_N^{(2)}\rangle_{++}\\ \\
&&-\frac{g^4\delta_{MN }}{m^2 L^2}\!\sum_{p,n=\half}^{\infty}\! \frac{e^{-ik_n \x}\langle 0|\,{ \stackrel {\;o}{\phi}}_L  \psi_R^\dag (0,\x)
C_{p }^+ (C_{p }^+)^\dag (C_{n}^+)^\dag \0}
{(2k_p  +m)^2 (2k_n +m)^2}
\eea
\lines
\bea
&&\hspace{-20pt}e^{-ik_n \x}\langle 0|\,{ \stackrel {\;o}{\phi}}_L  \psi_R^\dag (0,\x)
C_{p }^+ (C_{p }^+)^\dag (C_{n}^+)^\dag \0= \\
&&\hspace{-10pt}= e^{-ik_n \x}\langle 0|\,{ \stackrel {\;o}{\phi}}_L  \psi_R^\dag (0,\x)
 (C_{p }^+)^\dag C_{p }^+(C_{n}^+)^\dag \0
+(p-1)e^{-ik_n \x}\langle 0|\,{ \stackrel {\;o}{\phi}}_L  \psi_R^\dag(C_{n}^+)^\dag \0 \\
&&\hspace{-10pt}= e^{-ik_n \x}\langle 0|\,{ \stackrel {\;o}{\phi}}_L  \psi_R^\dag (0,\x)
 (C_{p }^+)^\dag \left((C^3_{n-p})^\dag +p\delta_{n,p} \right) \0 + (p-1) \langle 0| { \stackrel {\;o}{\phi}}_L  {\stackrel {\;o}{\phi}}_R \0 \\
&&\hspace{-10pt}=(p\delta_{n,p} +p-1) \langle 0| { \stackrel {\;o}{\phi}}_L  {\stackrel {\;o}{\phi}}_R \0 \\
&&\hspace{-10pt}\quad + e^{-ik_n \x}\theta\Big(n-p-\half\Big) \left[ \langle 0|\,{ \stackrel {\;o}{\phi}}_L
\left((C^3_{n-p})^\dag -e^{ik_{n-p} \x}\right)  \psi_R^\dag (C_{p }^+)^\dag \0 +
\langle 0|\,{ \stackrel {\;o}{\phi}}_L  \psi_R^\dag(C_{n}^+)^\dag \0 \right]\\
&&\hspace{-10pt}= \left( p\delta_{n,p} +p-1\right) \langle 0| { \stackrel {\;o}{\phi}}_L  {\stackrel {\;o}{\phi}}_R \0
\eea
so we have
\bea
&&\langle\Omega_M^{(2)}| \phi_L (0,\x) \phi_R (0,\x) |\Omega_N^{(2)}\rangle_{\pm\pm}=\\
&&\langle 0| { \stackrel {\;o}{\phi}}_L  {\stackrel {\;o}{\phi}}_R \0
\langle\Omega_M^{(2)} |\Omega_N^{(2)}\rangle_{\pm\pm} -\frac{g^4\delta_{MN }}{m^2 L^2}\!\sum_{p,n=\half}^{\infty}\! \frac{p-1 +p\delta_{n,p}}
{(2k_p  +m)^2 (2k_n +m)^2}\\ \\
\eea
$\langle\Omega_M^{(2)}| \phi_L (0,\x) \phi_R (0,\x)   | \Omega_N^{(2)}\rangle_{+\, -}$ is given by

\baselineskip15pt
\bea
\hspace{-20pt}\langle\Omega_M^{(2)}  |  \phi_L (0,\x) \phi_R (0,\x)| \Omega_N^{(2)}\rangle_{+\, -}
=\frac{g^4}{m^2 L^2}\! \sum_{n, p=\half}^{\infty}\!\Bigg\{\frac{\langle 0|\,C_{n}^- C_{p}^+
 \phi_L (0,\x) \phi_R (0,\x)
(C_{p}^+)^\dag (C_{n}^-)^\dag | \0}
{(2k_{p} +m)^2 (2k_n +m)^2 }\\ \\
+\frac{
\langle 0 |\,C^3_{p+n} \phi_L (0,\x) \phi_R (0,\x)    (C_{p}^+)^\dag (C_{n}^-)^\dag \0}
{4(k_{p +n} +m)^2 (2k_n +m)(2k_p+m)}
+\frac{\langle 0 |\,C_{n}^- C_{p}^+ \phi_L (0,\x) \phi_R (0,\x)   (C^3_{p+n})^\dag \0}
{4(k_{p +n} +m)^2 (2k_n +m)  (2k_p +m)}\\ \\
+\frac{\langle 0 |\,C_{n}^-  C_{p}^+  \phi_L (0,\x) \phi_R (0,\x)   (C^3_{p+n})^\dag \0  }{4(k_{p +n} +m)^2 (2k_p +m)^2  }
+\frac{\langle 0 |\,C^3_{p+n} \phi_L (0,\x) \phi_R (0,\x)  (C_{p}^+)^\dag  (C_{n}^-)^\dag \0 }
{4(k_{p +n} +m)^2 (2k_p +m)^2  }\\ \\
+\frac{\langle 0 |\,C^3_{p+n} \phi_L (0,\x) \phi_R (0,\x)   (C^3_{p+n})^\dag \0}
{4(k_{p +n} +m)^2 (2k_p +m)^2  }
+ \!\frac{\langle 0|\,C_{n}^- (C_{p}^-)^\dag \phi_L (0,\x) \phi_R (0,\x)   C_{p}^-  (C_{n}^-)^\dag | \0}
{4(k_{n-p} +m)^2 (2k_n +m)^2 }\\ \\
+\frac{\langle 0 |\,C_{p}^+ (C_{n}^+)^\dag  \phi_L (0,\x) \phi_R (0,\x)   C_{n}^+ (C_{p}^+)^\dag \0}
{4(k_{p-n} +m)^2 (2k_p +m)^2  }\Bigg\}
\eea
\lines

\bea
&&\langle 0|\,C_{n}^- C_{p}^+
 \phi_L (0,\x) \phi_R (0,\x)
(C_{p}^+)^\dag (C_{n}^-)^\dag  \0 =
-e^{-ik_p \x}\langle 0 |\,C_{n}^-  { \stackrel {\;o}{\phi}}_L \psi_R^\dag (C_{p}^+)^\dag (C_{n}^-)^\dag  \0 \\
&& \hspace{15pt}+e^{-ik_n \x}\langle 0 |\,{ \stackrel {\;o}{\phi}}_L \psi_R C_{p}^+ (C_{p}^+)^\dag (C_{n}^-)^\dag  \0
+\langle 0 |\,{ \stackrel {\;o}{\phi}}_L { \stackrel {\;o}{\phi}}_R C_{n}^- C_{p}^+ (C_{p}^+)^\dag (C_{n}^-)^\dag  \0 =\\
&&=-e^{-ik_p \x}\langle 0 |\,C_{n}^-  { \stackrel {\;o}{\phi}}_L \psi_R^\dag  (C_{n}^-)^\dag (C_{p}^+)^\dag \0
+e^{-ik_p \x}\langle 0 |\,C_{n}^-  { \stackrel {\;o}{\phi}}_L \psi_R^\dag
C^3_{n+p})^\dag \0 \\
&& \hspace{15pt}+e^{ik_n \x}\langle 0 |\,{ \stackrel {\;o}{\phi}}_L \psi_R (C_0^3 +p) (C_{n}^-)^\dag \0
+\langle 0 |\,{ \stackrel {\;o}{\phi}}_L { \stackrel {\;o}{\phi}}_R C_{n}^- C_{p}^+ (C_{p}^+)^\dag (C_{n}^-)^\dag  \0
\eea
and
\bea
&&e^{-ik_p \x}\langle 0 |\,C_{n}^-  { \stackrel {\;o}{\phi}}_L \psi_R^\dag (C^3_{n+p})^\dag \0 =
e^{-ik_p \x}\langle 0 |\,C_{n}^- \left( (C^3_{n+p})^\dag -e^{ik_{p+n} \x}\right)  { \stackrel {\;o}{\phi}}_L \psi_R^\dag \0=\\
&&=e^{-ik_p \x} \langle 0 |\,  (C_{p}^+)^\dag  { \stackrel {\;o}{\phi}}_L \psi_R^\dag \0
+\Big(e^{-ik_n \x}\langle 0 |\,{\stackrel {\;o}{\phi}}_L \psi_R (C_{n}^-)^\dag \0 \Big)^\ast =
 \langle 0 |\,{ \stackrel {\;o}{\phi}}_L { \stackrel {\;o}{\phi}}_R \0
\eea
so that
\bea
\langle 0|\,C_{n}^- C_{p}^+
 \phi_L (0,\x) \phi_R (0,\x)
(C_{p}^+)^\dag (C_{n}^-)^\dag  \0 &\!\!=&\!\!
-(n+p) \langle 0 |\,{ \stackrel {\;o}{\phi}}_L { \stackrel {\;o}{\phi}}_R \0\\
&&\!\! +\langle 0 |\,{ \stackrel {\;o}{\phi}}_L { \stackrel {\;o}{\phi}}_R C_{n}^- C_{p}^+ (C_{p}^+)^\dag (C_{n}^-)^\dag  \0 \;.
\eea
We then have
\bea
&&\langle 0 |\,C_{n}^- C_{p}^+   \phi_L (0,\x) \phi_R (0,\x)     (C^3_{p+n})^\dag \0 =
\langle 0 |\,{ \stackrel {\;o}{\phi}}_L   { \stackrel {\;o}{\phi}}_R C_{n}^- C_{p}^+   (C^3_{p+n})^\dag \0 \\
&& \hspace{15pt}+e^{-ik_n \x}\langle 0 |\,{\stackrel {\;o}{\phi}}_L  \psi_R  C_{p}^+   (C^3_{p+n})^\dag \0
-e^{-ik_p \x}\langle 0 |\,C_{n}^- { \stackrel {\;o}{\phi}}_L  \psi_R^\dag  (C^3_{p+n})^\dag \0=\\
&&=\langle 0 |\,{ \stackrel {\;o}{\phi}}_L   { \stackrel {\;o}{\phi}}_R C_{n}^- C_{p}^+   (C^3_{p+n})^\dag \0 - e^{-ik_n \x}\langle 0 |\,{\stackrel {\;o}{\phi}}_L  \psi_R (C_{n}^-)^\dag \0 -
 \langle 0 |\,{ \stackrel {\;o}{\phi}}_L { \stackrel {\;o}{\phi}}_R \0 \\
&&=\langle 0 |\,{ \stackrel {\;o}{\phi}}_L   { \stackrel {\;o}{\phi}}_R C_{n}^- C_{p}^+   (C^3_{p+n})^\dag \0 =\langle 0 |\,{ \stackrel {\;o}{\phi}}_L   { \stackrel {\;o}{\phi}}_R \0 \langle 0 |\,
C_{n}^- C_{p}^+   (C^3_{p+n})^\dag \0
\eea
Taking the adjoint we get
\bea
\langle 0 |\,C^3_{p+n}\phi_L (0,\x) \phi_R (0,\x)  (C_{p}^-)^\dag \0 (C_{n}^-)^\dag \0 =
\langle 0 |\,{ \stackrel {\;o}{\phi}}_L   { \stackrel {\;o}{\phi}}_R \0 \langle 0 |\,
\langle 0 |\,C^3_{p+n}(C_{p}^+)^\dag \0 (C_{n}^-)^\dag \0 \;.
\eea
Also
\bea
&&\langle 0 |\,C_{p}^\pm (C_{n}^\pm)^\dag  \phi_L (0,\x) \phi_R (0,\x)   C_{n}^\pm (C_{p}^\pm)^\dag \0=\\
&&= \langle 0 |\, (\pm C^3_{p-n}+n\delta_{nm}) \phi_L (0,\x) \phi_R (0,\x) (\pm C^3_{n-p}+n\delta_{nm})\0 =\\
&&=  \langle 0| { \stackrel {\;o}{\phi}}_L  {\stackrel {\;o}{\phi}}_R \0 \langle 0 |\,C_{p}^\pm (C_{n}^\pm)^\dag     C_{n}^\pm (C_{p}^\pm)^\dag \0
\eea
Therefore
\baselineskip15pt
\bea
\hspace{-20pt}\langle\Omega_M^{(2)}  |  \phi_L (0,\x) \phi_R (0,\x)| \Omega_N^{(2)}\rangle_{+\, -}
=\langle\Omega_M^{(2)} | \Omega_N^{(2)}\rangle_{+\, -} \langle 0| { \stackrel {\;o}{\phi}}_L  {\stackrel {\;o}{\phi}}_R \0 \\ \\
-\frac{g^4 \delta_{MN}}{m^2 L^2}\! \langle 0| { \stackrel {\;o}{\phi}}_L  {\stackrel {\;o}{\phi}}_R \0 \sum_{n, p=\half}^{\infty}\frac{2n}
{(2k_{n} +m)^2 (2k_p +m)^2 }\\
\eea
and $ \langle \theta^{(2)}| \phi_L (0,\x) \phi_R (0,\x)  | \theta^{(2)}\rangle$ is
\bea
 &&\langle \theta^{(2)}| \phi_L (0,\x) \phi_R (0,\x)  | \theta^{(2)}\rangle =\sum_{N=-\infty}^{\infty}
 \langle 0| { \stackrel {\;o}{\phi}}_L  {\stackrel {\;o}{\phi}}_R \0 \langle \theta^{(2)}| \theta^{(2)}\rangle\\ \\
&&+
 \frac{g^4}{m^2 L^2}\sum_{N=-\infty}^{\infty} \langle 0| { \stackrel {\;o}{\phi}}_L  {\stackrel {\;o}{\phi}}_R \0
\sum_{p=\half}^{\infty}\Bigg\{ -8m  \sum_{J=1}^{\infty}\frac{-\theta(p-J) -J+1}{(2k_J)^3 (2k_p+m)^2}
\\ \\
&&\hspace{20pt}+\sum_{J=1}^{\infty}
8m \sum_{J=1}^{\infty}\frac{1 }{(2k_p+m)^2 (2k_p +2k_J + m )^3}\\ \\
&&\hspace{20pt}+8m\sum_{J=1}^{\infty}\frac{\theta(J-p)}{(2k_J )^2 (-2k_p +2k_J +m )^2 }
\Bigg(\!\frac{1}
{2k_J }+\frac{1}{-2k_p +2k_J +m  }
\Bigg)\\ \\
&& \hspace{20pt}- \frac{1}{4(k_{p}+m)^2 (2k_p+m)^2}-\sum_{p,n=\half}^{\infty}\! \frac{2p-2 +2p\delta_{n,p}}
{(2k_p  +m)^2 (2k_n +m)^2}\\ \\
&&\hspace{20pt}-\sum_{n, p=\half}^{\infty}\frac{2n}
{(2k_{n} +m)^2 (2k_p +m)^2 }
\Bigg\}
\eea
\lines
It is easy to see that $ \langle \theta^{(1)}| \phi_L (0,\x) \phi_R (0,\x)  | \theta^{(2)}\rangle= 0$ .\\ \\
To normalize the real field contribution we  write
\bea
&&\frac{\langle \theta | \phi_L (0,\x) \phi_R (0,\x) |\theta \rangle}{\langle \theta | \theta \rangle}\simeq
\langle 0| { \stackrel {\;o}{\phi}}_L  {\stackrel {\;o}{\phi}}_R \0 \\
&&+
\left(\sum_{N=-\infty}^{\infty}1 \right)^{-1} \Bigg\{
\langle \theta^{(1)}| \phi_L (0,\x) \phi_R (0,\x)  | \theta^{(1)}\rangle
-\langle 0| { \stackrel {\;o}{\phi}}_L  {\stackrel {\;o}{\phi}}_R \0 \langle \theta^{(1)}| \theta^{(1)}\rangle\Bigg\}\\
&&-\left(\sum_{N=-\infty}^{\infty}1 \right)^{-2} \langle \theta^{(1)}| \theta^{(1)}\rangle
\Bigg[ \langle \theta^{(1)}| \phi_L (0,\x) \phi_R (0,\x) | \theta^{(1)}\rangle
-\langle 0| { \stackrel {\;o}{\phi}}_L  {\stackrel {\;o}{\phi}}_R \0
 \langle \theta^{(1)}| \theta^{(1)}\rangle
\Bigg] \\
&&+\left(\sum_{N=-\infty}^{\infty}1 \right)^{-1}  \Bigg\{
 \langle \theta^{(2)}| \phi_L (0,\x) \phi_R (0,\x)  | \theta^{(2)}\rangle
-\langle 0| { \stackrel {\;o}{\phi}}_L  {\stackrel {\;o}{\phi}}_R \0 \langle \theta^{(2)}| \theta^{(2)}\rangle\Bigg\}
\eea

so that

\baselineskip18pt
\bea
&&\frac{\langle \theta | \phi_L (0,\x) \phi_R (0,\x) |\theta \rangle}{\langle \theta | \theta \rangle}\simeq
 \langle 0| { \stackrel {\;o}{\phi}}_L  {\stackrel {\;o}{\phi}}_R \0   -2\langle 0|{\stackrel {\;o}{\phi}}_L  {\stackrel {\;o}{\phi}}_R \0
 \frac{g^2}{mL}  \sum_{n=\half}^{\infty}\frac{1}{(2k_n+m)^2} \\ \\
&&+\frac{g^4}{m^2 L^2 }\langle 0|{\stackrel {\;o}{\phi}}_L  {\stackrel {\;o}{\phi}}_R \0\Bigg\{\sum_{J=1}^{\infty}\left( \frac{-4mJ}{(2k_J^3}\right)
+\sum_{n=\half}^{\infty}\frac{2 n}{(2k_{n}+m)^2}\Bigg\}
  \sum_{p=\half}^{\infty}\frac{2}{(2k_p+m)^2} \\ \\
&&+
 \frac{g^4}{m^2 L^2} \langle 0| { \stackrel {\;o}{\phi}}_L  {\stackrel {\;o}{\phi}}_R \0
\sum_{p=\half}^{\infty}\Bigg\{ -8m  \sum_{J=1}^{\infty}\frac{-\theta(p-J) -J+1}{(2k_J)^3 (2k_p+m)^2}
\\ \\
&&\hspace{20pt}+
8m \sum_{J=1}^{\infty}\frac{1 }{(2k_p+m)^2 (2k_p +2k_J + m )^3}\\ \\
&&\hspace{20pt}+8m\sum_{J=1}^{\infty}\frac{\theta(J-p)}{(2k_J )^2 (-2k_p +2k_J +m )^2 }
\Bigg(\!\frac{1}
{2k_J }+\frac{1}{-2k_p +2k_J +m  }
\Bigg)\\ \\
&& \hspace{20pt}- \frac{1}{4(k_{p}+m)^2 (2k_p+m)^2}-\sum_{n=\half}^{\infty}\! \frac{2p-2 +2p\delta_{n,p}}
{(2k_p  +m)^2 (2k_n +m)^2}\\ \\
&&\hspace{20pt}-\sum_{n=\half}^{\infty}\frac{2n}
{(2k_{n} +m)^2 (2k_p +m)^2 }\Bigg\}
\\ \\
&=& \langle 0| { \stackrel {\;o}{\phi}}_L  {\stackrel {\;o}{\phi}}_R \0  -2\langle 0|{\stackrel {\;o}{\phi}}_L  {\stackrel {\;o}{\phi}}_R \0
 \frac{g^2}{mL}  \sum_{n=\half}^{\infty}\frac{1}{(2k_n+m)^2} \\ \\
&&+
 \frac{g^4}{m^2 L^2} \langle 0| { \stackrel {\;o}{\phi}}_L  {\stackrel {\;o}{\phi}}_R \0
\sum_{p=\half}^{\infty}\Bigg\{ -8m  \sum_{J=1}^{\infty}\frac{1}{(2k_{J+p-\half})^3 (2k_p+m)^2}
\\ \\
&&\hspace{20pt}+
8m \sum_{J=1}^{\infty}\frac{1 }{(2k_p+m)^2 (2k_p +2k_J + m )^3}\\ \\
&&\hspace{20pt}+8m\sum_{J=1}^{\infty}\frac{1}{(2k_{J+p-\half} )^2 (2k_{J-\half} +m )^2 }
\Bigg(\!\frac{1}
{2k_{J+p-\half} }+\frac{1}{2k_{J-\half} +m  }
\Bigg)\\ \\
&& \hspace{20pt}- \frac{1}{4(k_{p}+m)^2 (2k_p+m)^2}-\sum_{n=\half}^{\infty}\! \frac{-2 +2p\delta_{n}}
{(2k_p  +m)^2 (2k_n +m)^2}\Bigg\}
\eea
\lines
Finally, for the contribution of the real field we get
\baselineskip18pt
\bea
&&\frac{\langle \theta | \phi_L (0,\x) \phi_R (0,\x) |\theta \rangle}{\langle \theta | \theta \rangle}\simeq
\langle 0| { \stackrel {\;o}{\phi}}_L  {\stackrel {\;o}{\phi}}_R \0  \Bigg\{ 1 -2 \frac{g^2}{mL}  \sum_{n=\half}^{\infty}\frac{1}{(2k_n+m)^2} \\ \\
&&\hspace{40pt}+
 \frac{g^4}{m^2 L^2}  \Bigg[
8m\sum_{p=\half}^{\infty} \sum_{J=1}^{\infty}\frac{1 }{(2k_p+m)^2 (2k_p +2k_J + m )^3}\\ \\
&&\hspace{60pt}+8m\sum_{p=\half}^{\infty}\sum_{J=1}^{\infty}\frac{1}{(2k_{J+p-\half} )^2 (2k_{J-\half} +m )^3 }
\\ \\
&& \hspace{60pt}- \sum_{p=\half}^{\infty}\frac{1}{4(k_{p}+m)^2 (2k_p+m)^2}-\sum_{p,n=\half}^{\infty}\! \frac{-2 +2p\delta_{n,p}}
{(2k_p  +m)^2 (2k_n +m)^2}\Bigg] \Bigg\}
\eea
\lines
and it  is not hard to verify that it goes to zero in large-L limit,  as one might have expected considering
that this contribution is an artifact of the breaking of chiral invariance in free theory due to the zero modes. 

\chapter{CONCLUSIONS}

The following questions arose in the development of our study:
\begin{enumerate}
\item{ What is the algebra of the Lagrange multiplier fields and how do they determine the
physical subspace?\cite{bassetto}}
\item{ Can the symmetries of the classical Lagrangian be implemented in the quantized theory?
}
\item{Can one formulate the problem in such a way that the vacuum and the condensate can
be calculated using perturbation theory?}
\item{How many vacuum states does this model possess?}
\item{Can one estimate the condensate for large values of L, at least for some range of couplings?}
\end{enumerate}
We have given the following answers:
\begin{enumerate}
\item{The commutators relating the different colour components of the Lagrange multiplier 
have been calculated and the subsidiary condition has been clearly expressed in terms of 
Fock operators.}
\item{The existence of an operator implementing the residual gauge symmetry of the classical theory is not compatible with a representation of the gauge fields  in the indefinite-metric Fock
space of non-interacting gauge bosons. The residual symmetries can be implemented if $A^\pm$ are quantized in a positive-metric Fock space.}
\item{ While in the Schwinger model the nonperturbative contribution to the vacuum is given by the zero mode of
$A^3$ only, in the non-abelian case considered here the different symmetry properties of the free and interacting gauge theories do not allow a truly perturbative approach for the nonzero modes of the gauge field. On the other hand, if we pay our respects to the principle of gauge invariance when removing the ultraviolet divergences from Fermi operator products, we obtain, besides the normal-ordering prescription of the free fermion theory, an extra term reflecting the gauge symmetry of the interacting fermion. Including this term in the unperturbed Hamiltonian
allows us to evaluate the vacuum using perturbation theory, while the state we perturb about, together with the residual gauge transformations that give rise to the non-trivial structure, bring
along nonperturbative effects. 

The vacuum obtained in this way satisfies the subsidiary condition to the relevant order.
}
\item{In accordance with previous studies we found two vacuum states  \cite{witten}\cite{smilga}\cite{lenz}\cite{paper}.}
\item{We have a given a rough estimate of the condensate in the large-L limit. It would be interesting to be able to compare this result with a lattice calculation of the condensate.}
\end{enumerate}

It might be worthwhile to consider the theory with colour-symmetric boundary conditions, that is with
all three components of the gauge field periodic and all three components of the Fermi fields antiperiodic.
That would eliminate the undesired presence of the zero mode of the real Fermi field,  responsible for a nonzero condensate even in free theory. The treatment of the gauge 
fields would presumably be more complicated but it would be interesting to study the problem of their quantization with  a colour-symmetric approach. 

It might also be interesting to study the one-particle states perturbatively and, in particular,
verify whether the expected relation $M^2 \sim \mu \langle\theta | {\rm Tr}\bar{\Psi}\Psi | \theta \rangle$ ( where $\mu$ is the bare fermion mass) holds or not. 

\appendix
\include{sm}

\renewcommand{\bibname}{REFERENCES}

\end{document}